\documentclass{JHEP3}
\usepackage{amsfonts,amsbsy,latexsym,amssymb,amscd,amstext}
\usepackage{epsfig}
\usepackage{graphicx}
\newcommand{\be}{\begin{equation}}
\newcommand{\ee}{\end{equation}}
\newcommand{\ba}{\begin{array}}
\newcommand{\ea}{\end{array}}
\newcommand{\baa}{\begin{array}}
\newcommand{\eaa}{\end{array}}
\newcommand{\bea}{\begin{eqnarray}}
\newcommand{\eea}{\end{eqnarray}}
\newcommand{\half}{{1\over2}}

\newcommand{\sig}{\sigma}
\newcommand{\bsig}{\overline{\sigma}}

\newcommand{\OMF}{\omega}
\newcommand{\omeg}{\OMF}
\newcommand{\OME}{{\cal W}}

\newcommand{\EV}{e}
\newcommand{\ETV}{\widetilde{e}}
\newcommand{\II}{\mathbf{I}}
\newcommand{\I}{I\!\!I}

\newcommand{\Q}{Q}
\newcommand{\st}{^{\star}}
\newcommand{\U}{{\cal U}}
\newcommand{\zp}{z+\frac{1}{2}}
\newcommand{\pf}{{\boldsymbol{\psi}}}
\newcommand{\Texp}{{\rm Texp}}

\title{Adjoint fermion zero-modes for SU(N) calorons}

\author{Margarita Garc\'{\i}a P\'erez $^{a}$, Antonio Gonz\'alez-Arroyo $^{a,b}$
and Alfonso Sastre $^{a,b}$ \\
  $^a$ Instituto de F\'{\i}sica Te\'orica UAM/CSIC,  C-XVI \\
  $^b$ Departamento de F\'{\i}sica Te\'orica, C-XI \\
       Universidad Aut\'onoma de Madrid, E-28049--Madrid, Spain \\

E-mail: \email{margarita.garcia@uam.es, antonio.gonzalez-arroyo@uam.es, alfonso.sastre@uam.es
  }}

\abstract{
We derive analytic formulas for the zero-modes of the
Dirac equation in the adjoint representation in the background field
of $Q=1$ SU(N) calorons. Solutions with various boundary conditions
are obtained, including the physically most relevant cases of periodic
and anti-periodic ones. The latter are essential ingredients in a
semi-classical treatment of finite temperature supersymmetric
Yang-Mills theory. A detailed discussion of adjoint zero-modes in several
other contexts is also presented.
}


\preprint{IFT-UAM/CSIC-09-25\\ FTUAM-2009-07}

\begin{document}

{\vskip 1cm}

\section{Introduction}
\label{s.intro}

In recent years there has been renewed interest in the study of 
gauge theories on $R^3 \times S^1$ in situations that could be amenable to 
semi-classical treatment. This refers not only to thermal QCD but also to other 
QCD-like theories seemingly exhibiting exotic phase diagrams which depend on  
the representation and the boundary conditions for the fermionic fields. 
As an example, arguments are given to support the idea that adjoint QCD remains
confined for any value of the $S^1$ cycle when fermions are endowed with 
periodic boundary conditions~\cite{hollowood}-\cite{Cossu:2009sq}. 
This temperature independence, interpreted in the spirit of the Eguchi-Kawai 
reduction \cite{EK}, could allow for an analytical approach to confinement 
in this theory. Here, as well as in ordinary thermal QCD or in supersymmetric 
gluodynamics, the relevant semi-classical objects are BPS monopoles and finite 
temperature instantons (calorons). Remarkably, these two objects cannot be 
considered as independent entities. For non-trivial values of the Polyakov loop 
at infinity and high temperatures, calorons split into N (for SU(N)) constituent 
BPS monopoles~\cite{kvanbaal1}-\cite{lee2}, with the well 
known Harrington-Shepard caloron~\cite{HS} corresponding to the case in which 
N-1 of them become massless.  
Non-trivial holonomy calorons provide therefore a particular realisation of 
the idea of instanton quarks~\cite{quarks}-\cite{tonyfrac}, instrumental to many 
semi-classical attempts at explaining confinement in QCD. Constituent monopoles
within calorons have been successfully used for obtaining, yet another,
weak coupling computation of the gluino condensate in 4D ${\cal N}=1$ 
supersymmetric Yang-Mills theory~\cite{hollowood}. The success of this approach 
relies on the relevance of non-trivial holonomy configurations when the theory 
is defined on $R^3\times S^1$ with periodic boundary conditions for 
the adjoint fermions. On the contrary, in thermal QCD an old one-loop result 
by Gross, Pisarski and Yaffe~\cite{Gross:1980br} indicates the 
thermodynamic suppression of such configurations and 
consequently seems to point towards the irrelevance of non-trivial holonomy calorons  
for finite temperature QCD. 
Nevertheless, this result has recently been challenged on the basis of 
non-perturbative calculations~\cite{Diakonov} and lattice 
simulations~\cite{latcal}. Although 
the issue is far from being settled, see e.g.~\cite{bruckmann},  
it has revived the interest in applying semi-classical techniques to
the analysis of QCD in a thermal set up. Last but nor least, non-trivial 
holonomy calorons could as well be of relevance for settling 
issues at stake such as that of supersymmetry breaking at finite 
temperature~\cite{Das:1978rx}-\cite{Buchholz:1997mf}, 
or the previously discussed phase structure of adjoint QCD or other QCD-like 
gauge theories.

An essential ingredient in the semi-classical analysis of these theories are
the zero-modes of the Dirac operator in the background of the caloron gauge
field. Those in the fundamental representation have been previously obtained, 
first for $Q=1$ SU(2) calorons~\cite{fundamental_su2}, and later generalised 
for  SU(N)~\cite{fundamental_sun}. 
Remarkably they are supported on a single BPS monopole selected through the
choice of boundary conditions for the fermions on the thermal cycle. 
In the present paper we will focus on the derivation of the zero-modes of the
Dirac operator in the adjoint representation, relevant for supersymmetric 
gluodynamics as well as for adjoint QCD. The corresponding expressions for 
$Q=1$ SU(2) calorons have been obtained in Refs.~\cite{gluino_SU2} 
and~\cite{gluino_SU2_ap} for fermions with periodic and anti-periodic boundary 
conditions in time. Here we will generalise those formulas to gauge group SU(N).

Let us finally mention that adjoint zero-modes have also interest in the context 
of using eigenstates of the Dirac operator to trace topological structures in the 
QCD vacuum. In particular, the use of the so called supersymmetric zero-mode, whose
density profile matches that of the gauge action density, has been advocated 
by one of the present authors in Ref.~\cite{tonyadj}. First successful
tests of the proposal were presented both there and in Ref.~\cite{tonyadj2}.

The paper is organised as follows. Sections~\ref{s.adjoint} to~\ref{s.nahm} deal 
with the general construction of the adjoint zero-modes (the reader interested only in
the specific solutions for the calorons could go directly to 
section~\ref{s.deformations}). In section~\ref{s.adjoint} we introduce some
basic properties of adjoint zero-modes and discuss their relation to self-dual 
deformations of the gauge field. The derivation of the zero-modes
relies heavily on the ADHM formulation for self-dual solutions of the classical 
equations of motion which is revised in section~\ref{s.adhm}. There
we also describe the general construction of the adjoint modes as deformations of 
the ADHM gauge field. The ADHM formalism has been devised to provide self-dual
gauge fields on the sphere or $R^4$. To deal with more general manifolds one can make
use of the Nahm-ADHM formalism. A general introduction is presented in 
section~\ref{s.nahm}, starting with the simplest case of the 4-torus. 
We discuss there how the Nahm transform provides a mapping between the original 
self-dual gauge field and 
a self-dual gauge field living on the dual torus. We show how the 
adjoint zero-modes of the original Dirac operator are mapped onto those of 
the dual Dirac operator. In some cases the latter is 
much simpler than the original one and the Dirac equation can be easily
solved. This constitutes the basis of the construction. It provides adjoint
zero-modes with periodic boundary conditions in time. Anti-periodic
modes are derived by making use of the so called replica trick which is described
in section~\ref{s.replica}. The idea is to replicate the original thermal cycle
and solve for the Dirac equation on the replicated torus. 
The periodic zero-modes thus obtained can be easily decomposed
into periodic and anti-periodic zero-modes for the original, unreplicated torus.
The application of these ideas to the case of calorons is presented in 
section~\ref{s.calorons}. Explicit solutions for periodic and anti-periodic zero-modes
are derived in section~\ref{s.deformations} where we also discuss
the generalisation of our formulas to other periodicity conditions for the 
adjoint fermions. Section~\ref{s.prop} presents some  
illustrative examples and  discusses in detail some general 
properties of the solutions such as periodicity, normalisability and 
concordance with the expected number of zero-modes as dictated by the index 
theorem. Finally, we end in section~\ref{s.concl} with a brief summary of the results 
and a recollection of possible applications. 
The most technical details concerning the derivation of the adjoint 
zero-modes in sections~\ref{s.calorons} and~\ref{s.deformations} are deferred 
to an Appendix.

\section{Generalities about adjoint zero-modes}
\label{s.adjoint}

Our goal is that of studying euclidean adjoint fermionic zero-modes in the
background field of SU(N) calorons. In this section we will revise a
set of basic facts about adjoint modes. They are solutions of the
equation
\be
\gamma_\mu D_\mu  \psi=0\,, 
\ee
where the covariant derivative $D_\mu= \partial_\mu -i A_\mu$ is
expressed in terms  of the gauge potential $A_\mu$. This equation 
factorises into two Weyl equations for the 
two chiral components. The equation for the 
left-handed (or positive chirality $\gamma_5 \psi= \psi$) component is 
\be
\bsig_\mu D_\mu  \psi_+=0\,,
\ee
where the Weyl matrices are $\bsig_\mu =(\II, i \vec{\tau})$, and
$\tau_i$ are the  Pauli matrices. Multiplication formulas with 
the adjoint Weyl matrices ($\sigma_\mu =(\II, -i \vec{\tau})$)
\bea
\sigma_\mu \bsig_\nu = \eta^{\mu \nu}_\alpha \bsig_\alpha \\
\bsig_\mu \sigma_\nu = \bar{\eta}^{\mu \nu}_\alpha \sigma_\alpha
\eea
define the real `t Hooft symbols $\eta^{\mu \nu}_\alpha$. 

All the former equations are valid for all gauge groups and any
representation. This  dependence is hidden in the form of 
\be
A_\mu = \sum_a A_\mu^a T^{(R)}_a \,,
\ee
where $a$ runs over the generators of the group and the matrices
$T^{(R)}_a$ are their expression in representation $R$. In the adjoint
representation the generators  $(T_a)_{b c}=-i f_{ a b c}$ are expressed in terms 
of the structure constants of the group $f_{ a b c}$. Thus,  the
covariant derivative $D_\mu^{a b}$ is real. This, together with the
definition of $\bsig_\mu$ implies that given one solution $\psi_+$ one can
construct a linearly independent one $\psi_{+ C}$. In components 
\be
\psi^a_+=\pmatrix{\kappa^a_1\cr \kappa^a_2} \longrightarrow
\psi_{+ C}^{a }=\pmatrix{-\overline{\kappa}_2^a\cr
\overline{\kappa}_1^a} = \sigma_2 \overline{\psi}^a_+ \,,
\ee
where the bar denotes complex conjugation. Notice that the two vectors 
form the two columns of a $2\times 2$ matrix $\Psi^a=(\psi^a_+,
\psi_{+ C}^{a})$. This type of matrices define a representation of the
field of quaternions, and from now on we will make no distinction
among both sets. In this notation any linear combination of $\psi_+^a$
and $\psi_{+ C}^{a}$ can be simply represented by the multiplication by
an arbitrary constant quaternion $Q$ on the right
\be
\Psi^a \longrightarrow \Psi^a Q\,.
\ee
The density is the same for all vectors  in  this space.

Self-dual gauge fields $A_\mu^a$ are those for which the field strength 
$F_{\mu \nu}=i [D_\mu, D_\nu]$ satisfies 
\be
\bar{\eta}^{\mu \nu}_i F_{\mu \nu}=0\,,
\ee
for all $i=1,2,3$. In analysing the solutions of the Weyl equations for 
self-dual gauge fields, it is important to consider the possible
zero-modes $\kappa(x)$ of the  covariant Laplacian $D_\mu D_\mu$. It is clear that
the only zero-modes are functions that are simultaneously annihilated 
by all the $D_\mu$ ($D_\mu\kappa(x)=0$). This implies that 
\be 
[ F_{\mu \nu}(x), \kappa(x)]=0\,.
\ee
Except for very particular abelian-like gauge fields, this equation will
have no solution other than the trivial one $\kappa(x)=0$. From this
result it is easy to conclude that the right-handed Weyl equation will
have no zero-modes. From here  one can make use of the index
theorem~\cite{atiyah} to extract the number of zero-modes of the
left-handed Weyl equation. Notice also, that the absence of zero-modes 
would guarantee the invertibility of the covariant Laplacian.  

Another  very important result concerning adjoint zero-modes of self-dual
gauge fields is to realise the connection among self-dual deformations
and solutions of the left-handed Weyl equation. Given a self-dual
deformation $\delta A_\mu$, one can choose a representative within the
gauge trajectory satisfying the background gauge condition
\be
D_\mu \delta A_\mu=0\,.
\ee
The existence and  uniqueness (up to non-trivial gauge transformations
not connected with the identity)  of this representative relies on the
invertibility of the covariant Laplacian. Now, it is simple to see 
that the condition of self-duality, plus the background gauge condition 
amount to 
\be
\label{Weyl_eq}
\bsig_\mu D_\mu  \Psi=0\,,
\ee
where $\Psi=\delta A_\mu \sigma_\mu$. The relation works both ways, so
that there is a one-to-one correspondence between self-dual
deformations and left-handed zero-modes. This relation combined with 
the index theorem was used in Refs.~\cite{schwarz} to count the (dimensionality) 
number of moduli parameters of the manifold of self-dual gauge fields.

Notice that  if we transform one solution of Eq.~(\ref{Weyl_eq}) by
right  multiplication by a quaternion $\Psi \longrightarrow \Psi Q$
the corresponding deformation transforms as
\be
\label{transform}
\delta A_\mu \longrightarrow q_\nu \bar{\eta}^{\mu \rho}_\nu \delta
A_\rho=  q_\nu \eta^{\nu \rho}_\mu \delta A_\rho\,,
\ee
where $Q= q_\mu \bsig_\mu$. 

A particular class  of deformations is associated to symmetries
(conformal transformations of the metric) that do not leave the
solution invariant. Particularly interesting is the case of
translations. Translation in the $\rho$ direction gives a deformation 
$\partial_\rho A_\mu$, which in general does not satisfy the
background field gauge. However, if we write 
\be
\delta A_\mu = \partial_\rho A_\mu + D_\mu(-A_\rho) =F_{\rho \mu}\,,
\ee
the gauge condition is satisfied. This gives rise to  the famous
supersymmetric zero-mode present in conformally flat manifolds.

\section{ADHM construction  of adjoint zero-modes}
\label{s.adhm}
In this section we will revise the main formulas of the ADHM
construction~\cite{adhm} of self-dual gauge fields, and in particular the 
expression of adjoint zero-modes. We will restrict ourselves 
to the case of gauge group  SU(N).

 The ADHM expression for an SU(N) gauge field with topological charge
 $\Q$ is  given by:
\begin{equation}
A_\mu(x)= i N^\dagger \partial_\mu N\,,
\end{equation}
where $N$ is a $(2 |\Q|+N) \times N$  matrix satisfying  $N^\dagger
N =\I_N$. This vector is annihilated by the $(2|\Q|+N)\times 2 |\Q|$ matrix $M$:
\begin{equation}
 N^\dagger M= 0\,.
\end{equation}
The second index of the matrix $M$ is split into a index running from
1 to $ |\Q|$ and  an spinorial index $\alpha=1,2$. This matrix  has the form:
\begin{equation}
M= A + B \hat{x} \,,
 \end{equation}
where $A$ and $B$ are constant matrices of the same type as $M$, and 
$\hat{x}=x_\mu \sigma_\mu$ is an $x$-dependent quaternion which acts
on the spinorial index.  

One can show that $A_\mu$ constructed in such a way is self-dual
provided the matrix $R=M^\dagger M$ is 
invertible and commutes with the $\sigma_\mu$. Let us find out the implications 
of this condition for the matrices $A$ and $B$. We have that both 
$B^\dagger B$ and $A^\dagger A$ must commute with $\sigma_\mu$, and the matrix 
$B^\dagger A$ can be written as $S_\mu \sigma_\mu$ with $S_\mu$ hermitian.
To express this equation in a more compact way we might introduce some  notation,
which will turn out to play an important role in
what follows. 

Given a  $2m\times 2m$ matrix $S= S_\mu \sigma_\mu$, where 
$S_\mu$ are  $m\times m$ complex  matrices, we denote 
\[ \overline{S}= S_\mu \overline{\sigma}_\mu \]
The overline operation combined with the hermitian conjugate is
a particularly interesting operation (star-operation):
\be 
\label{eq:star_op}
S\st \equiv \overline{S^\dagger} =S_\mu^\dagger  \sigma_\mu\,,
\ee
which will turn out to be very useful in what follows. 
In particular, it satisfies 
\begin{eqnarray}
(S Q)\st = S\st Q  \,,\\
(Q S)\st = Q S\st  \,,
\end{eqnarray}
for any  square matrix with quaternionic entries $S$,
and an arbitrary quaternion $Q$. From here it follows, in the particular 
case of $2\times 2$ matrices (m=1), that $(TS)\st=T\st S\st$.
The subspace of quaternions is characterised by the condition $Q\st =Q$. 
Furthermore, if the  $2\times 2$ matrix is of dyadic form $S=\xi
\zeta^\dagger$, then 
\be
S\st  = \xi_C \zeta_C^\dagger\,,
\ee
where the subscript $C$ stands for the ``charge-conjugate'' introduced
in the last section.

Using the previous notation,  we can rewrite the ADHM conditions as 
\be
(B^\dagger A)\st = B^\dagger A \,,
\ee
and  
\be
\label{eq:BM_rel}
(M^\dagger M)\st  = M^\dagger M \,.
\ee
This type of equations will appear several times along the paper. 
If we write as before  $S= S_\mu \sigma_\mu$, the condition $S\st=S$
amounts to $S_\mu=S_\mu^\dagger$.

Another relation which plays an important role in the following
derivations is that, given  an arbitrary  $2|\Q|$ square matrix   $S=S_\mu
\sigma_\mu$, one obtains 
\[ \sigma_\mu S \sigma_\mu = \overline{\sigma}_\mu S \overline{\sigma}_\mu= -2
\overline{S} \,.\] 
Using this relation we can show that 
\begin{equation}
\label{relone}
(\partial_\mu(M^\dagger M) + (\partial_\mu M^\dagger)M)\overline{\sigma}_\mu =0\,.
\end{equation}
This formula will be important later. The proof is simple. We plug in 
the expression for $M$ and perform the derivatives and obtain:
$$ 2 \overline{\sigma}_\mu (B^\dagger M) \overline{\sigma}_\mu + M^\dagger B
\sigma_\mu \overline{\sigma}_\mu \,.$$
Using $\sigma_\mu \overline{\sigma}_\mu=4$ and Eq.~(\ref{eq:BM_rel}) one
easily verifies that it vanishes. 

Let us now proceed to study the zero-modes of the Dirac operator in terms of the
ADHM quantities introduced above.
The expression for the normalised zero-modes in the fundamental representation,
derived in Ref.~\cite{fundzm}, is given by: 
\begin{eqnarray}
 \pf = \frac{1}{\pi}N^\dagger B R^{-1}\tau_2 \, (B^\dagger B)^\half.
\label{eq:fundzm}
\end{eqnarray}
The quantity $\pf$ is a $N\times 2 Q$ matrix, with  the first index
being the ordinary colour index. The second index decomposes into an
spinorial one, and another labelling the $Q$ different linearly
independent solutions, in agreement with the index theorem. 

The adjoint zero-modes can be derived  by exploiting their
relation with  self-dual deformations. These can be obtained by
deforming the matrix $M$ and the vector  $N$ of the ADHM
construction~\cite{corrigan}. It
takes the form
 \begin{eqnarray}
 \label{Adeform}
 \delta A_\mu &=& i \delta N^\dagger \partial_\mu N +  i N^\dagger
\partial_\mu   \delta N \,,\\
\label{DNM}
\delta N^\dagger M &=& - N^\dagger \delta M\,,\\
\label{Dnorm}
\delta N^\dagger N + N^\dagger \delta N &=& 0 \,.
  \end{eqnarray}
Given a deformation $\delta M$, the variation of $\delta N$ is determined
from Eq.~(\ref{DNM}). The solution is not unique since the left-kernel of $M$ 
is non-trivial. This allows us to choose a solution  $\widetilde{\delta} N$ satisfying
\begin{equation}
\widetilde{\delta} N^\dagger N=0\,,
\end{equation}
which is also compatible with Eq.~(\ref{Dnorm}). A general solution
would then be 
\begin{equation}
\label{general_DN}
\delta N = \widetilde{\delta} N - i N \omega\,,
\end{equation}
with $w$ a hermitian matrix. Using the form of the 
orthogonal projector to the kernel of M:
\begin{equation}
P= M R^{-1} M^\dagger\,,
\end{equation}
we might write 
\begin{equation}
\widetilde{\delta} N^\dagger = \widetilde{\delta} N^\dagger P= - N^\dagger \delta M  R^{-1} M^\dagger\,.
\end{equation}
Plugging this expression into Eq.~(\ref{Adeform}) we arrive at:
\begin{equation}
\label{deformA}
 \widetilde{\delta} A_\mu = i N^\dagger \left( \delta M  R^{-1} \partial_\mu M^\dagger 
 - \partial_\mu M  R^{-1} \delta M^\dagger \right) N\equiv i N^\dagger (H_\mu
-H^\dagger_\mu)N \,,
\end{equation}
where we have moved the partial derivatives around in an obvious way. 
If instead of this solution we would have used the general one
Eq.~(\ref{general_DN}), the
result would be 
\begin{equation}
\delta A_\mu = \widetilde{\delta} A_\mu + D_\mu \omega \,,
\end{equation}
a gauge transform of the previous one. The quantity $H_\mu$ appearing in 
Eq.~(\ref{deformA}) can be written as 
\be
H_\mu= \delta M \bsig_\mu R^{-1} B^\dagger\,.
\ee
From here it is clear, by using the formula 
\be
\bsig_\nu \bsig_\mu = \eta^{\nu \rho}_\mu \bsig_\rho \,,
\ee
that multiplying $\delta M$ on the right by an
arbitrary quaternion $Q=q_\nu \bsig_\nu$ transforms the deformation as
in Eq.~(\ref{transform}), namely  another one corresponding to 
 the 2-dimensional space of adjoint zero-modes mentioned in the previous section.

We have obtained the expression of the self-dual deformation in terms
of  $\delta M$, the deformation of the $M$ matrix.
Notice that these deformations are not arbitrary, but should preserve
the form $\delta M = \delta A + \delta B \hat{x}$ and preserve the
invertibility (trivial) and the condition of commuting with
quaternions. The latter condition can be re-expressed as
\be
T-T\st = \overline{T-T\st}\,,
\label{eq:com_cond}
\ee
where $T=M^\dagger \delta M$. 

Now we should verify whether Eq.~(\ref{deformA}) satisfies the
background gauge condition. Acting with the covariant derivative and
after some algebra we obtain 
\begin{equation}
\label{eq:back_gauge}
D_\mu \widetilde{\delta} A_\mu =i4 N^\dagger\left( \, \delta B\, R^{-1} B^\dagger
-  B R^{-1}
\delta B^\dagger - B R^{-1} (M^\dagger \delta M - \delta M^\dagger M)_0
R^{-1} B^\dagger \right) N\,,
\end{equation}
where the sub-index $0$ amounts to extracting the part commuting with
quaternions: 
\begin{equation}
\label{eq:Szero}
(S)_0\equiv \frac{1}{2}(S+\overline{S})= \frac{1}{4} \sigma_\mu S
\overline{\sigma}_\mu\,.
\end{equation}

We leave to the reader the details of reaching to
Eq.~(\ref{eq:back_gauge}). As an aid,  we mention that we have used the 
form of the matrices $M$ and $\delta M$, the previous relations
Eq.~(\ref{relone}) and Eq.~(\ref{eq:Szero}), and the following formula for 
the covariant derivative of an object of the form $\omega=N^\dagger H
N$:
\begin{eqnarray}
\nonumber
&&D_\mu \omega = N^\dagger \partial_\mu \left( (\I-P) H (\I-P) \right) N =\\
&&N^\dagger \left( \partial_\mu  H  -\partial_\mu M  R^{-1}  M^\dagger H
- H
M  R^{-1} \partial_\mu  M^\dagger \right) N\,.
\end{eqnarray}

From Eq.~(\ref{eq:back_gauge}), we conclude that the covariant
derivative vanishes provided:
\begin{eqnarray}
\label{eq:first_condition}
\delta B\, R^{-1} B^\dagger -  B R^{-1} \delta B^\dagger =0 \,,\\
(M^\dagger \delta M - \delta M^\dagger M)_0=0\,.
\end{eqnarray}
The last equation can be combined with the condition on the
deformations $\delta M$, Eq.~(\ref{eq:com_cond}), to conclude:
\be
\label{eq:main_cond}
 T= T\st\,,
\ee
where $T=M^\dagger \delta M$.

Up to now we have not used the freedom to redefine $M$ without
altering the form of the gauge field  $A_\mu$.  There are two types of 
transformations of this kind.  The first is to replace 
$M \longrightarrow M L$ with invertible,  constant and commuting with 
quaternions $L$. The second
transformation is to modify $M \longrightarrow V M$ with $V$ unitary
and constant.  The same transformation has to be done to $N$ ($N \longrightarrow V N$).
Using these transformations Christ, Weinberg and Stanton \cite{christ}
were able to show that the $M$ matrix can be brought to the following
canonical form: 
\begin{equation}
M=\pmatrix{q^\dagger \cr \widetilde{A}-\hat{x}}\,.
\end{equation}
Then we have
\begin{equation}
N= \pmatrix{\I_N  \cr -u} F^{-1/2}\,,
\end{equation}
where $u$ is given by 
\begin{equation}
u\equiv\widetilde{M}^{-1\dagger} q=B^\dagger N F^{1/2}\,,
\label{eq:defu}
\end{equation}
with $\widetilde{M}=\widetilde{A}-\hat{x}$.
The $N\times N$ matrix
$F$ is fixed by the normalisation condition ($N^\dagger N= \I_N$):
\be
F= \I_N + u^\dagger u\,.
\label{eq:defF}
\ee

We will now give the expression of the gauge field and the vector
potential in this canonical form. Introducing  $\omega = R^{-1}q$, one
can show that 
\begin{equation}
u = \widetilde{M} \omega F\,.
\end{equation}
Another useful expression is 
\be
R^{-1} u= \frac{1}{2} (\hat{\partial}R^{-1})q F =\frac{1}{2}
\hat{\partial}\omega F\,,
\label{eq:defdw}
\ee
where $\hat{\partial} \equiv \sigma_\mu \partial_\mu$.

With this notation the main expressions for the ADHM fields are given
by:
\begin{equation}
F^{1/2}A_\mu(x)F^{1/2}= i u^\dagger \partial_\mu u - i F^{1/2}
\partial_\mu F^{1/2}
\equiv F^{1/2} {\cal A}_\mu F^{1/2} - i F^{1/2} \partial_\mu F^{1/2}\,,
\label{eq:pot}
\end{equation}
where
\begin{equation}
{\cal A}_\mu=  \frac{i}{2} F^{1/2} q^\dagger \bar{\sigma}^\mu
\hat{\partial}\omega F^{1/2}\,.
\end{equation}
The (chromo-)electric field is given by
\begin{equation}
E_i= - F^{-1/2} u^\dagger \tau_i \hat{\partial}\omega F^{1/2}\,,
\label{eq:elecf}
\end{equation}
and the zero-modes of the Dirac operator in the fundamental representation by: 
\begin{eqnarray}
 \pf = \frac{1}{2\pi}F^{1/2}\left(\hat{\partial}\omega\right)^\dagger\tau_2.
\label{eq:fcwszm}
\end{eqnarray}

Now we go back to our  main goal of obtaining the adjoint zero-modes. 
Notice that \mbox{$\delta B=0$} so that the first condition implying the
background gauge Eq.~(\ref{eq:first_condition}) is automatically satisfied.
The form of the self-dual deformation can be obtained by using 
\begin{equation}
\delta M^\dagger \,  N= (\delta q - \delta \widetilde{A}^\dagger u)
F^{-1/2}\,,
\end{equation}
and  replacing it 
in our general formula Eq.~(\ref{deformA}):  
\begin{equation}
\delta A_\mu= \frac{i }{2} F^{-1/2} (\delta q^\dagger - u^\dagger \delta
\widetilde{A)}
\bar{\sigma}^\mu \hat{\partial}\omega F^{1/2} + h.c. \label{defa}\,.
\end{equation}
The condition that this is a self-dual deformation in the background Lorentz gauge,
is simply  Eq.~(\ref{eq:main_cond}) where 
\begin{equation}
T\equiv M^\dagger \delta M=q \delta q^\dagger
+\widetilde{A}^\dagger \delta \widetilde{A}-\bar{x}\delta \widetilde{A}\,.
\end{equation}
The part proportional to $\bar{x}$ must satisfy  the equation by
itself ($\delta \widetilde{A} = (\delta \widetilde{A})\st$).
The remaining part can be rewritten as 
\be
\widetilde{A}^\dagger \delta \widetilde{A} - (\widetilde{A}^\dagger
\delta \widetilde{A})\st = - q \delta q^\dagger + (q \delta
q^\dagger)\st \,.
\ee

\section{Nahm-ADHM formalism}
\label{s.nahm}
The ADHM formalism is valid for gauge fields on the sphere or in $R^4$
with appropriate boundary conditions at infinity. For other
manifolds one can use the extension introduced by Nahm~\cite{nahm}. A
particularly symmetric case is that of the 4-torus $T^4$. Essentially,
one can embed the torus onto $R^4$ and impose the appropriate
periodicity conditions. If we look at the torus configurations  as solutions
in $R^4$, it is clear that their  topological charge $\Q$ would be
infinite. Then one can interpret the corresponding matrix indices of the basic
ADHM quantities, $M= A + B \hat x$ and $N$, as Fourier coefficients of 
functions or operators depending on 4 new {\em dual coordinates} $z_\mu$. 
In particular, the matrix $B$ can be
chosen as minus the identity operator ($B=-\I$) and $A=\hat{D}_\mu 
\sigma_\mu/(2 \pi i)$. The Nahm-dual covariant derivative $\hat{D}_\mu$ 
is given by 
 \be
 \hat{D}_\mu=\frac{\partial}{\partial z_\mu} -i \hat{A}_\mu(z)\,,
 \ee
expressed in terms of the Nahm-dual gauge field $\hat{A}_\mu(z)$. 
The operator $A^\dagger$ becomes simply proportional to the Nahm-dual left-handed
Weyl operator in the fundamental representation, and $N$ is replaced 
by the zero-modes $\hat\pf(z,x)$ of the modified Weyl equation
\be
\bsig_\mu (\hat{D}_\mu - 2 \pi i x_\mu) \hat\pf(z,x)=0 \,.
\ee
With this interpretation all the standard ADHM conditions and formulas 
adopt a simple interpretation. For example, the condition that $R$
commutes with quaternions, leads to the self-duality of the Nahm-dual
gauge field. The Weitzenb\"ock formula then gives   
\be
R=- \frac{ \hat D_\mu \hat D_\mu - 4 \pi i x_\mu \hat D_\mu - 4 \pi^2
x^2}{4 \pi^2} \, ,
\ee
which obviously commutes with quaternions. Its invertibility  
follows from the invertibility of the covariant Laplacian in Nahm-dual space. 

The formulas work  both ways, so that the Nahm-dual field can be
obtained from the original self-dual field by the standard formula 
\be 
\label{eq:nahm_tr}
\hat{A}_\mu(z)= i \int dx\,  \pf^\dagger(x,z) \frac{\partial}{\partial z_\mu}
\pf(x,z)\,,
\ee
in terms of the normalised zero-modes $\pf(x,z)$
of the  modified Weyl equation in the fundamental representation:
\be
\bsig_\mu (D_\mu - 2 \pi i z_\mu) \pf(x,z) = 0 \,.
\label{eq:weylz}
\ee
These zero-modes can also be expressed in terms of Nahm-dual quantities 
by replacing in Eq.~(\ref{eq:fundzm}) the quantities by their equivalents 
\be 
\pf(x,z)= -\frac{e^{2 \pi i zx}}{\pi} \int d^4 z' \, \hat{\pf}^\dagger(z',x)
\hat{G}(z',z;x) \tau_2
\ee
where $\hat{G}(z',z;x)$ is the Green function of the operator $R$. 

In Eq.~(\ref{eq:nahm_tr}) we have omitted the indices labelling the
different solutions of the Weyl equation. The index theorem tells us 
that there are $Q$ linearly independent solutions, implying that 
$\hat{A}_\mu(z)$ is an SU(Q) gauge field. It can be shown that the new
field has topological charge $N$.  Thus, we obtain a mapping between
SU(N) and SU(Q) self-dual gauge fields, which   is an involution~\cite{vanbaalb}. 
The Nahm-dual gauge field lives in the dual torus, parametrised by $z_\mu$,
whose periods are the inverse of those that define the original torus. 

Considering deformations, the aforementioned mapping implies 
$\delta M= - \delta \hat{A}_\mu\sigma_\mu/2 \pi$. Plugging this onto
Eq.~(\ref{eq:main_cond}) we get 
\be
\bar\sigma_\mu\hat D_\mu \hat\Psi = 0\,,
\label{eq:adjD}
\ee
where $\hat\Psi= \sig_\mu \delta \hat{A}_\mu$, which is the left-chirality
adjoint Dirac equation. Thus, adjoint zero-modes map onto adjoint
Nahm-dual zero-modes and viceversa. A explicit formula can be read off
from Eq.~(\ref{deformA}):
\be
\delta A^{i j}_\mu(x) = \frac{i}{2} \int d^4 z\, e^{-2 \pi i zx}\, \pf^{i a}(x,z) \tau_2
\sigma_\mu \bsig_\nu \delta \hat{A}^{a b}_\nu(z) \hat{\pf}^{b j}(z,x)
+ \mbox{\rm h. c.} 
\ee
where colour ($i=1 \ldots N$) and dual-colour ($a=1 \ldots Q$) indices are
explicitly shown and repeated indices summed over (spinorial indices
are not displayed).  

All general properties of the ADHM
construction are preserved, and in particular the fact that   the
mapping operates among the 2-dimensional spaces formed by each mode and
its CP transform.

\subsection{Replicas}
\label{s.replica}

There are a set of interesting relations which arise by embedding a self-dual 
solution on the 4-torus  onto that of a {\it replicated} torus with periods being
multiples of the original one. We call the new solution a replicated
solution or replica. The replicated solution has a topological charge
which is a multiple of the topological charge of the original one 
$Q$. Consider for example replicating the solution in one direction by a
factor $L$, then the topological charge of the replica is $QL$. If we
now take the Nahm  transform of this solution  it corresponds to an 
SU(LQ) self-dual configuration. How does this solution relate to the 
Nahm transform of the original configuration? This question was
addressed an answered in a previous paper by one of the
authors~\cite{AGAn}.  The answer is the following one:
\be
\label{eq:replica}
\hat{A}^R_\mu(z)= {\rm diag}\{\hat{A}_\mu(z),
\hat{A}_\mu(z+\frac{1}{L}), \ldots , \hat{A}_\mu(z+\frac{L-1}{L} )\}\,,
\ee
where the symbol diag constructs an $L\times L $ diagonal matrix with
its arguments.  An important point to take into account is that the Nahm transform 
of the replicated solution lives in a torus (the dual-torus) which is 
a fraction $\frac{1}{L}$ of the size of that of the original one. 
The transition matrices can be read off from
Eq.~(\ref{eq:replica}). In particular, if  $\hat{A}_\mu(z)$ is
strictly periodic with period 1, we have
\be
\hat{A}^R_\mu(z+\frac{1}{L})= P \hat{A}^R_\mu(z) P^\dagger  \label{rept}\,,
\ee
where $P$ is the $L\times L$ `t Hooft matrix ($P_{i j}= \delta_{j\, i+1}$).

Adjoint zero-modes of the Dirac equation in the background field of 
the replicated solutions can be derived once again from self-dual deformations
satisfying Eq.~(\ref{eq:main_cond}). They are connected to deformations of the 
SU(LQ) Nahm transform satisfying the left-chirality adjoint Dirac equation
on the dual-torus, i.e. Eq.~(\ref{eq:adjD}). The adjoint zero-modes derived
in this way are periodic in the replicated torus but  do 
not necessarily satisfy the same boundary conditions on the original one.
If applied to doubly replicated tori ($L=2$), this trick allows to obtain  
adjoint zero-modes which are anti-periodic in one direction. 
This technique was successfully applied in our previous paper~\cite{gluino_SU2_ap}
 to obtain adjoint zero-modes of the Dirac equation in the background of SU(2) calorons.
Here we will extend the derivation to general group SU(N).

\section{Application for SU(N) calorons}
\label{s.calorons}
In this section we will modify the previous formalism to make it valid
for the case of calorons. These are self-dual gauge fields living in 
$R^3 \times S^1$. The non-compact directions introduce modifications
to the  Nahm construction on the torus, which we will now specify. 
If we try to construct the Nahm-dual gauge field by the same formulas as
before, we should study the solutions of the Weyl equation in the fundamental 
representation:
\be
\bsig_\mu (D_\mu - 2 \pi i z_\mu) \pf(x,z) = 0 \,.
\ee
The non-compactness of the spatial directions is helpful in trivialising the 
dependence of $\pf(x,z)$ on $z_i$, which reduces to simple phase factors
of the form $\exp(2\pi i x_i z_i)$. With this, the Nahm-dual gauge field  
$\hat{A}_\mu(z)$ depends on a single variable $z_0$. There is, however, 
a problem that arises at specific values of $z_0$ for which some of the 
zero-modes become non-normalisable. This case depends on how fields
decay at large distances and this is related to the holonomy. 
The latter is characterised by the eigenvalues of the Polyakov loop at 
spatial infinity, i.e.:
\be
P_\infty = \exp\{i \, 2 \pi\, {\rm diag}(Z_1, Z_2, \cdots, Z_N)\} \,
,\quad {\rm with} \, \, 0\le Z_1 \le Z_2 \le \cdots Z_N<1, \, \,
\sum_{a=1}^N Z_a \in \mathcal{Z} \,,
\label{eq:ploop}
\ee
where we have made use of  gauge invariance to bring the Polyakov loop
to diagonal form and order the $Z_a$ in increasing values from 0 to 1.
Indeed, non-normalisable solutions can only occur when $z_0$ coincides with 
one of the $Z_a$. Since the Nahm construction is
local this means that the Nahm-dual fields would be self-dual except
at these isolated points in $z_0$ \cite{kvanbaal1}, \cite{kvanbaal2}. 

\subsection{Nahm data for Q=1 calorons}

The case of $Q=1$ calorons is particularly simple, since then the 
Nahm-dual field $\hat{A}_\mu(z)$ is abelian and, as already mentioned, depends
on a single variable $z_0\equiv z$. It is furthermore periodic with period 1.

In this case, the whole scheme can be fitted into the ADHM formalism by identifying, 
the matrix $\widetilde{A}$, appearing in the Christ-Stanton-Weinberg
canonical form, with the  Weyl-Dirac operator associated to the 
Nahm-dual field:
 \be
 \widetilde{A}=\frac{1}{2 \pi i}\left(\partial_z -i \sigma_\mu
 \hat{A}_\mu(z) \right)\,.
\label{eq:atilde}
 \ee
Since the Nahm-dual gauge field is self-dual except when $z$ equals one of the $Z_a$,
the ADHM condition of commutation of $R$ with the quaternions, 
can be interpreted as the condition of self-duality of $\hat{A}_\mu(z)$ up to
delta function singularities at $Z_a$. These follow from the $q(z)$
having the form:
\be
q(z)= \sum_{b=1}^N \delta(z-Z_b) \, \zeta^b \,,
\ee
where $\zeta^b$ is a $2\times N$ matrix, with spinor index $\alpha$
and SU(N) index $i$, such that $\zeta^b_{\alpha i} \propto \delta_{bi}$.
For the case of SU(N)~\cite{kvanbaal2,kraan} the Nahm-dual gauge field has the 
following form ($\widehat{A}_0=0$):
\be
\widehat{A}_i= -2 \pi \sum_{a=1}^N X_i^a \chi_a(z)\,,
\ee
which is discontinuous and constant at $N$ intervals. The symbol
$\chi_a(z)$ stands for  the characteristic function of the interval $(Z_{a-1}, Z_a)$, 
assumed to be periodic in $z$ with period 1.
As we will see, each interval is associated to one of the $N$ 
constituent monopoles of the caloron, and  $\vec{X}^a$ will represent
its  position  in 3-space.
The length of the intervals $(Z_{a}-Z_{a-1}) \equiv m_a /(2 \pi)$ 
define the masses of the constituent monopoles $M_a= 2 \pi m_a /g^2$.
Since these intervals  (their closure) provide  a covering of
the whole period, then  $\sum_a m_a= 2\pi$.

The ADHM condition implies  the following  one for  the spinors
$\zeta^b$:
\be
\label{eq:seta_X_rel}
\zeta^{\dagger b}\tau_i \zeta^b = \frac{1}{\pi} (X_i^{b+1} - X_i^b) \equiv \frac{1}{\pi} \Delta X_i^b\, .
\ee
Notice, that if we sum over all values of $b$ both sides of the
equation the right-hand side vanishes identically, so that there are
only N-1 independent spinors $\zeta^b$, which determine the relative
positions of the monopoles. The overall position of the centre of mass 
provides the remaining parameters of the solution. 

Given the form of the Nahm data, we can use the Nahm-ADHM formulas 
Eqs.~(\ref{eq:pot})-(\ref{defa}) to construct  the caloron vector potential, the 
field strength, and the fundamental and adjoint zero-modes.  
All these general expressions are written in terms the matrix functions $u$, $F$ and $\omeg$. 
A detailed evaluation of these quantities for the $Q=1$ caloron case is presented
in the Appendix. Here we will focus on deriving explicit
formulas for the adjoint zero-modes satisfying the periodicity
conditions: 
\be
 \Psi(\vec{x},x_0 + 1) = \pm P_\infty\Psi(\vec{x},x_0)P_\infty^{-1}\,,
\ee
in the gauge in which the caloron vector potential transforms as:
\be
 A_\mu(\vec{x},x_0 + 1) = P_\infty A_\mu(\vec{x},x_0)P_\infty^{-1}\,.
\ee
In order to derive the anti-periodic modes we will make use of the replica 
trick presented in section \ref{s.replica},
which we will describe below in detail for the case of the caloron.

\subsection{Nahm data for the replicated caloron}
\label{s.caloronrep}

In this subsection we will specialise the construction of replicas for
the case of the caloron. Let us concentrate upon the study of the
duplicate solution ($L=2$). The idea is that of regarding the
ordinary caloron solution presented previously as living in a double
torus in the time-direction. The topological charge, being an additive
quantity, is now equal to two. The Nahm-dual field is a U(2) field,
whose structure follows from Eq.~(\ref{eq:replica}). We have  
  \be
  \hat{A}^{R}_\mu(z)=\pmatrix{\hat{A}_\mu (z) & 0 \cr
  0 & \hat{A}_\mu (z+\frac{1}{2} )} \,,
  \ee
  where $\hat{A}_\mu(z)$ is the Nahm data of the ordinary
  caloron, and $\hat{A}^{R}_\mu(z)$ is the  Nahm data of the replicated
  caloron.
  
We still have to fix the    corresponding  $q$ for such a replica
  solution. We will argue that the solution is actually given by
  \be
  q^R(z)=\pmatrix{q(z)\cr q(z+\frac{1}{2})} \,.
  \ee
  Notice that each of the components of $q^R$ and $\hat{A}^R$ are periodic
with
unit period, but the whole set is periodic with  period  $1/2$
with a twist matrix given by $\tau_1$:
\be
\hat{A}^R_\mu (z+\frac{1}{2})=\tau_1 \hat{A}^R_\mu(z) \tau_1\,.
\ee
The quantity $q$ transforms by periodicity as follows:
\be
q^R(z+1/2)= \tau_1 q^R(z)\,.
\ee
From here it is possible to use the general formulas of the ADHM
construction to verify that indeed we obtain a replicated solution.
In particular we have that $u^R(z)$ is given by:
\be
u^R(z)=\pmatrix{u(z)\cr u(z+\frac{1}{2})}\, , 
\ee
in terms of the quantity  $u(z)$ for the normal (unreplicated) caloron.
Now 
\be
F^R-1=\int_0^{\frac{1}{2}} dz u^{R \dagger}(z) u^R(z) = \int_0^1 dz\,
u^{
\dagger}(z) u(z) \,,
\ee
which coincides with $F-1$ for the caloron. Similarly one can follow
the same steps as in our previous  SU(2) paper \cite{gluino_SU2_ap} 
to show that the replicated gauge potential coincides with the 
unreplicated one.

\section{Deformations of SU(N) calorons}
\label{s.deformations}
In order to obtain adjoint zero-modes we must analyse self-dual deformations of
the caloron vector potential. As described in section \ref{s.adhm}, this is 
achieved by deforming the corresponding ADHM-Nahm data. Imposing in addition 
the background gauge condition, we are led to Eq.~(\ref{eq:main_cond}), i.e:
\be
M^\dagger \delta M = (M^\dagger \delta M)\st\,.
\ee
Plugging into this formula the expression of 
$M$ and $\delta M$ in terms of Nahm data,  leads to 
\be
\bar\sigma_\mu\hat{D}_\mu \hat\Psi = 4 \pi^2 i (q \delta q^\dagger - (q \delta q^\dagger)\st)  \, ,
\ee
where $\hat\Psi= \delta \hat{A}_\mu \sig_\mu$ and $\hat{D}_\mu= \delta_{0\mu}\partial_z -i
\hat{A}_\mu $ is the Nahm-dual covariant derivative in the adjoint representation. The
previous equation should determine both $\hat\Psi$ and $\delta q$. The
latter should have the same delta-function singularity structure as $q$,
since that is fixed by the holonomy. 

If one wants to obtain periodic and anti-periodic zero-modes one can 
apply the same procedure to the duplicated caloron which has $Q=2$
topological charge. The formula now becomes 
\begin{eqnarray}
\bar\sigma_\mu\hat{D}^R_\mu\hat{\Psi}^R = 4\pi^2 i  (q^R \delta q^{R \dagger} - (q^R \delta q^{R\dagger})\st)\,,
\label{eq:eqini}
\end{eqnarray}
where the super-index $R$ specifies that the Nahm-dual of the
replicated caloron is used. This is an U(2) gauge field living in a
1-dimensional torus (circle) of period $\frac{1}{2}$, and 
which is self-dual except at isolated singularities. 
The deformations  take the form 
\begin{eqnarray}
 \hat{\Psi}^R = \left(\begin{array}{cc}
                  \hat\Psi_{11} & \hat\Psi_{12} \\
		                   \hat\Psi_{21} & \hat\Psi_{22}
				                  \end{array}\right)\,.
\label{eq:defrep}
						  \end{eqnarray}
The transition matrix is given by the first Pauli matrix $\tau_1$. Thus, the
boundary condition is $\hat\Psi^{R}(\zp) = \tau_1\hat\Psi^R(z)\tau_1$ 
which implies that the components satisfy 
\begin{eqnarray}
 \hat\Psi_{21}(z) = \hat{\Psi}_{12}(\zp) ; \;\;\; \hat\Psi_{22}(z) =
 \hat{\Psi}_{11}(\zp)\,.
\end{eqnarray}
The same conditions hold for $q^R$ and $\delta q^R$
\begin{eqnarray}
\delta  q^R(z) = \left(\begin{array}{c}
                  \delta \tilde{q}(z) \\ \delta {\tilde{q}}(\zp)
                 \end{array}\right)\,.
\label{eq:dqrep}
\end{eqnarray} 

To solve  Eq.~(\ref{eq:eqini}) several considerations are important. The first 
is that $\delta q^R$ is a linear combination of delta functions with
singularities at $Z_a$ and $Z_{\bar a}= Z_a+\frac{1}{2}$, which
correspond to the same point on the dual-circle, which has period $1/2$. The general form of the upper
component is:
\be
(\delta \tilde{q})_{\alpha i}(z)=\sum_{a=1}^N \left(\varphi^a_{\alpha i}
\delta(z-Z_a) + \varphi^{\bar{a}}_{\alpha i} \delta(z-Z_{\bar a})\right)\,,
\label{eq:deltaqt}
\ee
Furthermore, 
Eq.~(\ref{eq:eqini}) has to be understood as an identity among operators 
acting over two-component vectors functions of the form:
\begin{equation}
\left(\begin{array}{c}
       \phi(z) \\ \phi(\zp)
      \end{array}
\right)\,.
\end{equation}
Finally, it is easy to prove that each solution is associated to a
1-dim linear space in quaternions. More explicitly, given a solution 
$(\hat \Psi^R, \delta q^{R\dagger})$ a new solution is given by $(\hat\Psi^R Q, \delta q^{R\dagger} Q)$
for any constant quaternion $Q$.

In solving Eq.~(\ref{eq:eqini}) one must start by determining the form
of the  $\delta q^R$ coefficients $\varphi$, and later by solving the linear 
inhomogeneous equation for $\hat\Psi$ with fixed singularity structure.

Let us focus on the delta function structure of both sides of the
equation. The singularity structure of the left-hand side is of the
following form: 
\be
\sum_{b} \left[ \pmatrix{S_b & T_b \cr T'_b & S'_b} \, \delta(z-Z_b)\, +
\pmatrix{S'_b & T'_b \cr T_b & S_b} \, \delta(z-Z_{\bar b}) \right]\,.
\label{eq:sing}
\ee
This acts by multiplication on the appropriate space of two-component functions. 
On the other hand, the right-hand side of Eq.~(\ref{eq:eqini})
is a linear combination  of   tensor products of two delta function singularities 
at points $Z_a$ and $Z_{\bar a}$:
\be
\sum_{A\, B} Q_{A B} \pmatrix{\delta (z-Z_A) \, \delta(z'-Z_B) \,\,  & \,\,  \delta(z-Z_A)\,  \delta(z'-Z_{\bar B}) \cr
\delta(z-Z_{\bar A})\,  \delta(z'-Z_B) \,\,  &\,\,   \delta(z-Z_{\bar A})\,  \delta(z'-Z_{\bar B})  }\,,
\ee
where $A$ takes 2N values which run over the values of  $a$ and $\bar{a}$,
corresponding to  $Z_a$  and $Z_{\bar{a}}$ respectively. 
At first sight this seems quite different to the
structure displayed in Eq.~(\ref{eq:sing}). However, the  matching 
between both expressions  has to be deduced by identifying the result of
acting onto a vector of the appropriate space. The  delta
function on $z'$  acts over the argument function and the result is then   integrated
over $z'$ in the interval $[0,\frac{1}{2}]$. By comparing this action
with the result of acting with Eq.~(\ref{eq:sing}) we deduce that all
coefficients $Q_{A B}$ must vanish except $Q_{a a}$, $Q_{a \bar{a}}$, 
$Q_{ \bar{a} a}$ and  $Q_{\bar{a} \bar{a}}$ for $a$ running from 1 to
N. In more detail we conclude that $S'_a=0$ and 
\begin{eqnarray}
\label{Seq}
\delta_{ a b} S_a &=& Q_{a b}= 4 \pi^2 i \left( \zeta^a \varphi^{b\dagger} -
\zeta^b_C \varphi^{a \dagger}_C \right) \,,\\
\label{Teq}
\delta_{a b} T_{b} &=&  4 \pi^2 i (\zeta^a \varphi^{\bar{b} \dagger})\,, \\
\label{Tpeq}
\delta_{a b} T'_{b} &=&  -4 \pi^2 i (\zeta_C^b \varphi_C^{\bar{a} \dagger})\,, 
\end{eqnarray}
where the label $C$ stands for ``charge conjugate'' ($\zeta_C=\sigma_2
\bar \zeta$). 

It is convenient to realise that the set of equations involving $S_a$
coincide with those obtained for non-replicated calorons. Thus, it is
to be expected that the corresponding zero-modes are  periodic. 
On the other hand the equations for $T$ and $T'$ occur only for the
replicated caloron and should be  associated to anti-periodic zero-modes. This
is indeed the case as we will see in the next  two subsections dealing
with both cases respectively. 

Once the Nahm-dual deformations are obtained we can use our general
formula Eq.~(\ref{defa}) to obtain the ordinary deformations and
adjoint zero-modes:
\be
\delta A_\mu = \frac{i}{2} F^{-1/2} \int_0^{1/2} dz\,  \Big (\delta
q^{R
\dagger}+ \frac{1}{2 \pi} u^{R \dagger}(z) \hat \Psi^R(z)\Big ) \bsig_\mu
\hat{\partial} \omega^R(z) F^{1/2} + \rm{h.c.}\,.
\ee
This formula can be simplified using the periodicity properties of the
replicated functions. This will we done in the following two
subsections.

\subsection{Periodic adjoint zero-modes}
\label{s.periodic}
In this section we will deal with the 11 component of
Eq.~(\ref{eq:eqini}), and we will show that it gives rise to the periodic 
zero-modes. The resulting equation is 
\be
\partial_z \hat\Psi_{11} = \sum_a \delta(z-Z_a) S_a \,,
\ee
where $S_a$ is the solution of Eq.~(\ref{Seq}). Notice that $\hat\Psi_{11}$ is
constant at intervals, so that in order to have a solution which is
periodic in $z$ with period 1, one must have:
\be
\label{eq:nulljump}
\sum_a S_a= 0\,.
\ee

Now we should find the  solution of Eq.~(\ref{Seq}). Excluding
exceptional values of $\zeta^a$, which will be discussed at the end of this section,
 we can write 
\be
\label{varphi_sol}
\varphi^a = \bar Q_a \zeta^a \,,
\ee
where $Q_a$ are quaternions. This is easily seen to be a solution 
of Eq.~(\ref{Seq}), and furthermore 
\be
S_a = {\cal Q}_a Q_a \,,
\ee
where the quaternion ${\cal Q}_a$ is defined as 
\be
{\cal Q}_a= 4 \pi^2 i( \zeta^a \zeta^{a \dagger}-  \zeta_C^a
\zeta_C^{a \dagger})= -4 \pi \sigma_i \Delta X^{a}_i\,.
\ee
Notice  that the $Q_a$ are not independent, since they must
satisfy Eq.~(\ref{eq:nulljump}). 
\be
\label{eq:per_constraint}
\sum_{a=1}^N  {\cal Q}_a Q_a =0 \,.
\ee

We point that the solution Eq.~(\ref{varphi_sol}) coincides with the result 
of performing a variation of the $\zeta$ parameters describing the  non-replicated 
caloron solution. For non-vanishing $\zeta^a$ a general variation can
be written as 
\be
\delta \zeta^a= \bar{Q}_a \zeta^a\,.
\ee
Thus, the corresponding adjoint zero-modes are expected to be periodic in $z$. 
The connection  between the  $\zeta$-parameters and the position of the 
constituent monopoles, given by Eq.~(\ref{eq:seta_X_rel}),
automatically relates the deformation of these parameters with the
modification of relative position of the  monopoles. This relation
also  implies that the $\zeta^a$ are not all independent. 
Summing both sides of Eq.~(\ref{eq:seta_X_rel}) for all values of $b$
implies that 
\be
\sum_b \zeta^b \zeta^{b  \dagger}= \lambda \mathbf{I} \,,
\ee
which can be recast in the form
\be
\sum_b \zeta^b \zeta^{b  \dagger} = (\sum_b \zeta^b \zeta^{b
\dagger})\st\,.
\ee
Obviously, the deformed values $\zeta^a +\delta \zeta^a$ must also satisfy  this equation, so they are not
independent. It is a trivial exercise, that we leave to the reader, to
verify that Eq.~(\ref{eq:per_constraint}) automatically enforces this
constraint. 

We now turn back to solving Eq.~(\ref{eq:per_constraint}). The first
trivial solution amounts to taking $Q_a=0$ $\forall a$. Then $\delta
q=0$ and $\Psi_{11}$ is just a constant quaternion. This is easily
seen to correspond to  the supersymmetric
zero-mode, whose density coincides with the action density of the
caloron. It is obviously associated to an overall translation of the
solution, i.e. to a change of the  centre of mass of the monopoles. 

The remaining zero-modes are associated to non-vanishing $\delta \zeta^a$.
To solve the constraint equation we simply rewrite 
\be
Q_a= \frac{{\cal Q}^{a \dagger}}{{\cal Q}^{a} {\cal Q}^{a \dagger}}
S_a = \frac{\sigma_i (\Delta X^a_i)}{ 4 \pi ||\Delta \vec{X}^{a}||^2} S_a \,,
\label{eq:deltaqp}
\ee
in terms of the quaternions $S_a$,  which must  add up
to zero (Eq.~(\ref{eq:nulljump})).

Now that we have been able  to find the variations $\delta q$ and  $\delta
\tilde{A}$ which satisfy the background Lorentz gauge, we can simply
apply our general formulas for the adjoint modes:
\be
\delta A_\mu(x) = \frac{i}{2} F^{-\half} \Big \{ \int_0^1 dz
\delta \tilde q(z)^\dagger\bar{\sigma}_\mu\hat{\partial}\omega(z)
+\frac{1}{2\pi}\int_0^1 dz  u^\dagger(z)
\hat\Psi_{11}(z)\bar{\sigma}_\mu\hat{\partial}
\OMF(z)\Big \} F^\half  + \mbox{\rm h.c.}\,.
\label{eq:per}
\ee

For the supersymmetric mode we have $\delta \tilde q=0$ and  $\hat\Psi_{11}=2
\pi\mathbf{I}$. Thus, one obtains $\delta A_0=0$ and 
\begin{equation}
\delta A_i= E_i=  \sum_{a=1}^N  E^{(a)}_i\,,
\label{susy_zm}
\end{equation}
as expected. The second equality expresses the decomposition of the
electric field into the contribution of each constituent monopole
given by
\be
\label{eq:EMUa}
E_\mu^{(a)}= \frac{i}{2} F^{-1/2} \int_{Z_{a-1}}^{Z_a} dz\, u^\dagger(z) \overline{\sigma}^\mu
\hat{\partial} \OMF(z) F^{1/2}  + \mbox{\rm h.c.}\,,
\ee
where the quantities $u(z)$ and $\omega(z)$ are the ADHM functions for
the SU(N) calorons given in the Appendix. Furthermore, the integrals
can be performed analytically. Details are given in the Appendix.  

For the remaining modes one can choose a basis as follows. Take all
$S_b=0$ except for two: 
$S_{a-1}=2 \pi$ and $S_a=-2 \pi$. This gives
\be
\delta A_\mu^{(a)}(x) =  E_\mu^{(a)}(x)+ \widetilde{E}_\mu^{(a)}(x)\,,
\label{per_zm}
\ee
where the first term is given by Eq.~(\ref{eq:EMUa}) and follows from
the term involving $\hat\Psi_{ 1 1}$ in the deformation formula. The remaining piece 
$\widetilde{E}_\mu^{(a)}$  is associated to the $\delta q$ piece and
has a similar form to that of the vector potential:
\be
\label{Modes}
\widetilde{E}_\mu^{(a)}= -{i\over 4} \eta^{i \mu}_\nu
\left(\frac{\Delta X_i^a}{||\Delta \vec X^{a}||^2}
({\cal G}^a_\nu - {\cal G}^{a \dagger}_\nu) -
\frac{\Delta X_i^{a-1}}{||\Delta \vec X^{a-1}||^2}
({\cal G}^{a-1}_\nu - {\cal G}^{a-1 \dagger}_\nu)\right)\,,
\ee
where 
\be
{\cal G}^b_\alpha= F^{-1/2}  \zeta^{\dagger b}\overline{\sigma}_\alpha
\hat{\partial} \OME _b F^{1/2}\,,
\ee
with $\OME _b = \OMF(Z_b)$, and where all the quantities appearing in
the formulas are computed in the Appendix.

Notice that the $\delta A_\mu^{(a)}(x)$ 
provide a full basis of the space of periodic deformations. In
particular, the supersymmetric mode follows by addition of all these
deformations.

\subsubsection*{Special cases}
In our general discussion of solutions of Eq.~(\ref{Seq}) we excluded
several special cases. One of them is that of vanishing $\zeta^a$,
corresponding to $X_i^{a+1}=X^a_i$ (coinciding monopole positions). However,
there is another special case  which we excluded and gives rise to
additional solutions. This occurs whenever $\zeta^a$ is proportional to 
$\zeta^b_C$ for two different indices a and b. This condition leads 
to $\vec{X}^{a+1}-\vec{X}^a$ being anti-parallel to
$\vec{X}^{b+1}-\vec{X}^b$. When this situation occurs there are new
solutions to Eq.~(\ref{Seq}) having $S_a=0$ and therefore $\hat\Psi_{ 1
1}=0$ but $\delta q \ne 0$. Using the freedom to choose a
representative in the one-dimensional quaternionic space we might take
the solution to be
\be
\varphi^a_{\alpha i}= \delta_{ i b} \zeta^a_{\alpha  a}\quad \quad
\varphi^b_{\alpha i}= -\delta_{ i a} \zeta^b_{\alpha a} \,.
\ee
Plugging this solution onto the general formula for adjoint zero-modes 
we obtain:
\be
\delta A_\mu= \frac{i}{2} F^{-1/2} T F^{1/2}({\cal G}_\mu^a + {\cal
G}_\mu^b)+ \mbox{\rm h.c.}\,,
\label{eq:exceptionalp}
\ee
where $T$ is an $N\times N$ matrix whose only non-zero elements are
$T_{a b}=-T_{b a}=1$. As we will see in Section~\ref{s.prop} this 
solution does not satisfy neither periodic nor anti-periodic boundary 
conditions except for $Z_a=Z_b$ (periodic) and $|Z_a-Z_b|=\half$
(anti-periodic). One of the cases analysed in Section~\ref{s.prop} will
be of this type.  

\subsection{Anti-periodic adjoint zero-modes}
\label{s.antiperiodic}
To obtain the anti-periodic modes one has to solve the 12 component of 
Eq.~(\ref{eq:eqini}) with the singularity structure determined by
Eq.~(\ref{Teq})-(\ref{Tpeq}). Excluding the case in which $\zeta^a=0$, corresponding to 
monopoles having coincident locations, the solution is given by 
$\varphi^{\bar{a}}= \bar Q_{\bar{a}}\zeta^a$, where $Q_{\bar{a}}$ is an arbitrary quaternion. 
The equation now becomes 
\be
\label{eq:antipA}
\left(\bar\sig_\mu\hat{D}_\mu^R\hat\Psi^R\right)_{1 2}=  4 \pi^2 i \sum_{a=1}^N 
\left( \zeta^a \zeta^{a \dagger}
\delta(z-Z_a) -  \zeta_C^a \zeta_C^{a \dagger} \delta(z-\bar{Z}_a)\right)
Q_{\bar{a}}\,.
\ee
After  evaluating the
commutator, the  left-hand side of the previous equation becomes
\be
\Big(\partial_z -i \bsig_i \, [\hat{A}_i(z) - \hat{A}_i(\zp)]\Big)\, \hat\Psi_{ 1 2}\,.
\ee
We recall that $\hat{A}_i$ is constant at intervals separating the 
singular points located at  $Z_a$ and $Z_{\bar a}$. Excluding the
exceptional case of coinciding 
values (which can be treated as a limiting case), there are  altogether 2N intervals separating two contiguous 
singular points.  
Furthermore, by the construction, we know that there are N of these
singularities in the semicircle $[0, \frac{1}{2})$, and the remaining
ones are displaced by $\frac{1}{2}$. Let us reorder the values in
increasing order of $z$ (from 0 to 1)  and use capital letter subscripts as labels, 
running over  positive integers modulo 2N. The value of 
$(\hat{A}_i(z) - \hat{A}_i(\zp))$ in the interval separating $Z_{A-1}$
and $Z_A$ is constant and will be labelled $-2 \pi\Delta X_i^{A}$.  On the
other hand, the coefficient of the delta function $\delta(z-Z_A)$ on the
left hand-side of Eq.~(\ref{eq:antipA}) will be labelled $L_A Q_{\bar{A}}$, with $Q_{\bar{A}}$ a quaternion. From the 
periodicity properties under translations in $z$ by $\half$, we deduce 
$\Delta X_i^{A+N}=-\Delta X_i^{A}$, $L_{A+N}=L_A\st$ and
$Q_{\overline{A+N}}=Q_{\bar{A}}$.

With this notation it is easy to integrate $\hat\Psi_{ 1 2}$ in the interval
$(Z_{A-1},Z_A)$, giving 
\be
\hat\Psi_{ 1 2}(z)= \exp\{-i 2\pi (z-Z_{A-1}) \bsig_i \Delta X_i^{A} \}
\kappa^{A-1}_+, \quad  {\rm for} \, z\in (Z_{A-1},Z_A) \,.
\ee
At the edge of the interval this gives 
\be
\label{kappa_rel}
\kappa^{A}_-\equiv \lim_{z \rightarrow Z_A} \hat\Psi_{ 1 2}(z)  \equiv
\U_{A\, A-1}\kappa^{A-1}_+\,.
\ee
To match the solution at the different intervals one must use 
\be
\kappa^A_+-\kappa^A_-= L_A Q_{\bar{A}}\,.
\ee

Finally, by  imposing   periodicity (with period 1) in z, we obtain 
\be
\label{eq:jump}
\kappa^{2N+1}_-= W_1\st W_1\kappa^{1}_-+\sum_{A=1}^{2N} \U_{2N+1\, A}
L_A Q_{\bar{A}} = \kappa^{1}_-\,,
\ee
where we have introduced the following  $2\times 2$ matrices 
\be
\U_{B A}=\Texp\{i \int_{Z_A}^{Z_B} dz' \, \bar \sigma_i(\hat{A}_i(z')-
\hat{A}_i(z'+\frac{1}{2}))\}\,,
\label{eq:defU}
\ee
where $\Texp$ is the clockwise-ordered exponential, and $B>A$.
These matrices are like parallel transporters satisfying
\be
\U_{C A} = \U_{C B} \U_{B A}\,.
\ee
The symbol $W_A$ denotes $\U_{A+N\, A}$.
The elementary {\em
links} $\U_{A\,  A-1}$, appearing in Eq.~(\ref{kappa_rel}),  are simple exponentials having two important
properties: they are hermitian and have unit determinant. Thus, they are
elements of SL(2,C), namely of the type corresponding to boosts in the
$(\frac{1}{2},0)$ representation of the Lorentz group.
In addition, we have  $\U_{A\,  A-1}^{-1}=
 \U_{A\,  A-1}\st$. Obviously,
all matrices $\U_{B A}$ have unit determinant, but in general they
cease to be hermitian. From the periodicity in $z$ it follows that 
\be
\U_{B+N\, A+N}= \U_{ B A}\st\,.
\ee

In solving Eq.~(\ref{eq:jump}) one has to distinguish two cases. This
depends on whether $\mathbf{I} -  W_1\st W_1$ is invertible or not. 
Since $W_1\st W_1$ has unit determinant, the previous matrix is either
invertible  or zero. The second possibility is exceptional although 
it affects some particular arrangements of the monopoles. Thus, in the
remaining of this section we would concentrate on the generic case,
corresponding to invertibility of  Eq.~(\ref{eq:jump}), which allows the 
determination of $\kappa^A_+$ in terms of $Q_{\bar{A}}$. Later in this
section we will clarify the conditions under which invertibility does
not hold and provide the construction of adjoint zero-modes in that
case as well. 

To obtain a basis of the space of solutions we will proceed as
follows.  We construct the particular solutions for which  all the $Q_{\bar{b}}$ 
are set to zero except for one $b=a$, and such that $Q_{\bar{a}}=1$. Thus, the only
discontinuities in the solution take place at $z=Z_a$ and
$z=Z_{\bar{a}}$. The equation can then be solved in the two intervals
separating these two singularities. We have 
\begin{eqnarray}
\hat\Psi^{(a)}_{1 2}(z)  &=& \Texp\{ i \int_{Z_a}^z dz' \,  \bsig_i  (\hat{A}_i(z') -
\hat{A}_i(z'+\frac{1}{2}) )\}\kappa^a_+  \quad {\rm for } \, Z_a\le z \le Z_{\bar a} \,,
\label{solap1} \\
\hat\Psi^{(a)}_{1 2}(z)  &=& \Texp\{ i \int_{Z_{\bar a}}^z dz' \,  \bsig_i
 (\hat{A}_i(z') -
\hat{A}_i(z'+\frac{1}{2}) )\}\kappa^{a \star}_+ \quad {\rm for } \, Z_{\bar a}\le z \le Z_a \,.
\label{solap2}
\end{eqnarray}
The integrations have to be done clockwise along the unit circle. 
Now we should match the discontinuities from both sides of the
equation. This gives simply a particular case of Eq.~(\ref{eq:jump}): 
\be
(\mathbf{I} - W_a\st W_a)\kappa^a_+ = i 4 \pi^2(\zeta^a \zeta^{a\dagger} -
W_a\st \zeta_C^a  \zeta_C^{a\dagger})   \,.
\label{eq:defUa}
 \ee
We might now  collect  the value of $\delta q$ and $\hat\Psi_{1 2}(z)$ for this
solution. Plugging these values in
the general formula for the adjoint zero-modes for the replicated
caloron we obtain a particular anti-periodic  self-dual deformation:
\be
\label{antip_mode}
\delta A_\mu^{(\bar{a})}=\frac{i}{2} F^{-\half} \zeta^{a \dagger}
\bsig_\mu \hat{\partial} \OME_{\bar{a}} F^\half + 
\frac{i}{ 4 \pi}  F^{-\half} \int_0^1 dz\, u^\dagger(z+\frac{1}{2})
\hat\Psi^{(a)}_{1 2}(z+\frac{1}{2}) \bsig_\mu \hat{\partial} \OMF(z) F^\half\,,
\ee
where  $\OME_{\bar{a}}=\OMF(Z_{\bar{a}})$. 

The general anti-periodic zero-mode can be obtained by a linear
combination with quaternionic coefficients of these $N$ linearly
independent solutions:
\be
\Psi(x) \equiv \delta A_\mu \sigma_\mu = \sum_{a=1}^N \delta
A_\mu^{(\bar{a})} \sigma_\mu Q_{\bar{a}}\,.
\ee
 The counting agrees with the prediction of the index theorem. 
In the Appendix we show how to compute the integrals in
Eq.~(\ref{antip_mode}) analytically.

\subsubsection*{Special cases}
Our previous construction has focused in the solution of the problem
for the generic case, but there are several particular cases which
are interesting and do not fall into the previous category. In
particular, we have to address the case for which
$W_a\st W_a= \mathbf{I}$. First we should explain in which cases  
does this situation arise, and then find the general solution for
Nahm-dual deformations for it. 

Given the properties of the $\U _{ B A}$ matrices introduced earlier,
one easily concludes that the necessary and
sufficient condition for $W_a\st W_a= \mathbf{I}$, is that $W_a$ are 
hermitian. This  happens for all values of $a$ or for none. 
The hermiticity condition on $W_a$ amounts, via the connection to the
Lorentz group, to the problem of whether there exist a product of $N$
boosts which is itself a boost. Obviously, this occurs whenever the
boosts are collinear. In our case this occurs when all monopoles lie
along a straight line, and in cases where the appropriate relative
positions of monopoles are aligned. We do not know the answer to the
general case but we have investigated the SU(3) and SU(4) cases and
found that this in indeed the only solution for SU(3). For SU(4) a
necessary condition is that the
monopole relative positions must be coplanar and there are indeed 
an infinite set of solutions which are not collinear in which we will
encounter such a situation. 

In order to perform the construction of adjoint zero-modes in this case 
we examine the form of Eq.~(\ref{eq:jump}) in our case. We get
\be
0= \sum_{A=1}^{2N} \U_{2N+1\, A} L_A Q_{\bar{A}} =  \sum_{A=1}^{N} (W_1\st\,  
\U_{N+1\, A} L_A + \U\st_{N+1\, A} L_A\st) \, Q_{\bar{A}}\,.
\ee
To make this equation more clear we introduce the square root $S$ of $W_1$
($S^2=W_1$), with the property $SS\st=\mathbf{I}$. This is possible
given the properties of $W_1$. Multiplying the previous equation by
$S$ we arrive at 
\be
\label{per_cond}
0=\sum_{A=1}^N {\cal Q}_{\bar{A}} Q_{\bar{A}}\,,
\ee
where the quaternion ${\cal Q}_{\bar{A}}$ is given by 
\be
\label{per_cond2}
 {\cal Q}_{\bar{A}}= S\st\, \U_{N+1\, A} L_A + (S\st \,\U_{N+1\, A}
 L_A)\st\,.
\ee
Eq.~(\ref{per_cond}) is similar to the one appearing for the periodic
deformations, and the conclusions are similar. There are N-1
independent solutions 
with $Q_{\bar{A}}\ne 0$ in addition to the homogeneous $Q_{\bar{A}}=0$
one. 

The subsequent steps to obtain the explicit form of $\hat\Psi_{12}$ and 
$\delta \tilde q$ for each solution of Eq.~(\ref{per_cond}) are straightforward
and we will skip them.

\subsection{Adjoint zero-modes with more general boundary conditions}
\label{s.additional}
It is easy to generalise our construction to obtain adjoint zero-modes of the Dirac
operator with more general  periodicity conditions. These solutions  turn out to be useful in several 
contexts. For instance, they have been used to define the so-called 
dual quark condensate~\cite{Bilgici:2008qy}, which has been recently
used to the study of the SU(2) gauge theory with adjoint fermions~\cite{Bilgici:2009jy}.
They could also prove to be useful in probing the 
topological content of the QCD vacuum~\cite{tonyadj,tonyadj2}, in
analogy with the case  of  fundamental zero-modes~\cite{latcal}.

In this section, we will indicate how to make use of the replica procedure described in
section \ref{s.replica} to obtain solutions of the adjoint Dirac equation with
periodicity given by 
\begin{eqnarray}
 \Psi(\vec{x},x_0 + 1) = e^{i\frac{2\pi n}{L}} \, P_\infty\Psi(\vec{x},x_0)P_\infty^{-1},\quad n,L \in \cal{Z} \,. 
\label{eq:adper}
\end{eqnarray}
The construction is a straightforward generalisation of the one for 
anti-periodic zero-modes. It is based on replicating the caloron $L$ times in 
the time direction, and solving the adjoint Nahm-dual Dirac equation, Eq.~(\ref{eq:eqini}),
for the replicated solution. The Nahm-dual gauge connection is given by 
Eq. (\ref{eq:replica}), with transition matrices given in terms of the 't Hooft 
matrix $P$:
\be
\hat{A}^R_\mu(z+\frac{n}{L})= P^n \hat{A}^R_\mu(z) P^{n\dagger}  \label{reptn}\,.
\ee 
In what concerns $q^{R}(z)$, it is given by
\be
q^R(z)=\pmatrix{q(z)\cr q(z+\frac{1}{L})\cr . \cr . \cr  q(z+\frac{L-1}{L}) }\,,
\ee 
The deformations $\hat\Psi^R$ take the form of the $L\times L$ generalisation of
Eq. (\ref{eq:defrep}), with periodicity condition identical to that of 
$\hat{A}^R_\mu(z)$ given above. The structure of $\delta q^R$ is  also 
a straightforward generalisation of Eq.~(\ref{eq:dqrep}). It is given by an
L vector with components made up of a linear combination of delta functions 
with singularities at $Z_{a,q} \equiv Z_a+q/L$, with $q=0,\cdots, L-1$. The top 
component for example can be written as:
\be
(\delta \tilde{q})_{\alpha i}(z)=\sum_{a=1}^N \sum_{q=0}^{L-1}
\varphi^{a,q}_{\alpha i} \, \delta(z-Z_{a,q}) \,.
\ee

To obtain solutions with periodicity given by Eq.~(\ref{eq:adper}) one has
to solve the equations involving the $\hat\Psi_{ 1 n}$ component of $\hat \Psi^R$.
As in the case of the duplicated caloron, on must first determine
the singularity structure of Eq.~(\ref{eq:eqini}). The left hand side of the equation
is of the form:
\be
\sum_{b=1}^{N}\sum_{q=0}^{L-1}  \delta(z-Z_{b,q})\,  P^q \pmatrix{S_{b,0} & T_{b,0}^1 & \cdots & T_{b,0}^{L-1} \cr T_{b,1}^1 & S_{b,1} & \cdots & T_{b,1}^{L-1}\cr
.&.&\cdots&.\cr
T_{b,L-1}^1 & T_{b,L-1}^{2}&\cdots&S_{b,L-1}} P^{q\dagger}\,.
\label{eq:singrep}
\ee
An analysis analogous to that for the duplicated caloron allows to conclude that
$S_{b,q}$ and $T_{b,q}^l$ are all zero except for $S_{b,0}$, $T_{b,0}^q$ and 
$T_{b,q}^{L-q}$ for which:
\bea
\delta_{ a b} S_{a,0} &=& 4 \pi^2 i \left( \zeta^a \varphi^{b,0\dagger} -
\zeta^b_C \varphi^{a,0 \dagger}_C \right) \\
\delta_{a b} T_{b,0}^q &=&  4 \pi^2 i \, (\zeta^a \varphi^{b,q\, \dagger}), \quad   q=1,\,\cdots,\, L-1 \\
\delta_{a b} T_{b,q}^{L-q} &=&  -4 \pi^2 i \, (\zeta_C^b \varphi_C^{a,q\, \dagger}), \quad   q=1,\,\cdots,\, L-1 \,.
\end{eqnarray}
Note that the equation involving $S_{a,0}$ is identical to the one for the periodic 
zero-mode. The sought solutions with periodicity given by Eq.~(\ref{eq:adper}) involve
instead the equations depending on $T_{b,0}^n$ and $T_{b,n}^{L-n}$. As for 
anti-periodic zero-modes, excluding the case in which $\zeta^a=0$, the solution 
is given by setting all $\varphi^{a,q}$ to zero except for $q=n$ for which:
$\varphi^{a,n}= \bar Q_{a,n} \zeta^a$, with an arbitrary quaternion 
$Q_{a,n}$. The final equation becomes:
\bea
&&\Big(\partial_z -i \bsig_i \, [\hat{A}_i(z) - \hat{A}_i(z+{n \over L})]\Big)\, \hat\Psi_{ 1 n}=  T_{a,0}^n \, \delta(z-Z_{a,0}) 
+T_{a,n}^{L-n} \, \delta(z-Z_{a,n}) \nonumber \\
&& \equiv  4 \pi^2 i \, \sum_{a=1}^N
\left( (\zeta^a \zeta^{a\, \dagger})
\, \delta(z-Z_{a,0}) -   \zeta_C^a \zeta_C^{a\, \dagger} \, \delta(z-Z_{a,n})\right)
Q_{a,n}\,.
\eea
This equation is very similar to the one for the anti-periodic case and can be solved
using the same techniques. Once the solution has been obtained the adjoint zero-modes
with the required periodicity can be derived from:
\be
\label{addp_mode}
\delta A_\mu=\frac{i}{2} F^{-\half} \sum_{a=1}^{N} \zeta^{a \dagger} Q_{a,n}
\bsig_\mu \hat{\partial} \OME_{a,n} F^\half +
\frac{i}{ 4 \pi}  F^{-\half} \int_0^1 dz\, u^\dagger(z+\frac{n}{L})
\hat\Psi_{1 n}(z+\frac{n}{L})\,  \bsig_\mu \hat{\partial} \OMF(z) F^\half\,,
\ee
where $\OME_{a,n} = \OMF(Z_{a,n})$.

\section{Analyses of the solutions}
\label{s.prop}
In this section we will study several properties of the solutions
constructed in the previous sections. In particular, we will verify 
their  assumed periodicity properties and discuss their orthogonality
and normalisability. We will show several explicit examples that
illustrate the density profiles of the solutions and their relation to
the gauge field density and the position and masses of the constituent
monopoles.

\subsection{Periodicity}
Here we will argue that the solutions presented in the previous sections
have the required (anti)periodicity in time. This can be easily derived taking
into account the transformation properties of $u$ and $w$ under a time 
shift by one period. Using the formulas for $u$ and $w$ derived in the Appendix, 
it can be checked that:
\be
u(z,x_0 +1) = e^{i 2 \pi z} \, u(z,x_0) \, P_\infty^{-1}\, ,  
\ee
and that an identical expression holds for $w(z,x_0)$. Using in addition that 
$\delta q$ and the Nahm data do not depend on $x_0$, and that $F= \I_N + u^\dagger u$, 
the appropriate periodicity is obtained 
for periodic, Eq.~(\ref{eq:per}), and anti-periodic, Eq.~(\ref{antip_mode}), solutions.
To show that the term in $\delta q$ satisfies the required periodicity
in time, one must realise that 
\be
\zeta^{a \dagger} e^{2 \pi i Z_a} = P_\infty \zeta^{a \dagger}
\ee
These formulas also serve to substantiate the claim that the adjoint
zero-modes associated to Eq.~(\ref{addp_mode}) satisfy the boundary
conditions given in Eq.~(\ref{eq:adper}). The additional zero-modes
given in Eq.~(\ref{eq:exceptionalp}) fail to satisfy simple boundary
conditions unless  $P_\infty^{-1}T P_\infty=\pm T$.

\subsection{Normalisability}

To analyse the normalisability of our solutions one can make use of the 
general formulas for the norm and the scalar products of the solutions derived
in \cite{kvanbaal1,kraan} in terms of the Nahm data. 
Following \cite{kraan} one can compute the scalar products of the zero-modes in
terms of the quaternionic quantities:
\bea
\langle \Psi, \Psi' \rangle &\equiv& \int d\vec x \int_0^1 dt  \, {\rm Tr} \Big 
(\Psi^\dagger(x) \Psi' (x)\Big )=   
 \int_0^1 \! dz \,  \Big (  
\hat \Psi^\dagger_{11}(z)\,  \hat {\Psi'}_{11}(z) +\hat \Psi^\dagger_{12}(z)\,  \hat {\Psi'}_{12}(z)  \Big ) \nonumber \\ 
 &+&\, 4\pi^2\, \int_0^\half dz \, \int_0^\half dz'   \Big (\delta \tilde q(z)\,   
\delta {\tilde{q}\, '}^{\dagger}(z') \,  +\, (\delta \tilde q(z)\, \delta {\tilde{q}\,'}^{\dagger}(z') )\st \Big )\\
 &+&\, 4\pi^2\, \int_\half^1 \, dz \, \int_\half^1 \, dz'   \Big (\delta \tilde q(z)\,  \delta {\tilde {q}\,'}^{\dagger}(z') \,  +\, (\delta \tilde q(z)\, \delta {\tilde {q}\,'}^{\dagger}(z') )\st
 \Big ) \nonumber
\label{scalarp}
\eea
where $\Psi = \delta A_\mu \sigma_\mu$, Tr denotes the trace over
colour indices, 
and $\delta \tilde q$ is given by 
Eq.~(\ref{eq:deltaqt}).  Notice that, given the shift by $\half$ in the
singularity position in $\delta \tilde q$ for periodic and anti-periodic solutions, the formula automatically 
gives the orthogonality of both sets of solutions. In the following
two subsections we will analyse the scalar products within each set
separately (periodic and anti-periodic).

\subsubsection{Periodic zero-modes}
 
The generic periodic zero-modes correspond to solutions where $\hat \Psi_{12}=0$ 
and  $\delta \tilde{q}(z)=\sum_{a=1}^N \varphi^a \delta(z-Z_a)$. Accordingly:
\be
\langle \Psi, \Psi' \rangle = \int_0^1 \! dz \,  \Big (
\hat \Psi^\dagger_{11}(z)\,  \hat {\Psi'}_{11}(z)  \Big ) + 
\, 4\pi^2\, \sum_{a=1}^N \Big ( \varphi^a {\varphi'}^{a \dagger} 
+ (\varphi^a {\varphi'}^{a\dagger})\st \Big)\,,
\label{scalarp1}
\ee
where we have used the fact that $\varphi^a {\varphi'}^{b \dagger}$ is proportional
to $\delta_{a,b}$ for our choice of variations: $\varphi^a= \bar Q_a \zeta^a$, 
${\varphi'}^a= \bar Q'_a \zeta^a$. 

Let us start with the supersymmetric CP-pair of zero-modes Eq.~(\ref{susy_zm}), 
given by $\varphi^a=0, \, \forall a$, 
and $\hat \Psi_{11} (z) $ a constant  quaternion. From eq. (\ref{scalarp})  
it follows that the solution is normalisable with norm $4\pi^2$. This    
can also be derived using the already discussed proportionality between the 
density of the supersymmetric zero-modes and the action density of the caloron, 
i.e. $\delta A_0= 0$ and  $\delta A_i= E_i$, with $E_i$ the electric field of the self-dual caloron.  

We now proceed to analyse the normalisability of the remaining zero-modes.  
Consider the  solution given by $\Psi^{(a)} = \delta A_\mu^{(a)} \sigma_\mu$, with
$\delta A_\mu^{(a)} =  E_\mu^{(a)}+ \widetilde{E}_\mu^{(a)}$ as in 
Eq~(\ref{per_zm}).  
They correspond to setting $\hat \Psi_{11}^{(a)} = 2 \pi \chi_a(z)$, and
all $\varphi^b$ equal to zero except for $b=a-1$ and $b=a$, for which they are  
given by Eqs.~(\ref{varphi_sol}) and (\ref{eq:deltaqp}), with $S_{a-1}=-S_{a}= 2\pi$.
Using formula (\ref{scalarp1}) we obtain the following expression for the scalar products:
\be
\langle \Psi^{(a)}, \, \Psi^{(b)} \rangle 
= 2\pi  \, m_a \, \delta_{a b}  + {\pi  \over  ||\Delta \vec X^a||} \, (\delta_{a b} 
	- \delta_{a+1\, b})+ {\pi \over  ||\Delta \vec X^{a-1}||} \, (\delta_{a b} - \delta_{a-1\, b})\,.
\ee
The set becomes orthogonal in the limit in which the $N$ constituent monopoles of
the caloron are infinitely separated, i.e. $||\Delta \vec X^a|| \rightarrow \infty$, for all $a$.
As mentioned in section~\ref{s.periodic}, the supersymmetric zero-modes can be 
obtained as the linear combination $\Psi = \sum_{a=1}^N \Psi^{(a)}$.
This correctly reproduces the fact that 
in the limit of large separation the energy density of the caloron decomposes 
in the sum of N BPS monopoles with respective masses equal to $2 \pi
m_a/g^2$. Each 
of them carries a CP-pair of periodic zero-modes 
given by  $\Psi^{(a)} = (\psi^{(a)}, \psi_{C}^{(a)})$, in accordance to the Callias 
index theorem for BPS monopoles \cite{Callias}.  

\subsubsection{Anti-periodic zero-modes}

In a similar way we can derive the expression for the scalar products of the
anti-periodic zero-modes. They correspond to solutions with $\hat \Psi_{11}=0$
and  $\delta \tilde{q}(z)=\sum_{a=1}^N \varphi^{\bar a} \delta(z-Z_{\bar a})$,
with $\varphi^{\bar a} = \bar Q_{\bar a} \zeta^a$. The general formula
for the scalar products reduces in this case to:
\be
\langle \Psi, \Psi' \rangle = \int_0^1 \! dz \,  \Big (
\hat \Psi^\dagger_{12}(z)\,  \hat {\Psi'}_{12}(z)  \Big ) +
\, 4\pi^2\, \sum_{a=1}^N \Big ( \varphi^{\bar a} {\varphi'}^{\bar a \dagger}
+ (\varphi^{\bar a} {\varphi'}^{\bar a\dagger})\st \Big)\,.
\label{scalarp2}
\ee

We will only provide here an explicit expression for the
generic case presented in section \ref{s.antiperiodic}, non-generic situations
are left to the reader. The generic solutions are given by $\Psi^{(\bar a)} = 
\delta A_\mu^{(\bar a)} \sigma_\mu$, with $\delta A_\mu^{(\bar a)}$ as in
Eq.~(\ref{antip_mode}). They are obtained by 
setting all $\varphi^{\bar b}$ to zero except for $\varphi^{\bar a}= \zeta^a$, and by taking $\hat \Psi_{12}$ as
in  Eqs.~(\ref{solap1}), (\ref{solap2}). Inserting these expressions into 
the formula for the the scalar products, we obtain:
\be
\langle \Psi^{(\bar a)}, \, \Psi^{(\bar b)} \rangle = 4 \pi\,  ||\Delta \vec X^a|| \, \delta_{ab}    
+ \, \sum_{A=1}^{2N}  { \delta_A \over \, g(2\pi \delta_A \, ||\Delta \vec X^A||)} \,
 O_A^{a \dagger} \,  e^{-i 2\pi \bar \sigma_i \delta_A \Delta X^A_i } \, O_A^b\,,
\ee
where $\delta_A = Z_{A}-Z_{A-1}$, and
\bea
O_A^c &=& \Texp\{ i \int_{Z_c}^{Z_{A-1}} dz' \,  \bsig_i (\hat{A}_i(z') -
\hat{A}_i(z'+\frac{1}{2}))\}\kappa^c_+,\quad {\rm for}\,\, (Z_{A-1},Z_A)\subset (Z_c,Z_{\bar c})\,,\nonumber\\ 
O_A^c &=& \Texp\{ i \int_{Z_{\bar c}}^{Z_{A-1}} dz' \,  \bsig_i (\hat{A}_i(z') -
\hat{A}_i(z'+\frac{1}{2}))\}\kappa^{c \star}_+,\quad {\rm for}\,\, (Z_{A-1},Z_A)\subset (Z_{\bar c},Z_c)\,.\nonumber
\eea

\subsection{Zero-mode density profiles}
\label{examples}

We proceed now to discuss some illustrative examples of periodic and anti-periodic
zero-modes for different gauge groups. In order to obtain the density profiles we
have developed two independent numerical codes which work for general
SU(N) gauge group and give matching results. We will discuss separately the cases corresponding to
different periodicity.

\subsubsection{Periodic zero-modes}
\label{exp}

Figure \ref{fig1} displays results for the gauge group SU(3)
and for three different choices of the monopole masses: $m_1=m_2=m_3=2\pi/3$;
$m_1=m_2=\pi/2$, $m_3=\pi$; and $m_1=m_2=\pi/3$, $m_3=4\pi/3$.
In all cases the monopoles are located on the vertices of an equilateral triangle
of side 2. We plot both the density of the supersymmetric CP-pair of zero-modes and the
3 CP-pairs denoted previously by $\Psi^{(a)}$. The figure exemplifies how in the regime
in which $||\Delta \vec X^a|| > 1, \, \forall a$,
the action density of the caloron, given by the density profile of the supersymmetric
zero-mode, decomposes into three constituent BPS monopoles. Each of 
\clearpage
\FIGURE[h]{
\centerline{
\psfig{file=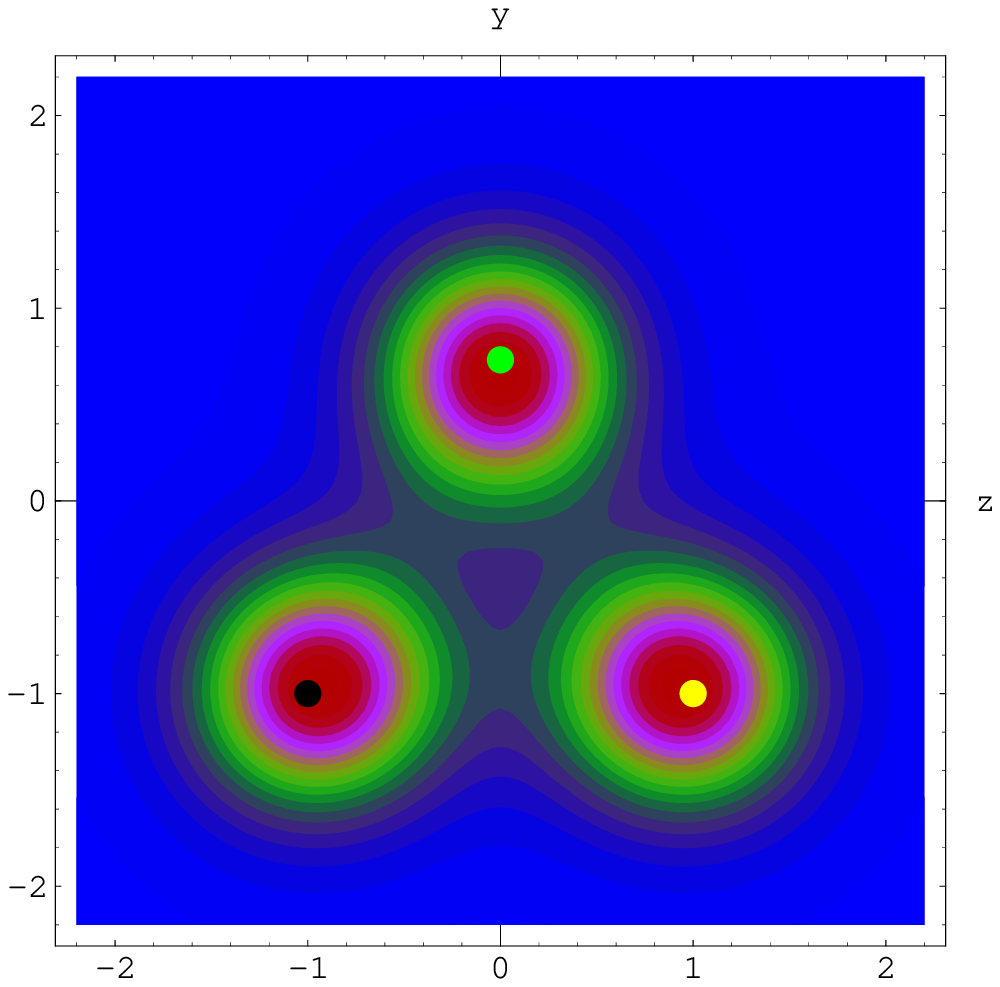,angle=0,width=4.cm}
\psfig{file=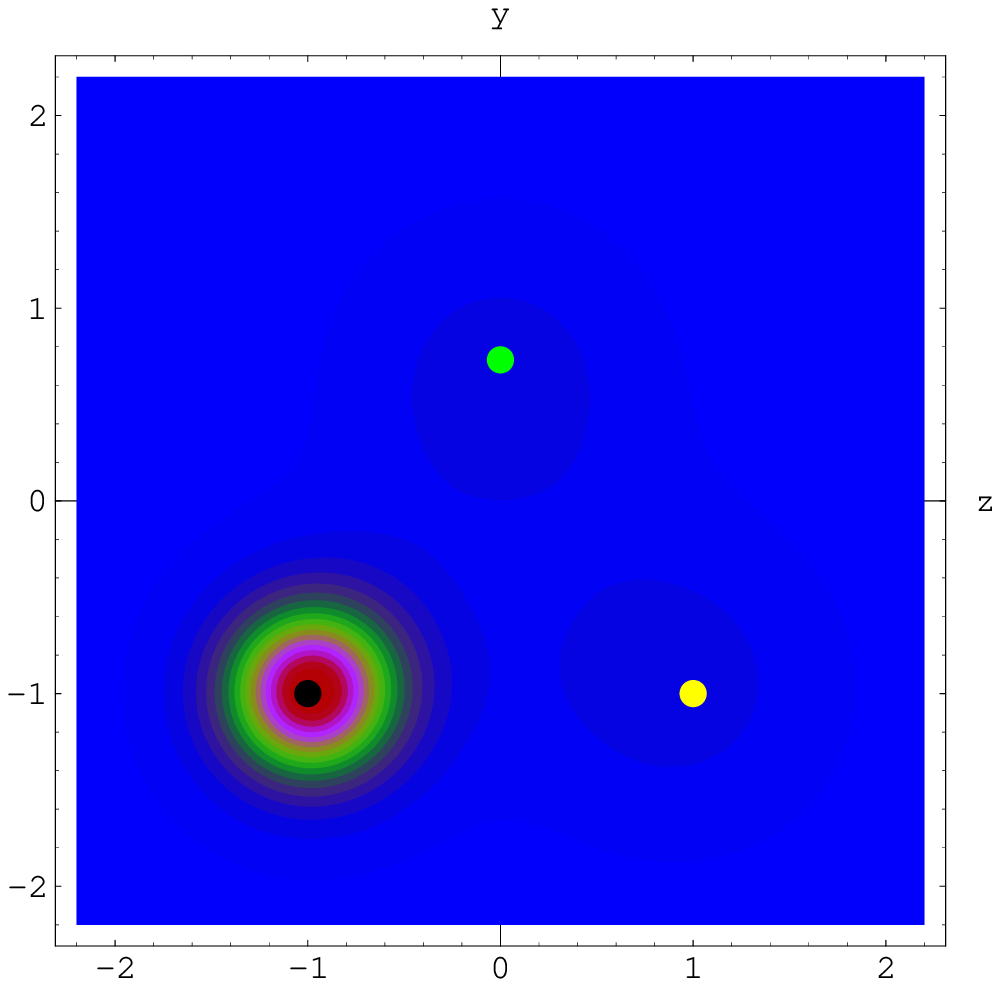,angle=0,width=4.cm}
\psfig{file=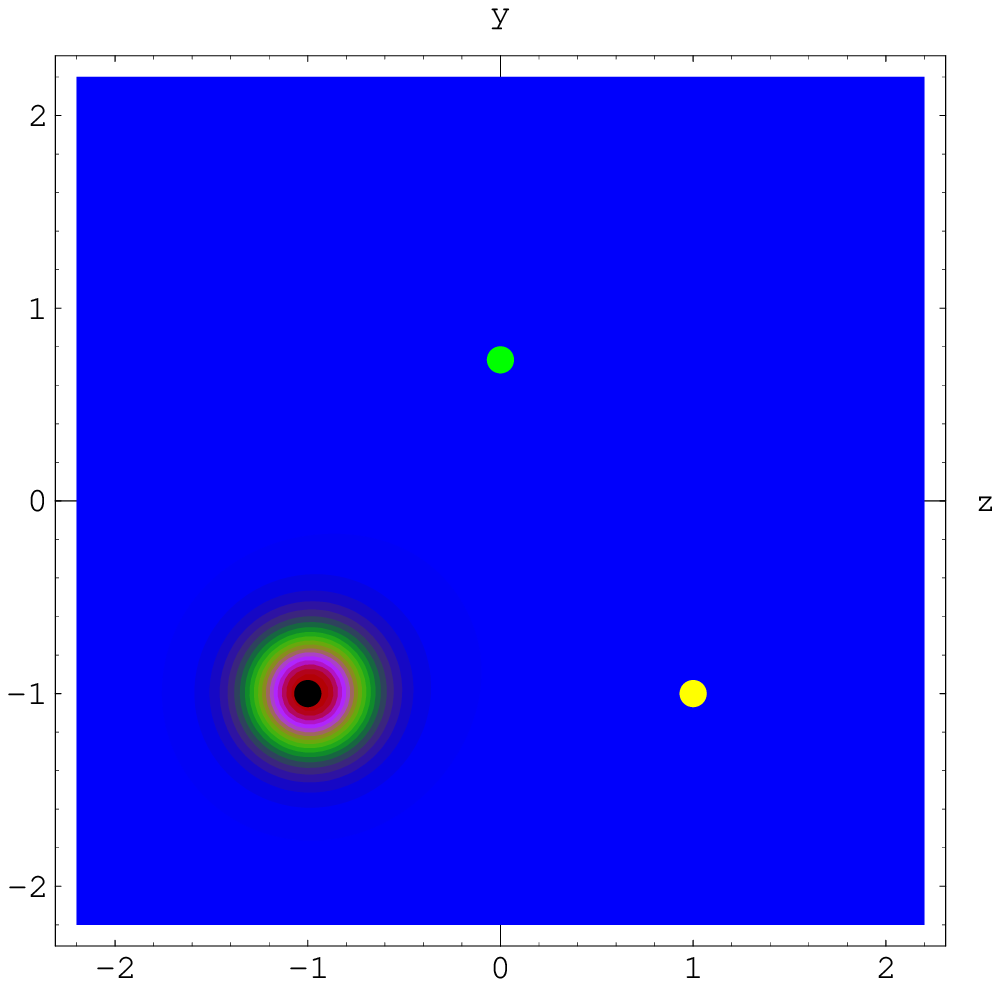,angle=0,width=4.cm}}
\centerline{
\psfig{file=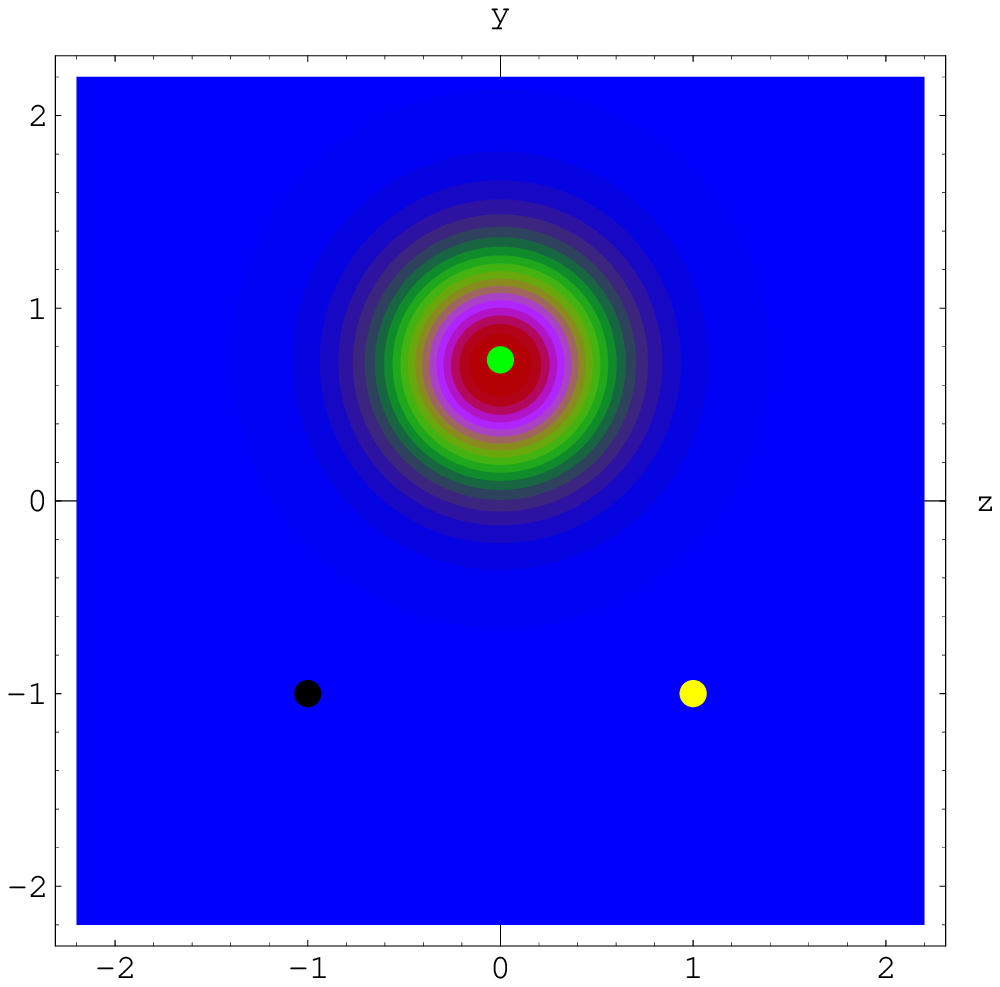,angle=0,width=4.cm}
\psfig{file=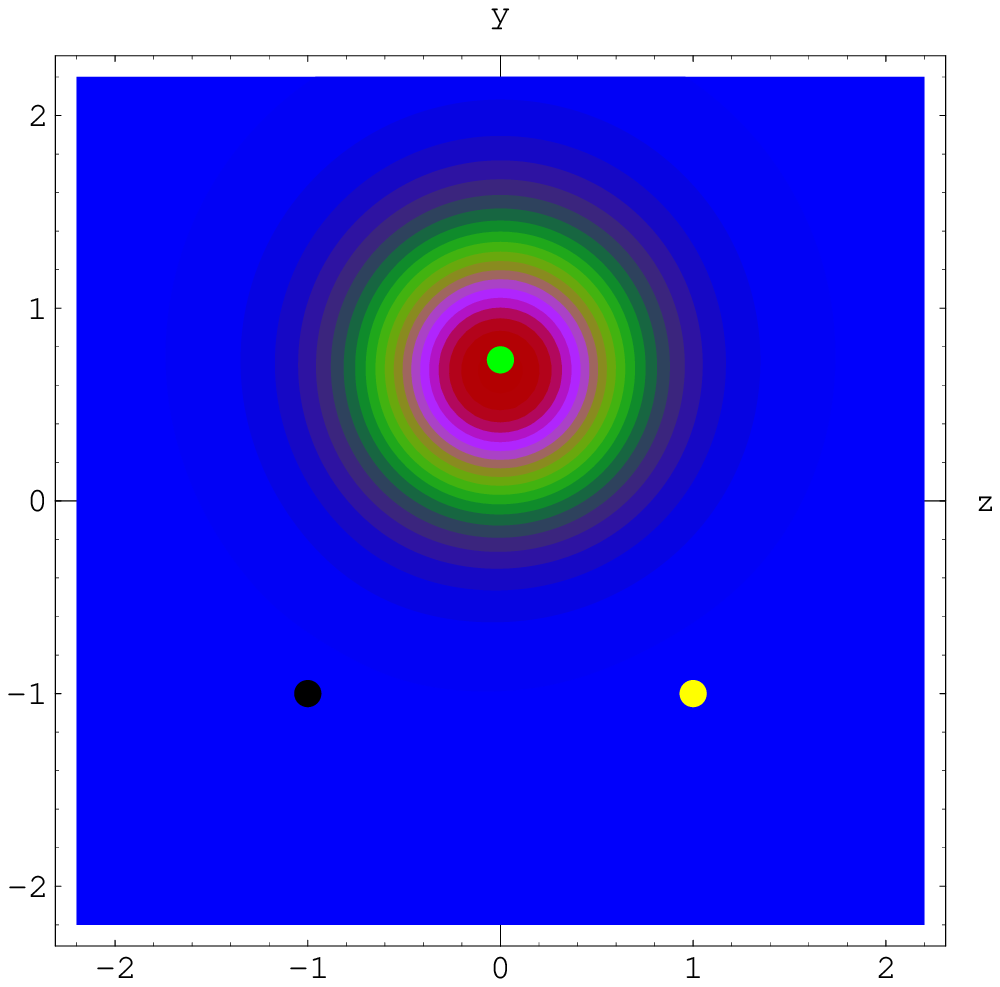,angle=0,width=4.cm}
\psfig{file=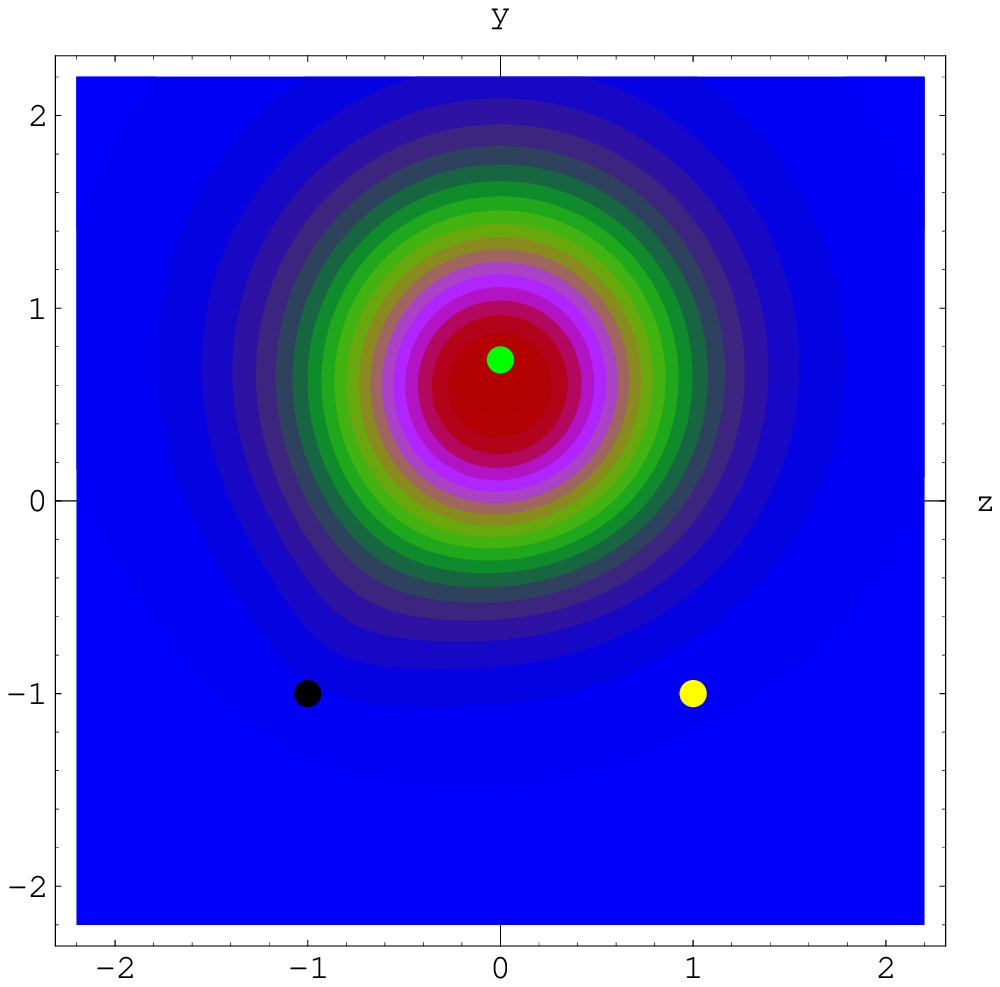,angle=0,width=4.cm}}
\centerline{
\psfig{file=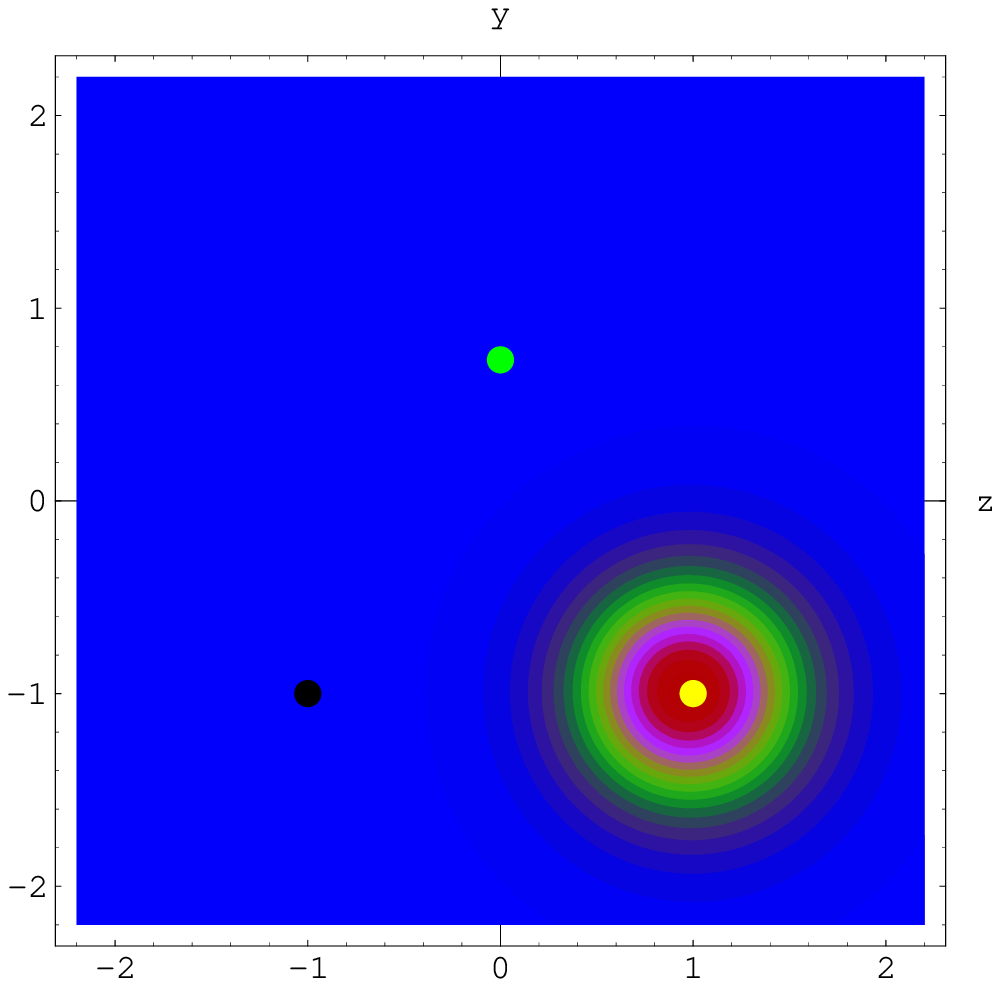,angle=0,width=4.cm}
\psfig{file=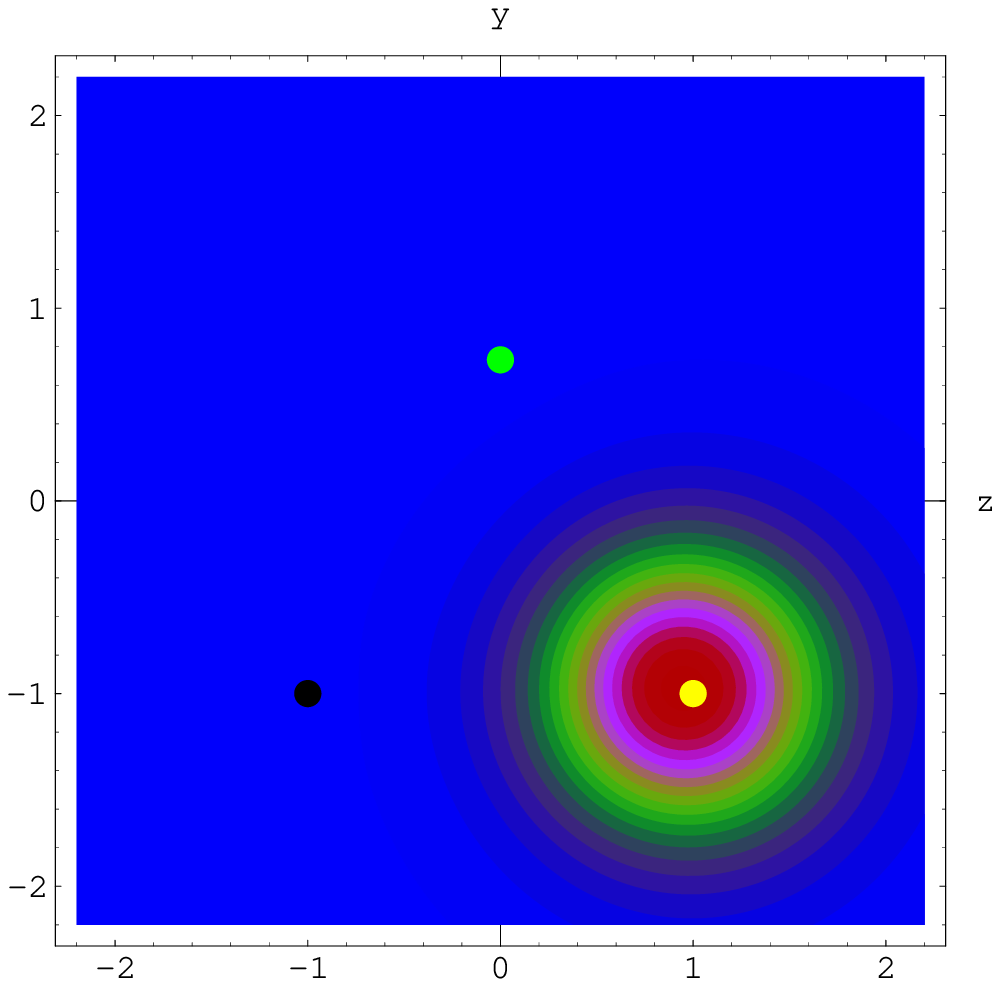,angle=0,width=4.cm}
\psfig{file=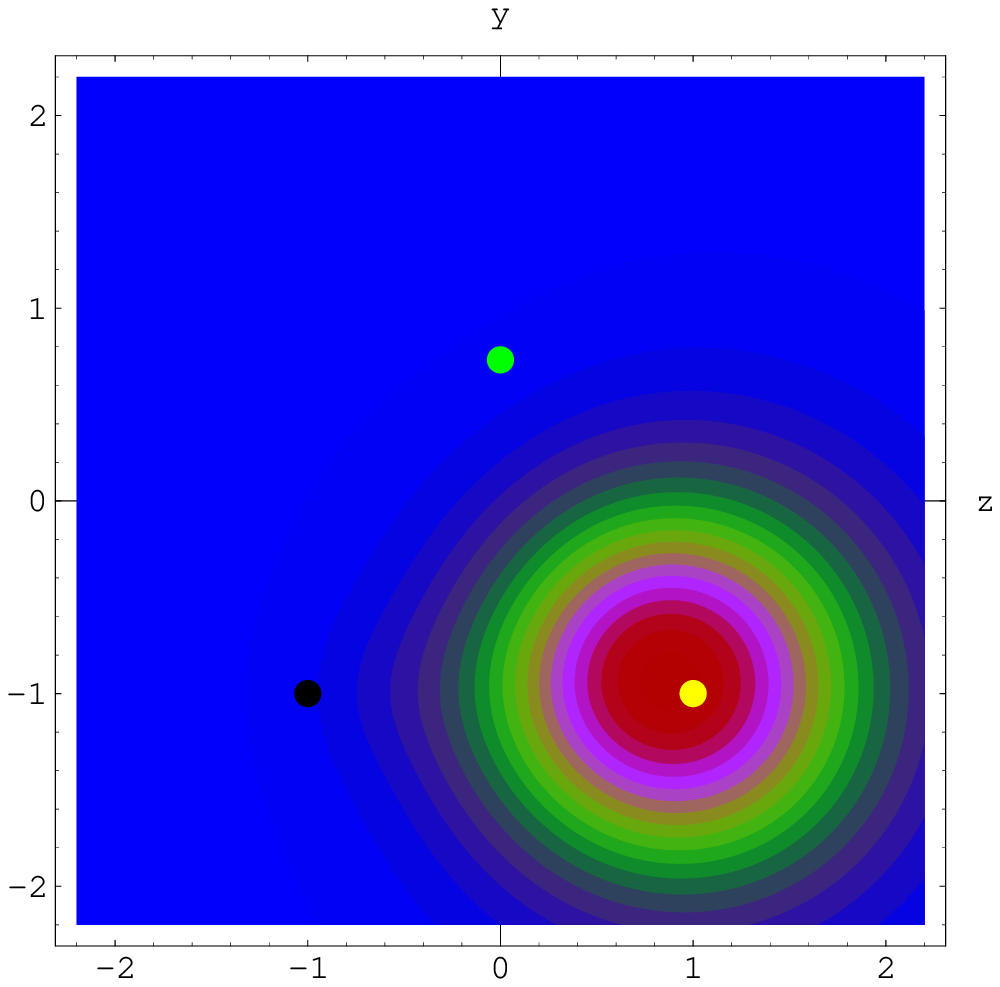,angle=0,width=4.cm}}
\centerline{
\psfig{file=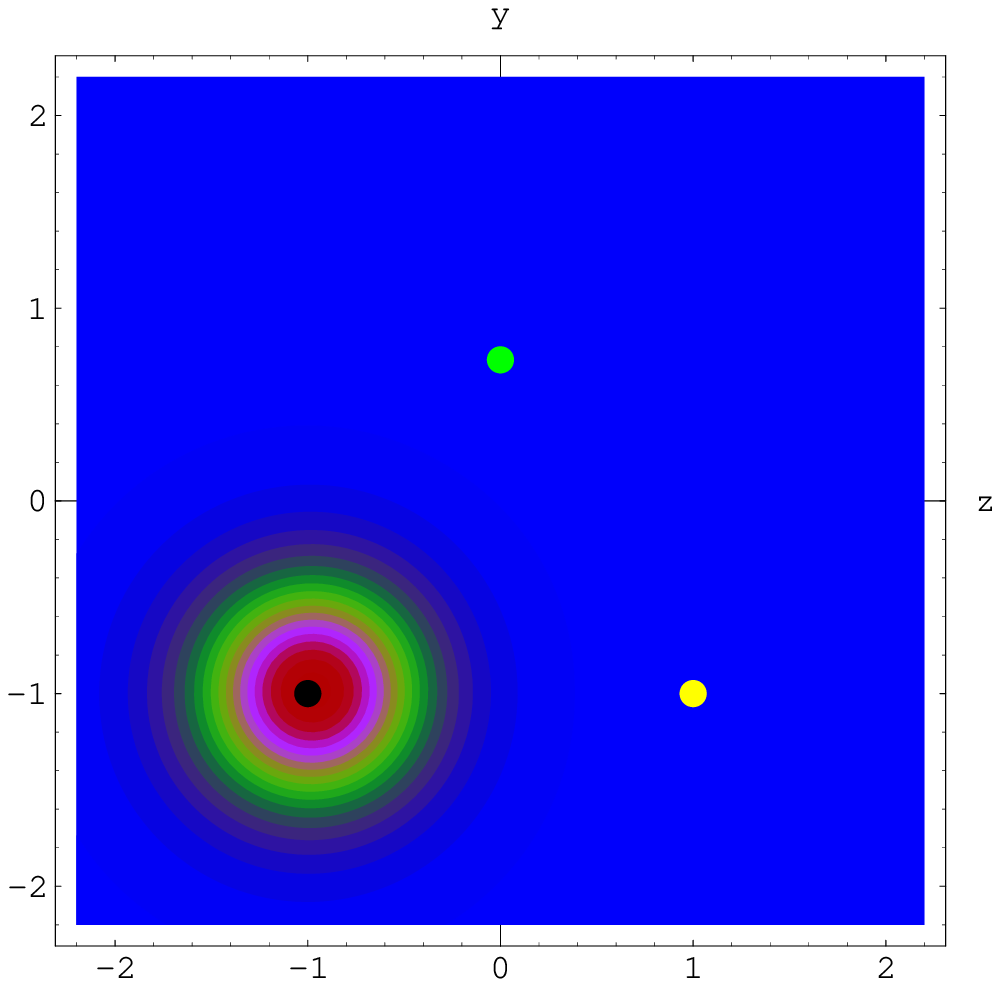,angle=0,width=4.cm}
\psfig{file=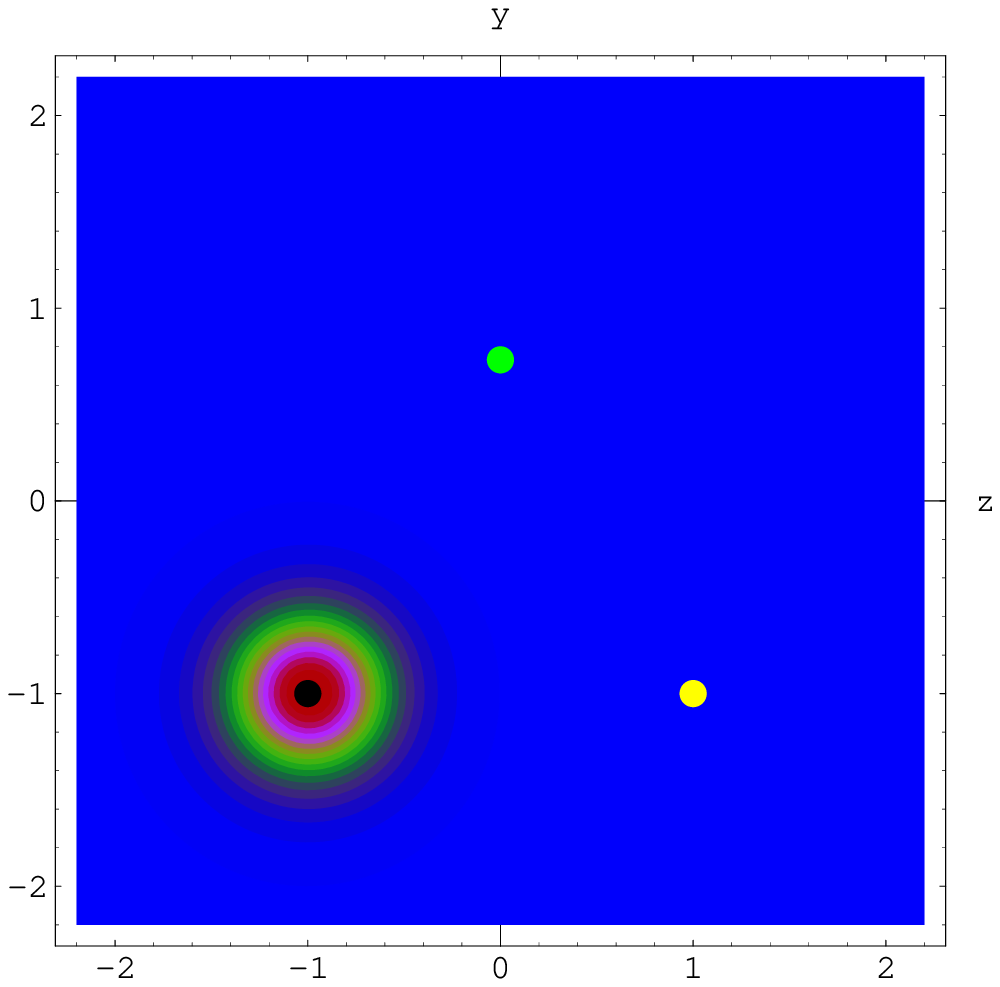,angle=0,width=4.cm}
\psfig{file=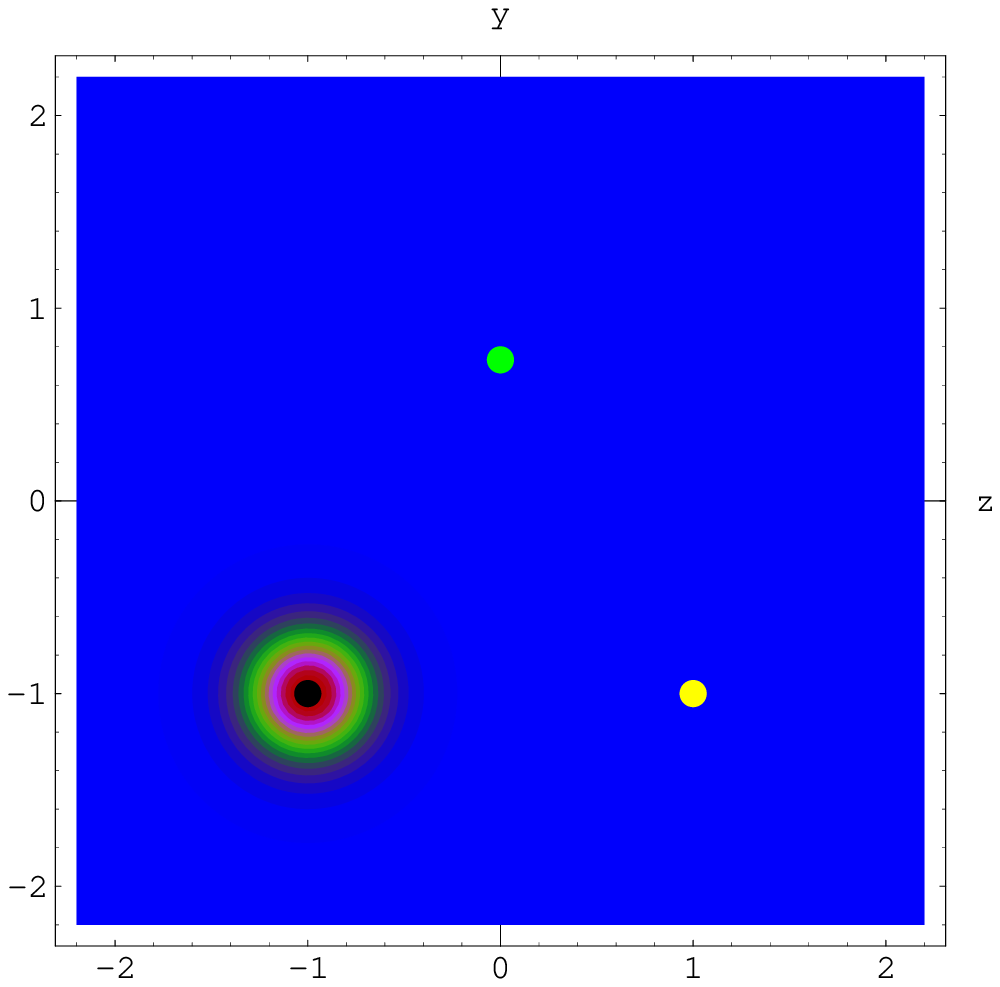,angle=0,width=4.cm}}
\caption{Contour plots of the periodic zero-mode densities in the
$x=t=0$ plane for gauge group SU(3). 
The three columns display,
from left to right, the zero-modes corresponding to monopole masses:
$m_1=m_2=m_3=2\pi/3$; $m_1=m_2=\pi/2$, $m_3=\pi$; and $m_1=m_2=\pi/3$, $m_3=4\pi/3$.
The top row displays the supersymmetric zero-modes and the lower three the
non-supersymmetric modes, i.e. $\delta A_\mu^{(a)}$ as in Eq.~(\ref{per_zm}),
with, from top to bottom, $a=1,2,3$.
Monopoles are localised on the vertices of an equilateral triangle of side 2,
on the $x=t=0$ plane.  The small filled circles
indicate the monopole positions, starting from the top monopole in each plot and in clockwise sense:
$\vec X^1$-green, $\vec X^2$-yellow, $\vec X^3$-black.}
\label{fig1}}

\noindent them carries a
CP-pair of adjoint zero-modes given in this limit by the solutions parametrised by
$\Psi^{(a)}$.

In the opposite limit in which $||\Delta \vec X^a|| << 1, \, \forall a$, the caloron
becomes an ordinary BPST instanton and the action density recovers spherical symmetry.
For the equal mass case and with monopoles located on the vertices of
an equilateral triangle, the symmetry properties allow us to obtain
three orthogonal modes,  given by the supersymmetric mode and two
other  linear combinations. The latter are obtained in terms of the 
following linear combinations of the Nahm data

\bea
\label{comb}
\hat \Psi^{(1)}_{11}+\hat  \Psi^{(2)}_{11}\, e^{-{2\over 3}\pi\sigma_3}+ \hat \Psi^{(3)}_{11}\,e^{\,{2\over 3}\pi\sigma_3}\,,\nonumber\\
\hat \Psi^{(1)}_{11}+\hat \Psi^{(2)}_{11}\, e^{\,{2\over 3}\pi\sigma_3}+ \hat \Psi^{(3)}_{11}\,e^{-{2\over3}\pi\sigma_3}\,,\nonumber
\eea
with analogous combinations for the $\delta \tilde{q}^\dagger$ terms. The resulting density
profiles, for  an equilateral triangle of side 0.1, are presented
in figure \ref{fig1c}.

\FIGURE[ht]{\centerline{
\psfig{file=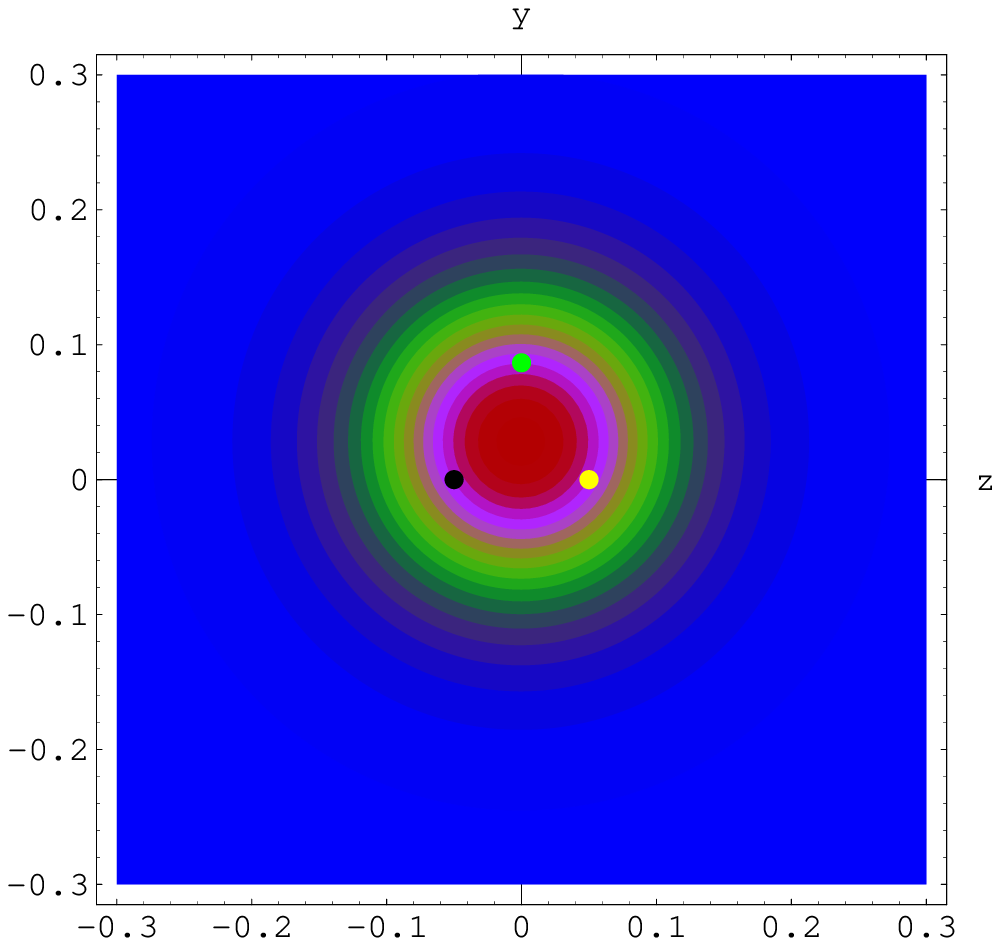,angle=0,width=5cm}
\psfig{file=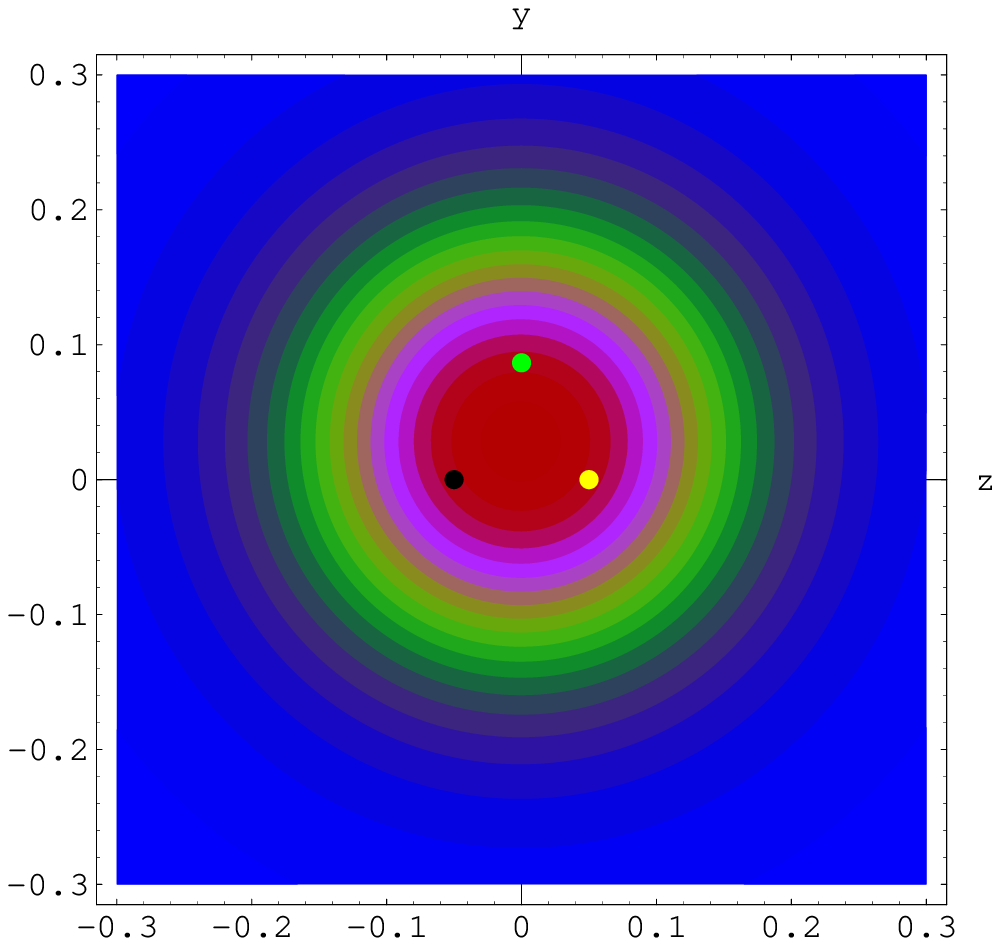,angle=0,width=5cm}
\psfig{file=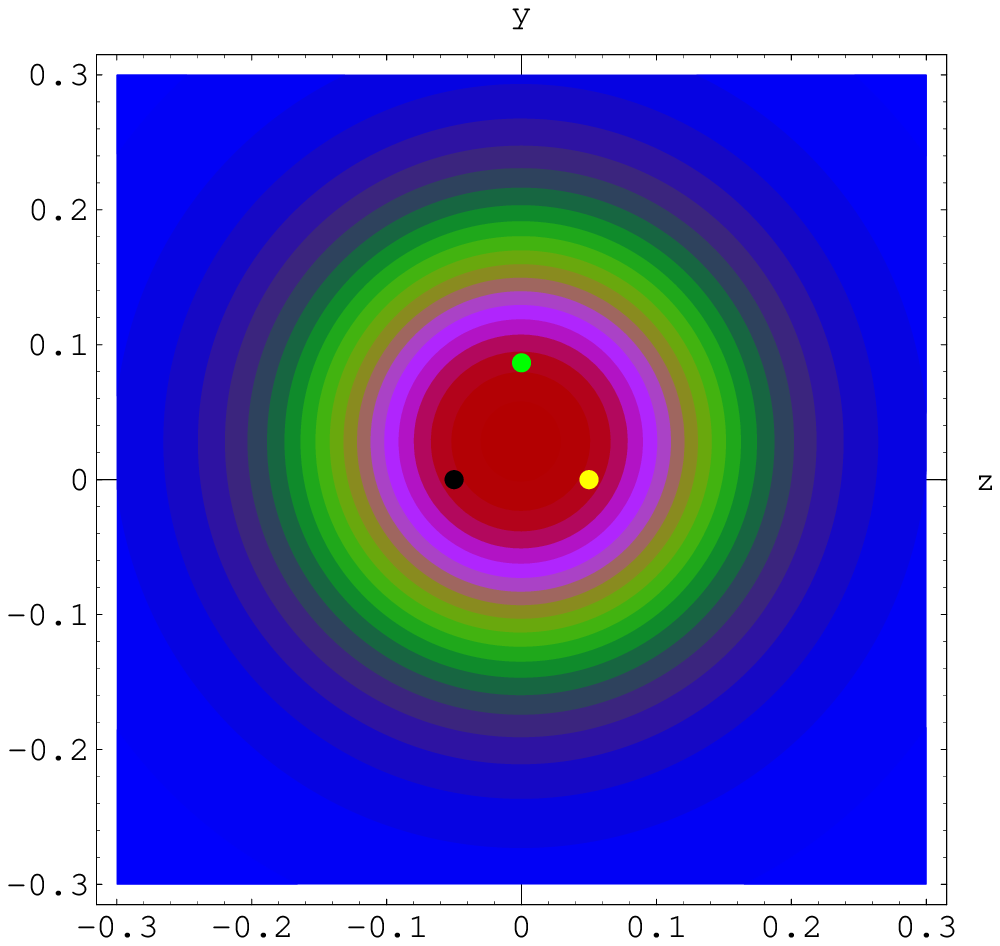,angle=0,width=5cm}}
\caption{Contour plots of the periodic zero-mode densities at the
$t=x=0$ plane for gauge group SU(3) and the equal mass case. The zero-modes have been constructed from
orthogonal combinations of the $\Psi^{(a)}$  described in
Eq.~\ref{comb}. The plot on
the left corresponds to the supersymmetric mode, the ones on the right display
the other two orthogonal modes.
Monopoles are localised on the vertices of an equilateral triangle of side
0.1, on the  $x=t=0$ plane. 
The small filled circles indicate the monopole positions, starting from the top monopole in each plot and in
clockwise sense: $\vec X^1$-green, $\vec X^2$-yellow, $\vec X^3$-black.}
\label{fig1c}}

Figure~\ref{fig3} displays the periodic zero-modes for gauge group SU(4)
and for equal mass monopoles arranged on the vertices of a square of side 2.
Again each non-supersymmetric CP-pair of zero-modes has support on a single
constituent monopole. The limit of small separation reproduces again the situation
for the BPST instanton.

\subsubsection{Anti-periodic zero-modes}
\label{exap}

In this subsection we will focus on describing a few representative examples of
anti-periodic zero-modes for SU(3) and SU(4).
We have already discussed in section~\ref{s.antiperiodic} that one has to distinguish
two main cases depending on the invertibility of $\mathbf{I} -  W_1\st W_1$. We will only
focus on the generic situation which correspond to the invertible case. This covers
all the possible cases for SU(3) although not for SU(4). 

\FIGURE[ht]{
\centerline{
\psfig{file=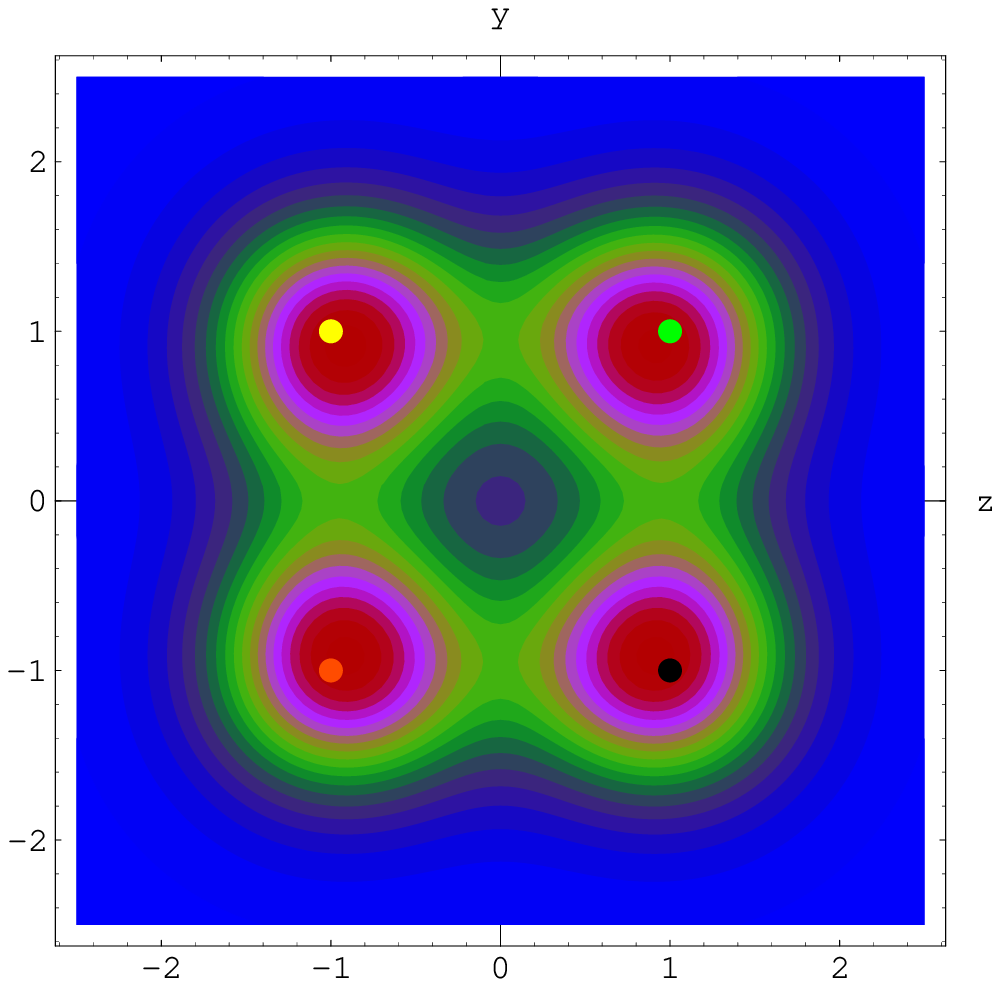,angle=0,width=6cm}
\psfig{file=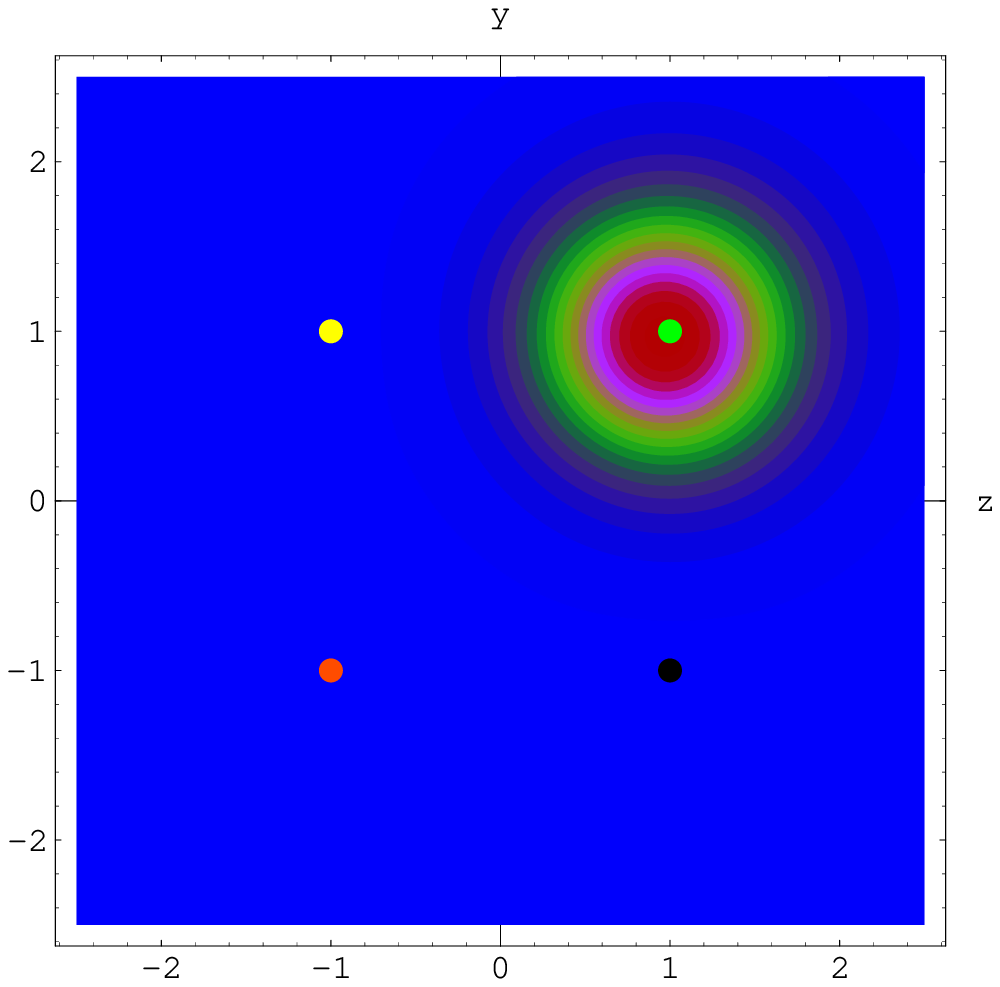,angle=0,width=6cm}}
\caption{Contour plots of the supersymmetric (left) and non-supersymmetric (right) periodic zero-mode densities
on the $x=t=0$ plane for gauge group SU(4) and the equal mass case. We only show one of the non-supersymmetric
zero-modes, $\Psi^{(1)}$. The rest look alike but rotated by 90, 180 and 270 degrees
along the $x$ symmetry axis. Monopoles are localised on the vertices
of a square of side 2, on the  $x=t=0$ plane. 
The small filled circles indicate the monopole positions, starting from the top-right monopole in each plot and in clockwise sense:
 $\vec X^1$-green, $\vec X^2$-black, $\vec X^3$-red, $\vec X^4$-yellow.}
\label{fig3}}

\FIGURE{
\centerline{
\psfig{file=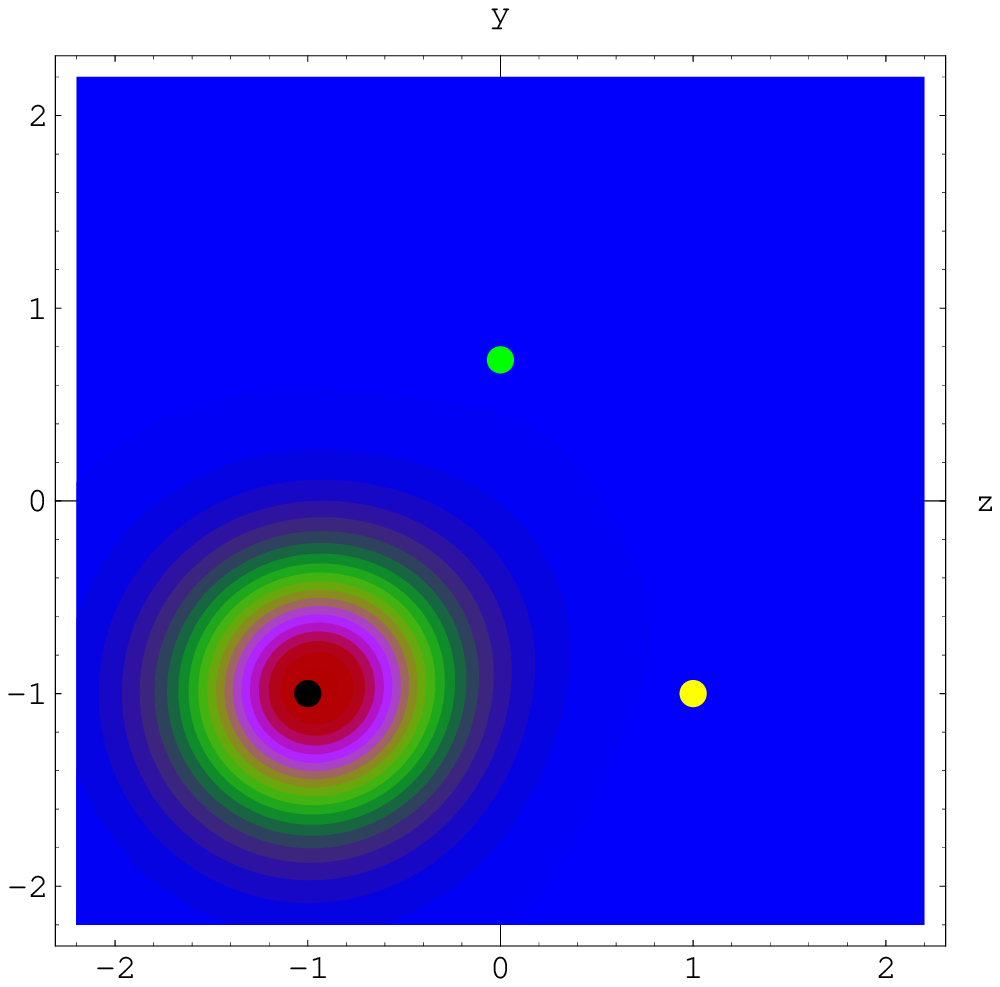,angle=0,width=4.cm}
\psfig{file=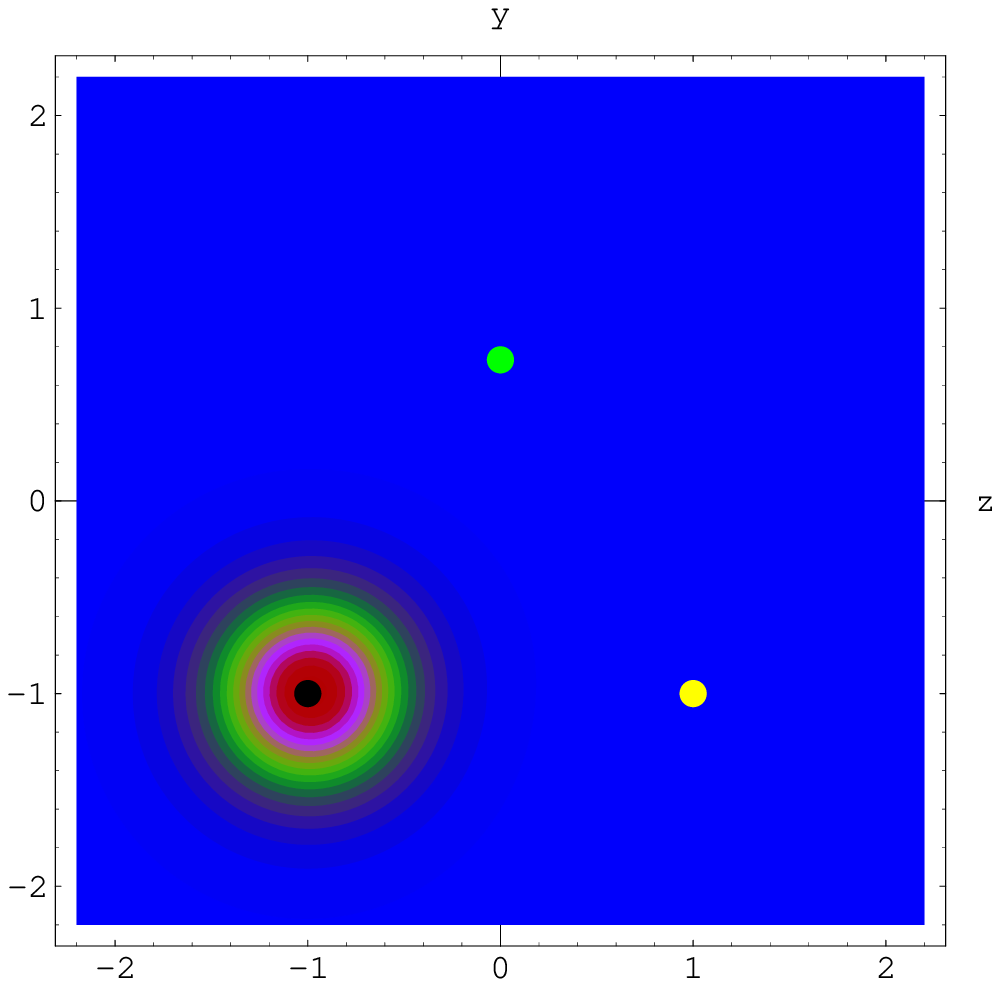,angle=0,width=4.cm}
\psfig{file=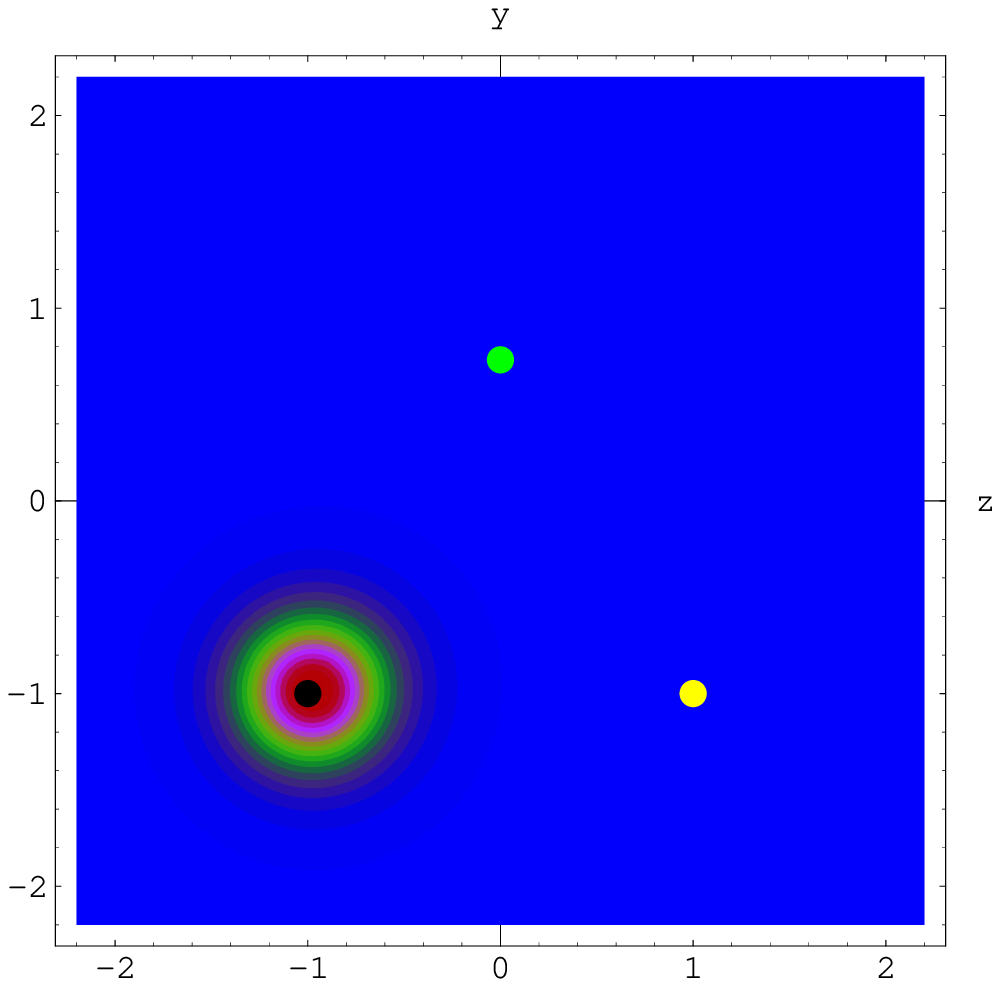,angle=0,width=4.cm}}
\centerline{
\psfig{file=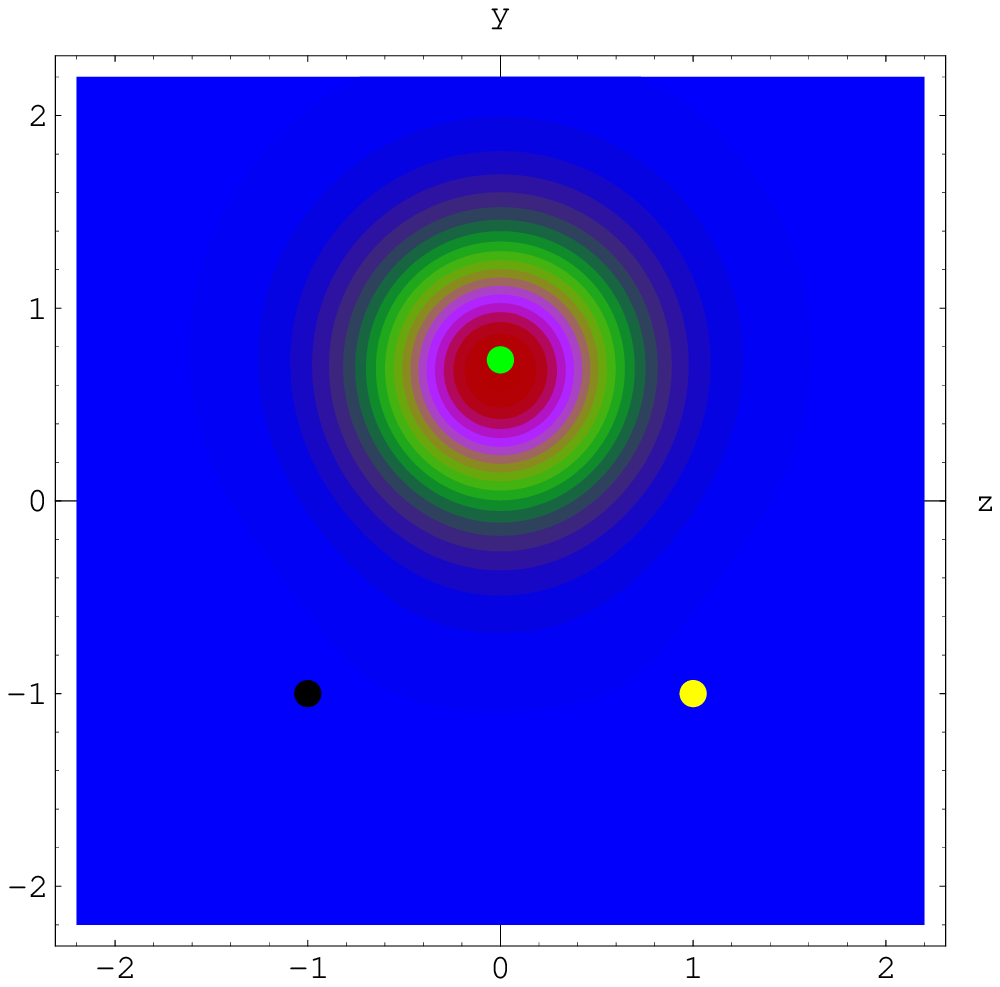,angle=0,width=4.cm}
\psfig{file=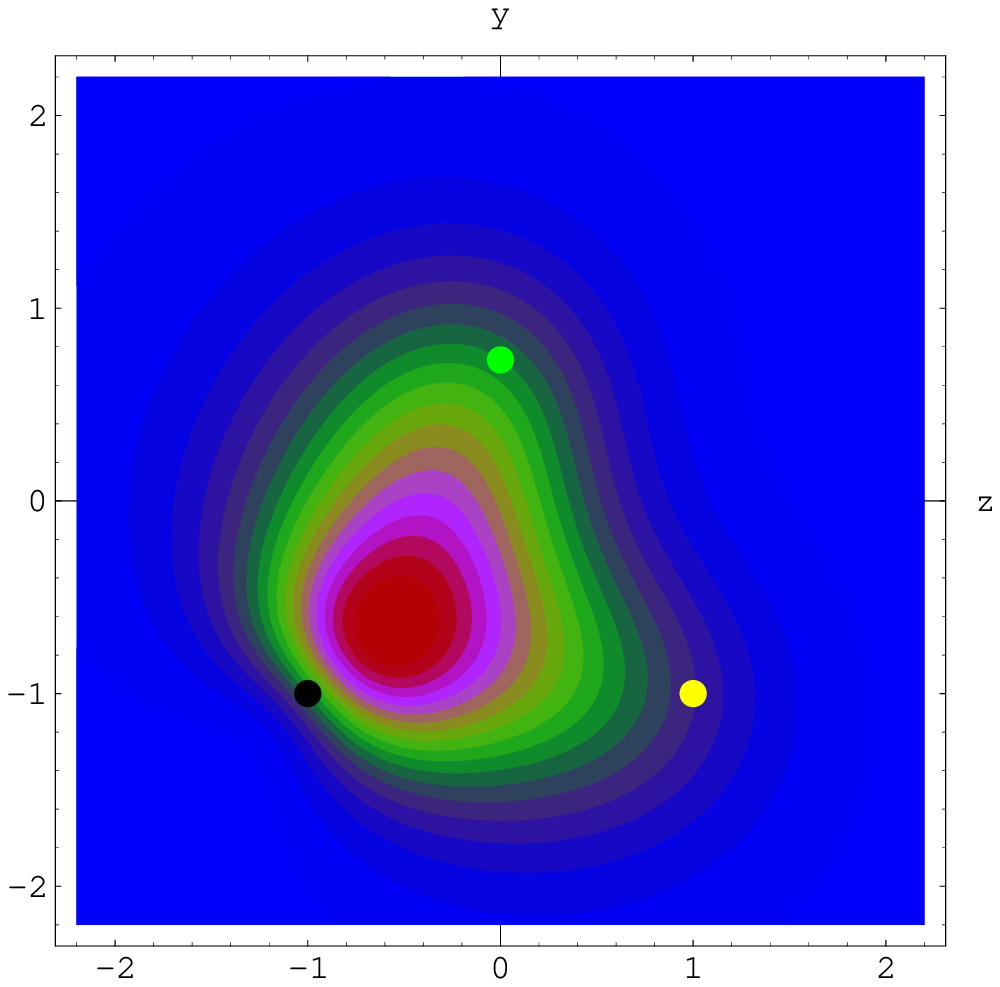,angle=0,width=4.cm}
\psfig{file=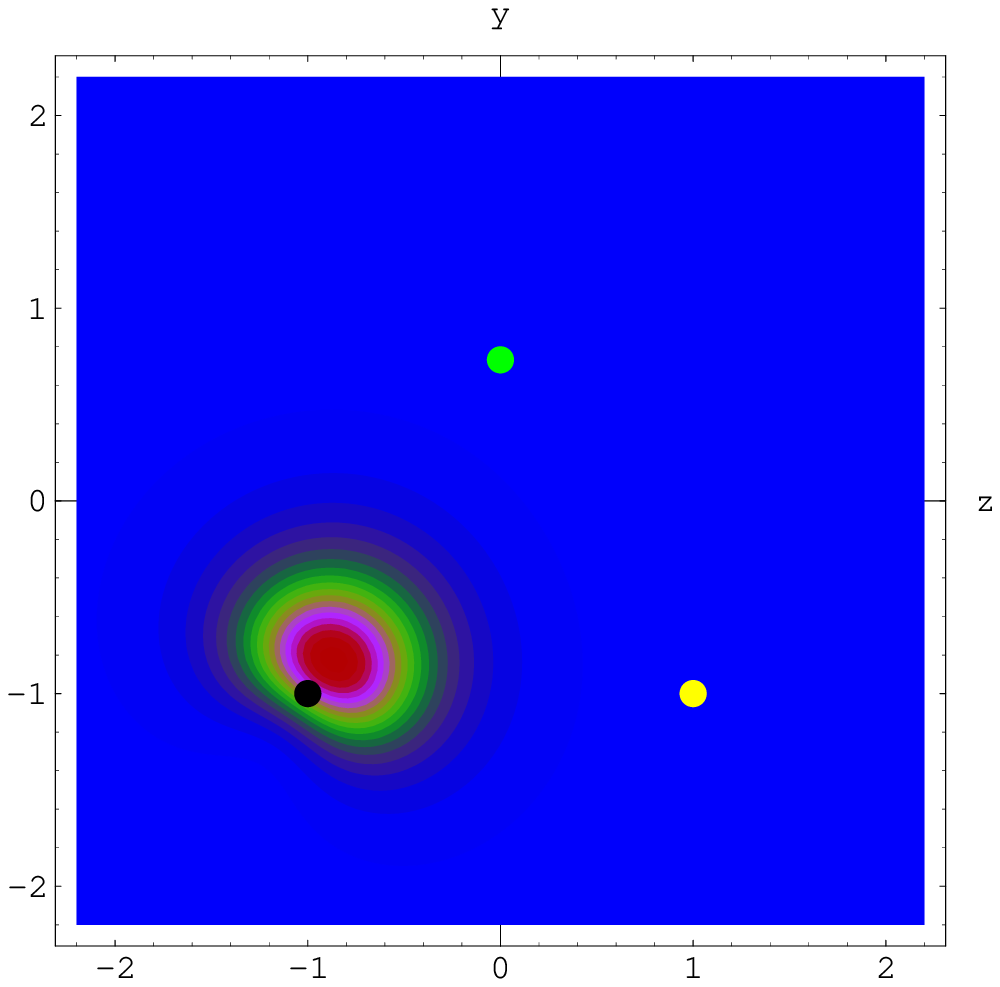,angle=0,width=4.cm}}
\centerline{
\psfig{file=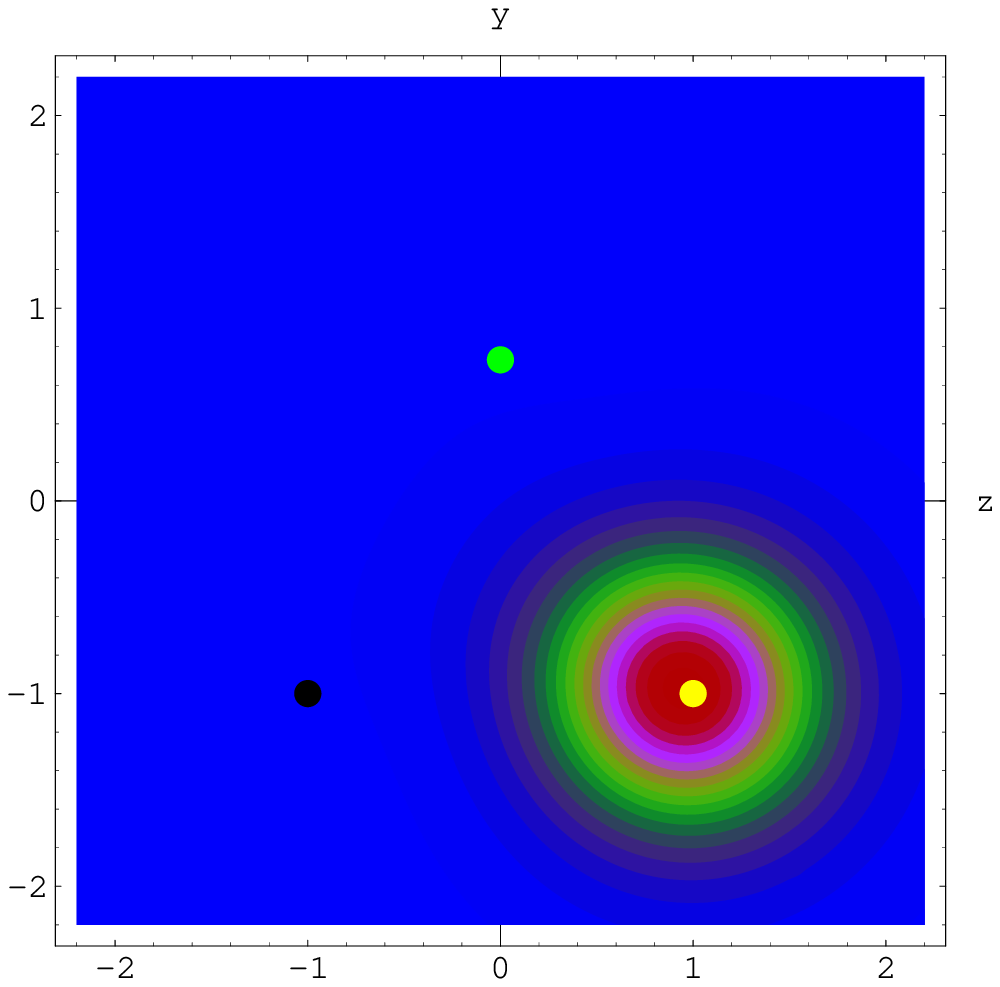,angle=0,width=4.cm}
\psfig{file=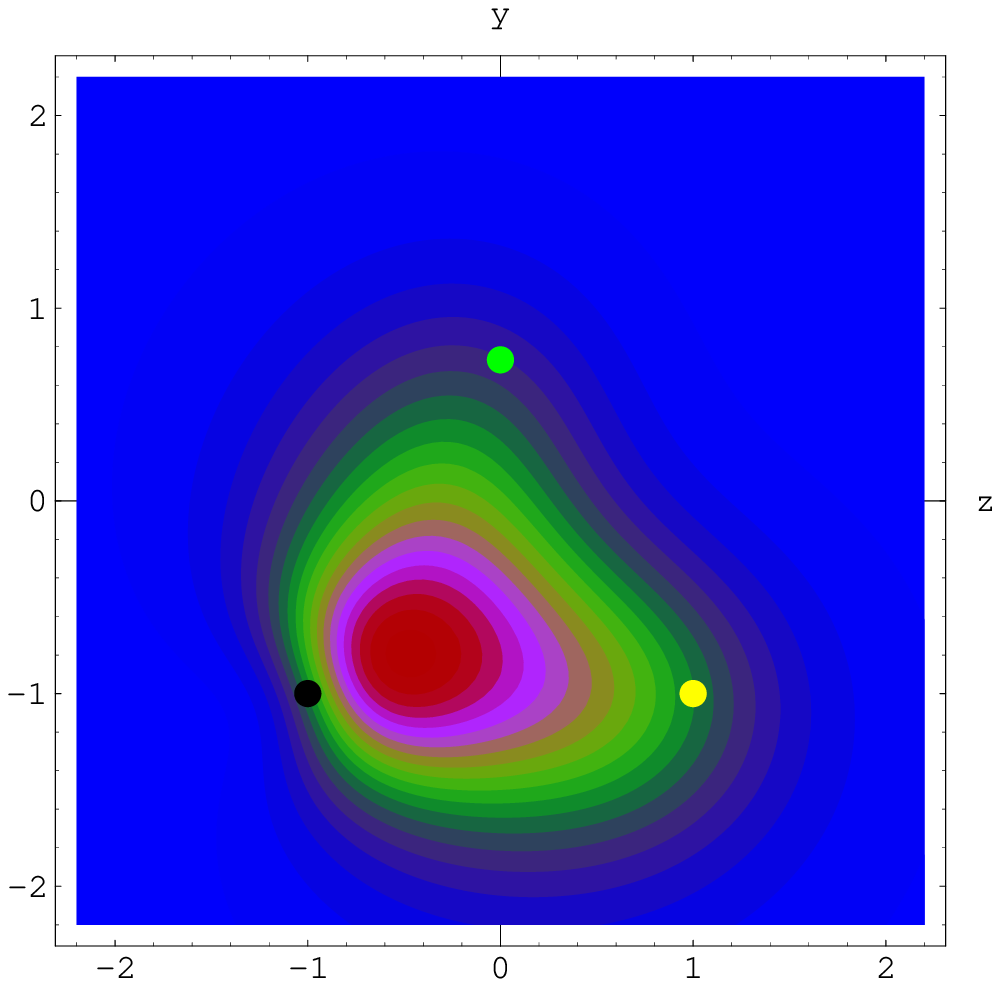,angle=0,width=4.cm}
\psfig{file=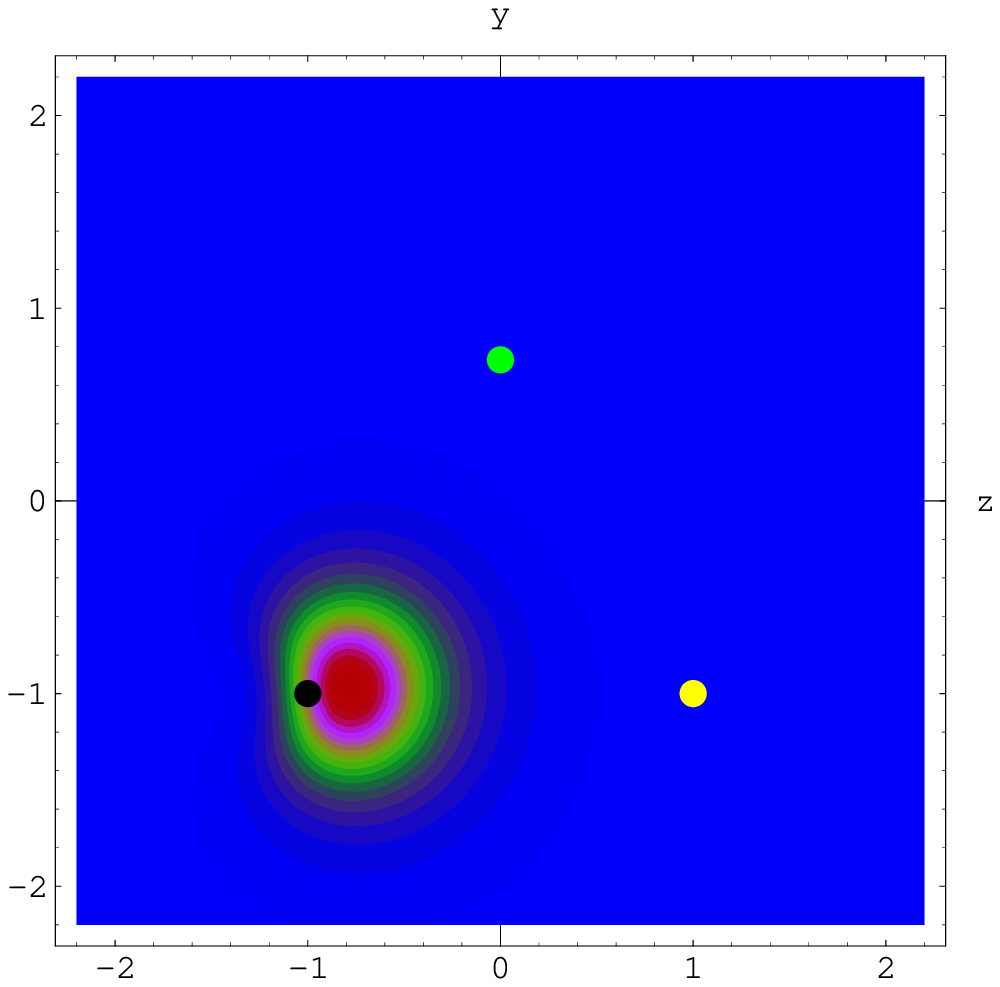,angle=0,width=4.cm}}
\caption{
Contour plots of the anti-periodic zero-mode densities on the  $x=t=0$
plane for gauge group SU(3). The three columns
display, from left to right, the zero-modes corresponding to monopole masses:
$m_1=m_2=m_3=2\pi/3$; $m_1=m_2=\pi/2$, $m_3=\pi$; and $m_1=m_2=\pi/3$, $m_3=4\pi/3$.
The three rows display the modes corresponding to  $\delta A_\mu^{(\bar a)}$
as in Eqs.  (\ref{solap1})-(\ref{antip_mode}), with, from top to bottom, $a=1,2,3$.
Monopoles are localised on the vertices of an equilateral triangle of side 2,
on the $x=t=0$ plane.  The small filled circles indicate the monopole positions, 
starting from the top monopole in each plots and in clockwise sense: 
$\vec X^1$-green, $\vec X^2$-yellow, $\vec X^3$-black.} \label{fig4}}

\FIGURE{
\centerline{
\psfig{file=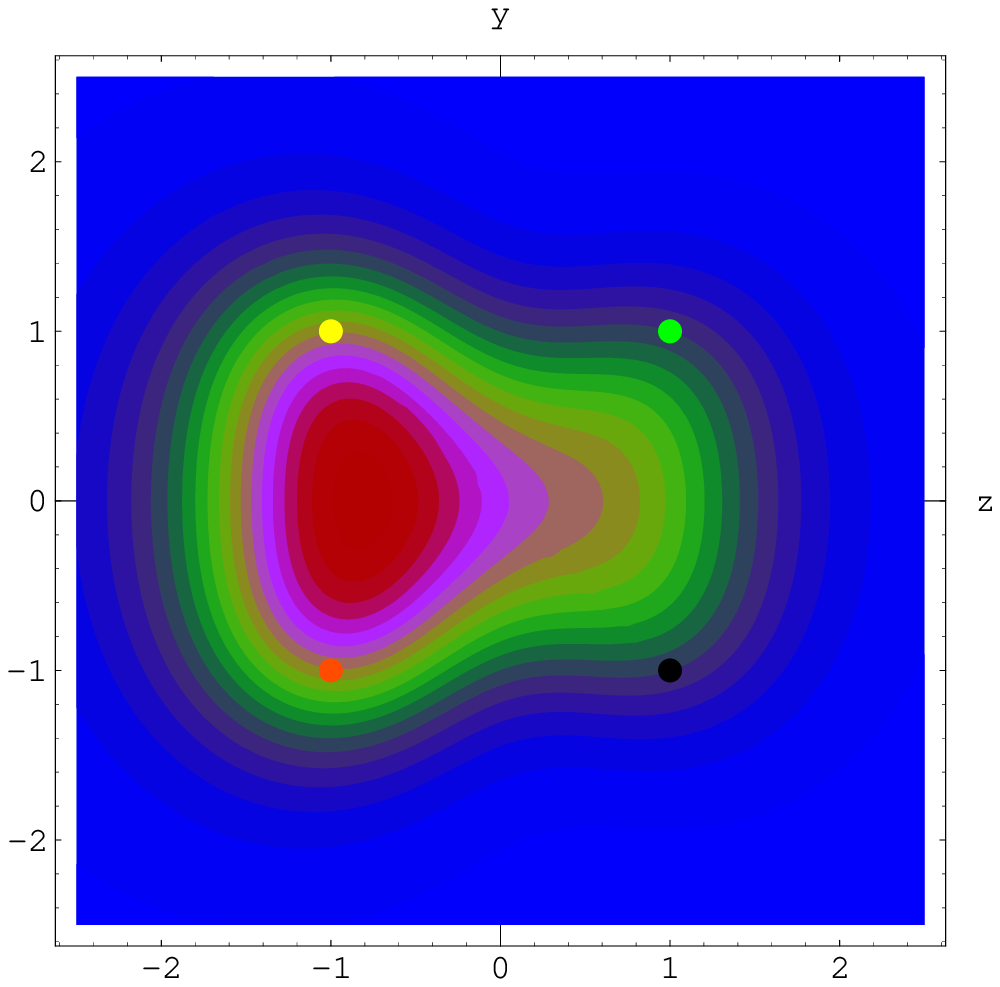,angle=0,width=6cm}
\psfig{file=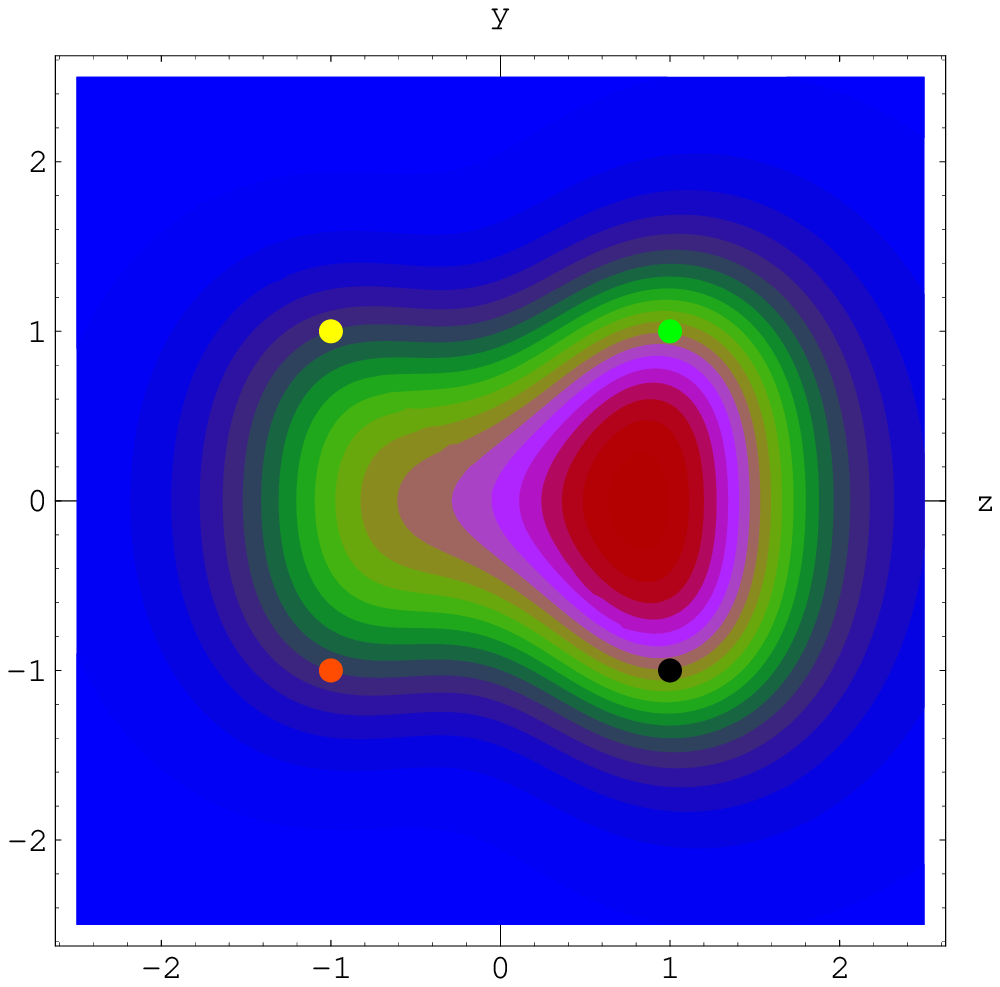,angle=0,width=6cm}}
\centerline{
\psfig{file=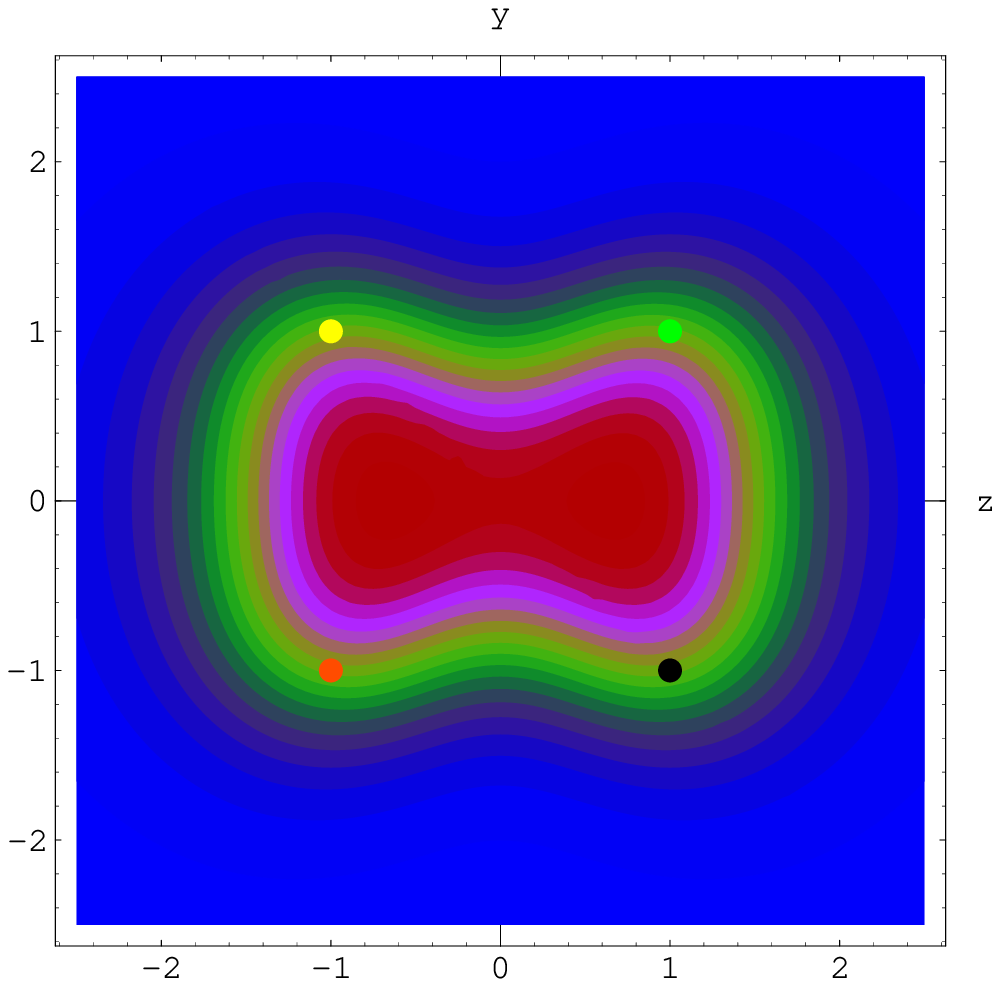,angle=0,width=6cm}
\psfig{file=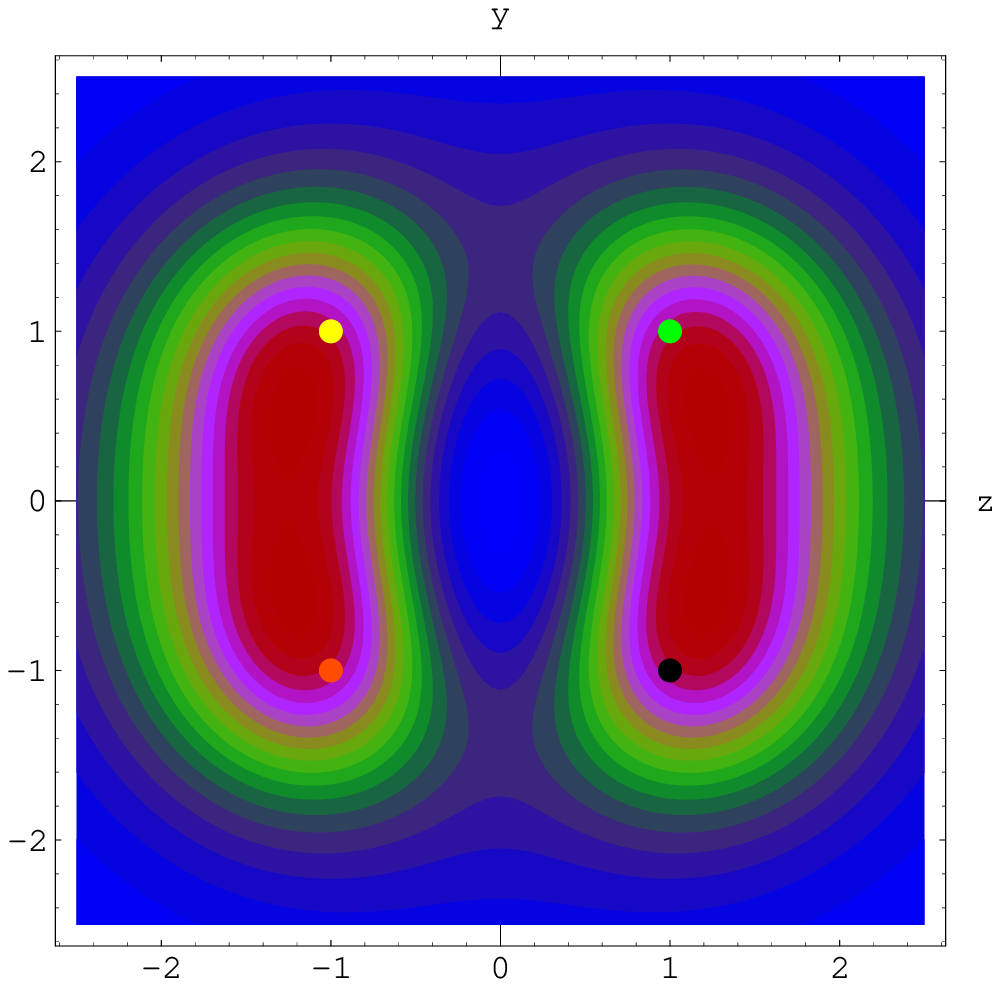,angle=0,width=6cm}}
\caption{Contour plots of the anti-periodic zero-mode densities on the
$x=t=0$ plane for SU(4) and the equal mass case. The relative positions of the monopoles are
such that $\Delta \vec X^2=-\Delta \vec X^4= -2 \hat k$.
The upper plots display $\Psi^{(\bar 1)}$ (left)
and $\Psi^{(\bar 3)}$ (right). The lower plots  display the linear combinations
$\Psi^{(\bar 1)} - \Psi^{(\bar 3)}$ (left) and $\Psi^{(\bar 1)} +
\Psi^{(\bar 3)}$
(right).
The remaining modes can be obtained by rotating each of these plots by 90 degrees.
Monopoles are localised on the vertices of a
square of side 2, on the  $x=t=0$ plane.
The small filled circles indicate the monopole positions, starting from the top-right 
monopole in each plot and in clockwise sense: $\vec X^1$-green, $\vec X^2$-black, $\vec X^3$-red, $\vec X^4$-yellow.}
\label{fig5}}

Let us start with the SU(3) case for which the three CP-pairs of solutions, i.e. 
$\Psi^{(\bar a)}\equiv \delta A_\mu^{(\bar a)}\sigma_\mu$, are given by 
Eqs.~(\ref{solap1})-(\ref{antip_mode}). 
The first thing that singles out anti-periodic as compared to periodic zero-modes is 
their localisation with respect to the constituent monopole positions. For well 
separated monopoles the  $\Psi^{(\bar a)}$ CP-pair of zero-modes has support on 
the monopole attached to the interval $[Z_{b-1},Z_b]$ containing
$Z_{\bar a}=Z_a+\frac{1}{2}$. 
This gives rise to different possibilities depending on the monopole masses.
Fig.~\ref{fig4} displays three characteristic cases corresponding to:
$m_1=m_2=m_3=2\pi/3$; $m_1=m_2=\pi/2$, $m_3=\pi$; and $m_1=m_2=\pi/3$, $m_3=4\pi/3$.
The first one is analogous to the periodic case (Fig.~\ref{fig1}) in
that each monopole supports one 
CP-pair of zero-modes. This changes, however, when one of the monopole masses 
exceeds $\pi$.  In that case the more massive monopole 
{\it attracts} the full set of anti-periodic zero-modes and the rest support none.
The intermediate case is a limit between the two situations in which 
two of the zero-modes become delocalised \footnote{
An analogous result was obtained for SU(2) in our previous paper~\cite{gluino_SU2_ap}.}. 
These results match the predictions of the index theorem      
for SU(N) self-dual configurations on $R^3\times S^1$.
We will make here a brief interlude, following Ref.~\cite{index_th}, to show that this is 
indeed the case. We will not give  the general formula for the index
but we will directly analyse the case for SU(3), relevant for the discussion of 
Figs.~\ref{fig1} and  \ref{fig4}. The reader is referred to Ref.~\cite{index_th} for the
general treatment.

The index in the adjoint representation can be written in terms of the topological
charge $Q$, the eigenvalues of 
the holonomy $Z_a$  (equivalently the monopole masses) and the magnetic charges $(M_1,\cdots,M_{N-1})$. The latter are derived from 
the asymptotic behaviour of the magnetic field at spatial infinity, i.e.
\be
\vec B(|\vec x|\rightarrow \infty) = {\vec x \over |\vec x|^3} \,{\rm diag}(n_1-n_N,n_2-n_1,\cdots,n_N-n_{N-1})\,,
\ee
with magnetic charges given by $M_i=n_i-n_N$.  
Note that one can fit both calorons and BPS monopoles into this description. For 
the latter, there are N elementary types of monopoles obtained by setting
$n_i=1$,  with $n_{j\ne i}=0$ and $i=1,\cdots,N$. The monopole that corresponds to 
taking $i=N$  is usually called the Kaluza-Klein monopole and
has magnetic charges $(-1,\cdots,-1)$. Calorons with topological charge $Q=1$
have zero magnetic charge and are obtained by setting all $n_i$ equal to one, 
reproducing therefore the picture of the caloron as a composite of $N$ BPS monopoles.

In terms of the $n_i$ and the  monopole masses the index  in the adjoint 
representation of SU(3), giving the number of periodic zero-modes,
is~\cite{index_th}:
\be
I_{\rm adj} (Q,n_1,n_2,n_3) = 2 N Q \, n_3 - \sum_{i=1}^3 n_i \,\,\Big ( 2\, [-v_i]  - 2 \,[v_i] -\sum_{i\ne j} ([-v_j]-[v_j]) \Big )\,
\label{eq:adj_index}
\ee
where $v_i=m_i/2\pi$, for $i=1,2$,  $v_3=m_3/2\pi-1$, and  
$[v] \equiv {\rm max}\{ n\in {\cal Z} \, | \, n\le v\}$.
Note that the Kaluza-Klein monopole is peculiar in that it receives a
contribution from the topological charge term. 
It is easy to check from the formula that, for monopole masses smaller than $2\pi$
and $Q=1$, each monopole supports 2 periodic zero-modes in agreement with 
the index theorem by Callias. The formula also gives a correct counting of the 
number of periodic zero-modes for the $Q=1$ caloron, i.e. 2N. To compute the number of 
anti-periodic zero-modes we have to resort again to the replica trick. After  replicating
once in time  we obtain that $Q \rightarrow 2Q$ and $v_i \rightarrow 2v_i$. Accordingly
the number of zero-modes for the replicated caloron is $4N$, out of which $2N$ are periodic
and $2N$ anti-periodic on the original torus. In what refers to monopoles, it
is easy to check that the formula for the index in the replicated torus gives: 
\begin{itemize}
\item
If $m_1=m_2=m_3=2\pi/3$, then: 
$I_{\rm adj}^R (Q=2,n_1,n_2,n_3) = 4 n_1 + 4 n_2 + (4 N - 8 ) n_3$.
Each monopole supports 4  periodic zero-modes in the replicated torus.
This implies two periodic plus two anti-periodic zero-modes in the original
configuration as observed in the corresponding plots of Figs.~\ref{fig1} and
 \ref{fig4}.
\item
If $m_1=m_2=\pi/3$, $m_3=4\pi/3$, then:
$I_{\rm adj}^R (Q=2,n_1,n_2,n_3) = 2 n_1 + 2 n_2 + (4 N - 4 ) \, n_3$.
Monopoles 1 and 2 have only two zero-modes in the replicated torus,
while the more massive monopole, attached to $n_3$, has 8 zero-modes. Out of those
2 for each monopole correspond to periodic zero-modes of the original
configuration. We obtain therefore the result advanced in Figs.~\ref{fig1} and
 \ref{fig4}:
monopoles 1 and 2 support no anti-periodic zero-modes, while 
monopole 3 carries 6 anti-periodic zero-modes.
\end{itemize}
More general situations can also be analysed using the expression for the index
given in Eq.~(\ref{eq:adj_index}).

We proceed now to present results for the gauge group SU(4). We will restrict the discussion 
to the case of equal mass monopoles
localised on the vertices of a square. Given that the monopole
relative positions are coplanar this admits special solutions as the ones discussed
in section~\ref{s.antiperiodic}. However, we  will only consider the
case having  $\Delta \vec X^2=-\Delta \vec X^4= -2 \hat k$, which
belongs to the generic situation  in which  $\mathbf{I} -  W_1\st W_1$ is invertible.
Thus,  the zero-modes are given by  Eqs.~(\ref{solap1})-(\ref{antip_mode}). 
The alternative choice having  $\Delta \vec X^2=\Delta \vec X^4= -2\hat k$, 
gives rise to $W_1\st W_1 = \mathbf{I}$.  The solutions can be obtained by the 
construction following Eqs.~(\ref{per_cond}), (\ref{per_cond2}), but
will not be presented here. 

Figure~\ref{fig5} displays the invertible case with monopoles localised on the 
vertices of a square of side 2. The fact that zero-modes are not attached to a 
single BPS monopole is related to a singularity of the equal mass case for SU(4). 
It happens that every $Z_{\bar a}$ coincides with one of the eigenvalues 
of the holonomy ($Z_b$ for some $b$). Consequently  it lies between two monopole
intervals and the zero-mode density spreads among both. The $a=1$ case
for instance has the singularity at $Z_{\bar 1}=Z_3$, therefore the density profile
should be attached to monopoles 3 and 4 and aligned with the 3-4 relative
position: $\Delta \vec X^4= 2 \hat k$. The upper-left plot of Fig.~\ref{fig5}
shows that this is indeed the case. 
Note in addition that the anti-parallelism of $\Delta \vec X^2$ and $\Delta \vec X^4$ 
corresponds to the condition for the existence of spurious solutions
with modified periodicity described at the end of section~\ref{s.periodic}. 
According to the discussion presented there 
whenever this condition holds it should be possible to find a pair of zero-modes
for which  $\hat \Psi_{11}=0$ and only the term in $\delta \tilde q$
survives. As already discussed, for equal mass monopoles these spurious solutions
do give rise to anti-periodic zero-modes. Indeed,  two such  solutions can be obtained 
from the linear combinations $\Psi^{(\bar 2)} + \Psi^{(\bar 4)}$ and
$\Psi^{(\bar 1)} + \Psi^{(\bar 3)}$. The latter and the alternative combination  
$\Psi^{(\bar 1)} - \Psi^{(\bar 3)}$  are displayed in Fig.~\ref{fig5}.

\section{Summary and Outlook}
\label{s.concl}

In this paper we have addressed the analytical  construction  of adjoint 
zero-modes of the Dirac equation in the background field of SU(N) $Q=1$ 
calorons, thus extending our previous SU(2) results~(\cite{gluino_SU2}-\cite{gluino_SU2_ap}). 
We have covered both periodic and anti-periodic boundary conditions in
time. The latter is useful for the semi-classical study of ${\cal N}=1$
SUSY Yang-Mills at finite temperature and the longstanding issues of
Supersymmetry breaking. In this paper we have made an effort to make a
complete study of the problem of adjoint zero-modes in
SU(N) Yang-Mills theory, presenting general formulas and derivations 
for the ADHM construction, the Nahm transform on the torus, and,
finally, for the caloron case. Thus, we hope this paper could serve 
as a useful reference for researchers  interested in  gluino zero-modes. 

Although the formulas for SU(N) contain those for  the SU(2) case, it is
found out that the latter case is quite exceptional in several aspects, so
that the generalisation is far from being trivial. Ultimately, given 
the generality of our formulas, we hope that they will be instrumental
in addressing several open issues, including  theoretical
problems about higher charge calorons, the finite temperature
behaviour of SUSY Yang-Mills theory, which the authors intend to address
themselves in future publications, and other applications of adjoint
zero-modes mentioned in the Introduction of this paper.

\section*{Acknowledgements}
We acknowledge financial support from  Comunidad Aut\'onoma
de Madrid under the program  HEPHACOS P-ESP-00346 and  from CICYT grants  
FPA2006-05807, FPA2006-05485 and FPA2006-05423. The authors
participate in the Consolider-Ingenio 2010 CPAN (CSD2007-00042).
We acknowledge the use of the IFT cluster for part of our numerical
results.

\appendix
\section{Appendix.}
\label{a.uandw}
In this appendix we will collect different formulas and calculations
of a more technical type needed to complete the analytical computations 
of zero-modes described in the text. We will start by generalising the
calculation of $u$ and $\OMF$ for SU(N) calorons.

\subsection{Computation of $\OMF $ and $u$}
In this subsection  we derive explicit formulas for the basic 
ADHM quantities $\OMF$ and $u$ in terms of our Nahm data. Here we will
follow the same procedure given in Ref.~\cite{gluino_SU2} for SU(2).  
Most of our formulas are generalised in a simple way, so we will 
be very sketchy in the derivation and invite the readers to consult 
that reference. 

All interesting functions belong to the space of linear combinations 
of the following $2N$ functions:
\be
\Psi^{\pm}_a(z) \equiv \Psi^{(\pm)}(z-Z_a^M,x-X^a,\frac{m_a}{4 \pi})\,,
\ee
where $m_a$ and $X^a$ are the mass and position of the a-th constituent 
monopole,  $Z^M_a$ are the midpoints of the intervals in $z$, 
$Z_a^M = (Z_a+Z_{a-1})/2$, $x$ is the  space-time coordinate and 
\bea
\Psi^{(+)}(z,x,\delta)=\chi(-\delta,\delta) \,
e^{i 2 \pi \bar{x}z}
\\
\Psi^{(-)}(z,x,\delta)=\chi(-\delta,\delta) \,
e^{i 2 \pi \hat{x}z}
\eea
Using the form of the operator $\widetilde{M}\equiv \widetilde{A}- \hat{x}\, 
\I_Q$, see Eq.~(\ref{eq:atilde}), and the equations fulfilled
by the functions $u$ and $\omeg$, i.e.
Eqs.~(\ref{eq:defu})-(\ref{eq:defdw}), it is easy to deduce that they 
can be expanded as:
\bea
u(z)&=&\sum_a u_a(z) =\sum_a (\Psi_a^{+}(z) A_a) \label{equ}\\
\OMF(z) &=& \sum_a \OMF_a(z) = \sum_a (\Psi_a^{+}(z) D_a^{(+)} + \Psi_a^{-}(z) D_a^{(-)})\\
\hat{\partial}\OMF &=& \sum_a (4 \pi i (z-Z^M_a) \Psi_a^{+}(z) D_a^{(+)}
+\Psi_a^{+}(z)
S_a^{(+)} + \Psi_a^{-}(z) S_a^{(-)}) \label{eqdw}
\eea
where the  coefficients  ($A_a$, $D_a^{(\pm)}$, $S_a^{(\pm)}$) are $2\times N$
matrix functions of space-time $x$. These can be determined by matching 
the value of the functions at the edges of the intervals to satisfy
the delta function part of the equations. For that we would need to
introduce the $2\times 2$ matrix $E'$ defined as:
\be
E^{\prime \epsilon' \epsilon}_a=
\Psi_a^{\epsilon}(z=Z^M_a+\epsilon'\frac{m_a}{4 \pi})
\ee
where $\epsilon$ and $\epsilon'$ take the values $\pm 1$. The inverse
matrix is given by
\be
E_a^{\prime -1}= \frac{g(m_a r_a) \hat{n}_a}{2 m_a r_a} \times
E_a^\dagger
\ee
where $E_a$ is a new matrix. 
The symbol  $\hat{n}_a$ stands  for a hermitian unitary traceless matrix
defined through the decomposition
\be
(\vec{x}-\vec{X}^a)\vec{\tau}= r_a \hat{n}_a
\label{eq:defna}
\ee
where $r_a$ is the distance to the corresponding constituent monopole.
The expressions also contain the function $g$:
\be
g(u)=u/\sinh(u)
\label{eq:defg}
\ee
evaluated at the product of the mass and the
distance.

The first step in the determination of the 
coefficients $D^{(\pm)}_a$ and $S_a^{(\pm)}$  will be to express them 
in terms of the values of $\OMF$ at the points separating the different 
intervals: $\OMF(Z_a)\equiv \OME _a$. 
Imposing the continuity in $z$ of the function $\OMF$ we obtain:
\bea
\OME_a &=& \sum_{\epsilon} E^{\prime + \epsilon }_a D_a^{(\epsilon)} =
\sum_{\epsilon} E^{\prime - \epsilon }_{a+1} D_{a+1}^{(\epsilon)} \\
\hat{\partial}\OME_a&=& i m_a  E^{\prime + + }_a D_a^{(+)}+ E^{\prime +
\epsilon}_a  S_a^{(\epsilon)} = -i m_{a+1}  E^{\prime - +}_{a+1}
D_{a+1}^{(+)}+
E^{\prime - \epsilon}_{a+1}  S_{a+1}^{(\epsilon)}
\eea
These  equations can be rewritten as a vector equation allowing
to solve for $D_a^{(\pm)}$ in terms of $\OME_a$ and $\OME_{a-1}$.
Using  our previous definitions  of the matrices $E_a$ and $E'_a$
we can write:
\bea
D_a^{(\epsilon)}&=& \frac{  \hat{n}_a g(m_a r_a)}{ 2 m_a r_a} (E_a^{
\dagger})^{\epsilon \epsilon'} \OME_{a+(\epsilon'-1)/2}\\
S_a^{(\epsilon)}&=& \frac{  \hat{n}_a g(m_a r_a)}{ 2 m_a r_a} (E_a^{
\dagger})^{\epsilon \epsilon'} \hat{\partial}\OME_{a+(\epsilon'-1)/2} +
\frac{ i g^2(m_ar_a)}{2 m_a r_a^2} \widetilde{E}^{\epsilon} (E_a^{
\dagger})^{+ \epsilon'} \OME_{a+(\epsilon'-1)/2}
\eea
where $\widetilde{E}^+= \cosh(m_a r_a)$ and $\widetilde{E}^-=-1$.
The  coefficient $A_a$ appearing in the expansion of $u$ can be related
to $D_a^{(+)}$ by the equation $\widetilde{M} \OMF =uF^{-1}$. Hence, we get
\be
A_a= 2 i r_a  \hat{n}_a D_a^{(+)} F
\ee

Notice that  the  functions $u_a$ and $\OMF_a$ can be regarded as the 
contribution of the a-th constituent monopole to the function $u$
and $\OMF$, since they only depend on the distance $r_a$ and the mass
$m_a$ of the corresponding monopole. Nonetheless, the  mixing and
interaction among the constituent monopoles is hidden in the 
expression of $\OME_a$, which we will now derive. The main equation
satisfied by $\OME_a$ comes from
the equation:
\be
q=\widetilde{M}^{\dagger} u
\ee
leading to
\be
\zeta_\alpha^a P_a = \frac{-1}{2 \pi i} ( E_a^{\prime  + +} A_a -
E_{a+1}^{\prime - +} A_{a+1})
\ee
where $P_a$ is an N-vector with components $\delta_{a i}$. 
To write the previous equation in a more compact form we arrange the
right hand side into an N-component column vector of $2\times N$ matrices.
The unknown $\OME_a$ are also arranged  as a column vector of the same
kind. Finally, we introduce the N-component column  vectors of quaternions
$\EV_a$ and $\ETV_a$ whose
components are $2 \times 2 $ matrices and such that the only non-zero
components are the rows $a-1$ and $a$, given by
\begin{eqnarray}
\EV^{a-1}_a=&&E_a^{\prime -+} \ ; \quad \EV^a_{a}=-E_a^{\prime ++} \\
\widetilde{e}^{a-1}_a = &-&E_a^{\prime --} \ ; \quad \widetilde{e}^{a}_a =\ \ E_a^{\prime +-}
\end{eqnarray}
Then the previous  equations can be re-written as follows:
\be
\label{zeta_eq}
\zeta = \sum_a \frac{g(m_a r_a)}{2 \pi m_a} \EV_a \EV_a^{\dagger} \OME F
\ee
The projector $\EV_a \EV_a^{\dagger}$ is an $N\times N$ matrix of
quaternions whose only non-zero components correspond to  the $a-1$ and $a$ rows
and columns.  Restricting to this non-zero  $2\times 2$ matrix we have
\be
\frac{g(m_a r_a)}{2 \pi
m_a} \EV_a \EV_a^{\dagger}\big|_{\mbox{\rm \tiny restricted}}
=\frac{g(m_a r_a)}{2 \pi
m_a} \pmatrix{\cosh(m_a r_a)
& - e^{-i m_a x_0} \cr - e^{i m_a x_0} & \cosh(m_a r_a)} +\frac{1}{2
\pi}\vec{\tau}\pmatrix{\vec{x}-\vec{X}^a & 0 \cr 0& -\vec{x}+\vec{X}^a}
\ee
The first term on the right-hand side can be completed (with zeroes) 
to an $N\times N$ complex matrix
that we will call $U_a$. The second piece when summed over ``a'' can be
rewritten in terms of the projector $\zeta \zeta^\dagger$. In this way
Eq.~(\ref{zeta_eq}) can be rewritten as
\be
\zeta= V^{-1} \OME F -\zeta \zeta^\dagger \OME F
\ee
where
\be
V^{-1}=\sum_a U_a  +
\frac{1}{2\pi}\mbox{diag}(||\vec{X}^{a+1}-\vec{X}^{a}||)
\ee
Using the definition of $F$
\be
F^{-1}=\mathbf{I}-\zeta^\dagger \OME
\ee
we finally arrive to
\be
\zeta = V^{-1} \OME
\ee
whose solution is
\be
\OME= V \zeta
\ee
The matrix $V$ is an $N\times N$ complex hermitian matrix. The
elements of $F^{-1}$ can be obtained in terms of this matrix
\be
(F^{-1})_{a b} = \delta_{ a b} -V_{a b} \zeta^\dagger_a \zeta_b
\ee

\subsection{Field strength and  periodic zero-modes}

In section \ref{s.periodic} we expressed the zero-modes in terms of 2
sets of functions $E_\mu^{(a)}$ and  $\tilde E_\mu^{(a)}$ where the
label $a$ runs from 1 to N. With the aid of the formulas of the
previous subsection one can compute these functions. 

Let us start with the computation of $E_\mu^{(a)}$  defined in
Eq.~(\ref{eq:EMUa}). Using the definitions of $u_a$ and $\OMF_a$ of the
previous subsection it can be re-expressed as
\be
E_\mu^{(a)}= \frac{i}{2}F^{-1/2} \int dz \, u_a^\dagger(z) \bar{\sigma}_\alpha
\hat{\partial}\OMF_a(z)  F^{1/2} +  {\rm h.c.}
\ee
To obtain the corresponding expression one must perform the
integration in z. Notice that all the $z$ dependence of $u_a$ and
$\OMF_a$ enters through the $ \Psi_a^{\pm}$ functions. Their integrals
can be performed analytically leading to the expression 
\be
I^{\pm}_\alpha = \int dz \Psi_b^{+ \dagger} \bar{\sigma}_\alpha \Psi_a^{\pm}=
\delta_{a b} \frac{m_a}{2 \pi} ({\cal P}^{ a\,  \alpha}_{\pm} \frac{1}{g(m_a r_a)} +
{\cal P}^{a \, \alpha}_{\mp})\, ,
\ee
One needs in addition the following integral 
\be
\widetilde{I}_\alpha = 4 \pi i \int dz  (z-Z_M^a) \Psi_b^{+ \dagger} \bar{\sigma}_\alpha
\Psi_a^{+}= \delta_{ab}\frac{m_a^2}{2 \pi} \left(\frac{g'(m_ar_a)}{g^2(m_ar_a)}\right) {\cal P}^{ a\,
\alpha}_{+}(i \hat{n}_a)\, ,
\ee
In the final  expressions we have introduced  the quaternions
\be
{\cal P}^{ a\,  \alpha}_{\pm} =\frac{1}{2} (\bar{\sigma}^\alpha \pm
\hat{n}_a \bar{\sigma}^\alpha \hat{n}_a)\, , 
\ee
where $\hat{n}_a$ and $g(u)$ have been defined in Eqs.~(\ref{eq:defna}) and 
(\ref{eq:defg}).

Using the previous formulas and those of  the first
subsection of the appendix we arrive at 
\be
\label{maineq}
E_\mu^{(a)}=
\frac{i}{2}F^{1/2}\OME^\dagger {\cal L}_{a
\alpha} \OME F^{1/2}+ \frac{i}{2}  F^{1/2} \OME^\dagger \widetilde{{\cal L}}_{a}
\bar{\sigma}_\alpha \hat{\partial}\OME F^{1/2} +  {\rm h.c.}
\ee
where 
\be
\label{Leq}
{\cal L}_{a \alpha} = -\frac{ i  m_a g(m_a r_a)}{2 \pi } \EV_{ a} \left( -i \frac{g^2(m_a
r_a)-1}{2 m_a^2 r_a^2}  {\cal P}^{ a\,  \alpha}_{+} -i 
\frac{g'(m_a r_a)}{2 m_a r_a}  {\cal P}^{ a\,  \alpha}_{-}  \right)
\EV_{ a}^\dagger 
\ee
and 
\be
\label{LPeq}
\widetilde{{\cal L}}_{a }=\frac{-i g(m_a r_a) }{4 \pi m_a r_a} \EV_{ a}
\hat{n}_a \left(  g(m_a r_a) \ETV_{a}^\dagger  + \EV_{ a}^\dagger  \right)
\ee
Using the relation between the quantity inside parenthesis in
Eq.~(\ref{Leq}) with the gauge field of a SU(2) BPS monopole we  can write 
this first term as 
\be
\sum_a E^{\rm  BPS}_i(x-X^a; m_a) I_{i}^{a}
\ee
where we have introduced the $N\times N$ hermitian matrices
$I_{i}^{a}$ given by 
\be
I_{i}^{a}= \frac{   g(m_a r_a)}{2 \pi m_a} F^{1/2}\OME^\dagger \EV_a
\tau_i \EV_{a}^\dagger \OME F^{1/2}
\ee
they provide an embedding of the SU(2) group in SU(N). 

In this way all our formulas are expressed in terms of the values of
$\OMF(Z_b)=\OME_b$. The latter can be expressed in terms of the matrix
$V$ defined in the first subsection of the appendix. The same
quantities also appear in the expression of  $\tilde E_\mu^{(a)}$, as
shown in Eq.~(\ref{Modes}).

Notice that the electric field strength is a particular case of the previous
formula, since $E_i= \sum_a E^{(a)}_i$. 

\subsection{Anti-periodic zero-mode formulas}
The corresponding expressions for the anti-periodic zero-modes are
considerably more involved, although again the integrals appearing in 
Eq.~(\ref{antip_mode}) can also be performed analytically. The main
difference with respect to  the periodic case is  that in the latter each mode
in the basis depends on a single region $(Z_{a-1}, Z_a)$ associated to a
given constituent monopole. The formula only depends on the remaining
monopole positions through the matrix $F$. In the anti-periodic case,
the expressions involve pairs of monopoles, namely those pairs such
that 
\be
(Z_{a-1} , Z_a) \cap (Z_{b-1} +\half , Z_b +\half) \ne \emptyset
\ee
The condition can be easily implemented if we introduce 
$\chi_{\bar{b}}$ as the characteristic function of the interval 
$(Z_{b-1} +\half , Z_b +\half)$. Then, one can rewrite 
\be
 \hat\Psi^a_{12}(z) \equiv \sum_{b\,  c}  \hat\Psi^{a b c }_{12}(z)= \sum_{b\,  c} \chi_b 
 \chi_{\bar{c}} \exp\{2 \pi
 \tau_i (X_i^b -X_i^c) z\} \kappa_{a b c} 
\ee
where the coefficient matrices $\kappa_{a b c}$ follow from simple
linear relations in terms of $\U_{A B}$ and $\kappa_{\pm}^A$. 
Notice then that,  due to the characteristic functions, the integrals
appearing in the expression of the anti-periodic zero-modes
Eq.~(\ref{antip_mode}) only involve 
integrals of $ \hat\Psi^{a b c }_{12}(z)$ with $u_b$ and $\OMF_c$.
Using appropriate matrix projections, these integrals reduce to those of
ordinary exponentials. The details are somewhat cumbersome and we will skip
them here. 


\end{document}